\documentclass[a4paper,useAMS,usenatbib]{mn2e}

\usepackage{graphicx}
\usepackage{cleveref}
\usepackage{float}
\usepackage{amsmath}
\usepackage{amssymb}
\usepackage{subfig}
\usepackage{wrapfig}
\usepackage{rotating}
\usepackage{array}
\usepackage{url}

\title[The origins of post-starburst galaxies]{The origins of post-starburst galaxies at $z<0.05$.}

\author[M. M. Pawlik et al.]{M. M.~Pawlik$^{1}$, L.~Taj Aldeen$^{1,2}$, V.~Wild$^{1}$, J.~Mendez-Abreu$^{1,3,4}$, N. ~Lah\'{e}n$^{5}$,\newauthor
P. H.~Johansson$^{5}$, N.~Jimenez$^{1,6}$, W.~Lucas$^{1}$, Y.~Zheng$^{1}$, C. J.~Walcher$^{7}$ \newauthor
 and K.~Rowlands$^{1,8}$ \\ 
$^{1}$School of Physics and Astronomy, University of St Andrews, North Haugh, St Andrews, KY16 9SS, U.K. (SUPA)\\
$^{2}$Department of Physics, College of Science, University of Babylon, Hillah , Babylon, Iraq \\
$^{3}$Instituto de Astrof\'{i}sica de Canarias
C/ V\'{i}a L\'{a}ctea, s/n E38205 - La Laguna (Tenerife), Spain \\
$^{4}$Departamento de Astrof\'{i}sica, Universidad de La Laguna, Calle Astrof\'{i}sico Francisco S\'{a}nchez s/n, E-38205 La Laguna, Tenerife, Spain \\
$^{5}$Department of Physics, University of Helsinki, Gustaf H\"allstr\"omin katu 2a, FI-00014 Helsinki, Finland \\
$^{6}$Unbound, Unit 18, Waterside, 44-48 Wharf Road, London N1 7UX, UK \\
$^{7}$Leibniz-Institut f\"ur Astrophysik Potsdam (AIP), An der Sternwarte 16, D-14482 Potsdam, Germany \\
$^{8}$Department of Physics $\&$ Astronomy, Johns Hopkins University, Bloomberg Center, 3400 N. Charles St., Baltimore, MD 21218, USA}

\makeatletter

\newcommand{\Rmnum}[1]{\expandafter\@slowromancap\romannumeral #1@}
\makeatother

\newcommand{\gadget}{\textsc{Gadget-3}}

\defcitealias{Johansson+2009a}{Johansson et al. 2009a}
\defcitealias{Johansson+2009b}{Johansson et al. 2009b}

\begin{document}

\parindent=15pt

\date{}

\pagerange{\pageref{firstpage}--\pageref{lastpage}} \pubyear{0000}

\maketitle

\label{firstpage}

\begin{abstract}

Post-starburst galaxies can be identified via the presence of prominent Hydrogen Balmer absorption lines in their spectra. 
We present a comprehensive study of the origin of strong Balmer lines in a volume-limited sample of 189 galaxies with $0.01<z<0.05$, $\log(\mbox{M}_{\star}/\mbox{M}_{\odot})>9.5$ and projected axis ratio $b/a>0.32$. We explore their structural properties, environments, emission lines and star formation histories, and compare them to control samples of star-forming and quiescent galaxies, and simulated galaxy mergers. Excluding contaminants, in which the strong Balmer lines are most likely caused by dust-star geometry, we find evidence for three different pathways through the post-starburst phase, with most events occurring in intermediate-density environments: (1) a significant disruptive event, such as a gas-rich major merger, causing a starburst and growth of a spheroidal component, followed by quenching of the star formation (70\% of post-starburst galaxies at  $9.5<\log(\mbox{M}_{\star}/\mbox{M}_{\odot})<10.5$ and 60\% at $\log(\mbox{M}_{\star}/\mbox{M}_{\odot})>10.5$); (2) at $9.5<\log(\mbox{M}_{\star}/\mbox{M}_{\odot})<10.5$, stochastic star formation in blue-sequence galaxies, causing a weak burst and subsequent return to the blue sequence (30\%); (3) at $\log(\mbox{M}_{\star}/\mbox{M}_{\odot})>10.5$, cyclic evolution of quiescent galaxies which gradually move towards the high-mass end of the red sequence through weak starbursts, possibly as a result of a merger with a smaller gas-rich companion (40\%). Our analysis suggests that AGN are `on' for $50\%$ of the duration of the post-starburst phase, meaning that traditional samples of post-starburst galaxies with strict emission line cuts will be at least $50\%$ incomplete due to the exclusion of narrow-line AGN.

\end{abstract}





\section{Introduction}

About one percent of the local galaxies within the Sloan Digital Sky Survey (SDSS) have optical spectra featuring unusually strong Balmer lines in absorption accompanied by weak emission lines \citep{Goto+2008, Wong+2012}. Such features have been interpreted as a signature of a rapid decrease in the star-formation activity, likely following a recent starburst \citep{DresslerGunn1983, Nolan+2007,Balogh+2005, Wild+2007, vonderLinden+2010}. In this picture, shortly after the starburst, the balance of the newly born stars in a galaxy varies on short timescales as the most massive stars come to the end of their lives and move off the main sequence. This variation is imprinted on the galaxy spectrum in the form of increasing strength of the Balmer absorption lines, which peak for stars of spectral type A. As these stars have main-sequence lifetimes between 0.1 and 1 Gyr, strong Balmer lines in absorption should indicate a starburst not older than $\sim$ 1 Gyr. A coincidental lack of measurable emission lines indicates no ongoing star-formation on a detectable level. This means that these galaxies, often referred to as post-starburst galaxies, have abruptly quenched their star formation in their recent past and could be caught in transition between the star-forming blue cloud and the quiescent red sequence. As such, they offer a unique insight into galaxy evolution and may provide a means of constraining the origin of the bimodality in the population of massive galaxies: blue star-forming gas-rich systems with prominent disks and `red and dead' gas-poor spheroids (e.g. \citealt{Strateva+2001, Kauffmann+2003b, Bell+2004, Baldry+2004, Baldry+2006, Bundy+2005}). 

The evolutionary scenario in which galaxies migrate from the blue cloud over to the red sequence is supported by observations which reveal that the stellar mass and number density of galaxies on the red sequence has doubled since $z\sim1$, during which time the mass density of the blue cloud has remained nearly constant (see e.g. \citealt{Bell+2004, Arnouts+2007, Faber+2007}).
The physical processes governing this transition have not yet been determined and it is unlikely that all star-forming galaxies follow the same pathway to the red sequence. The evolutionary path of a galaxy may be determined by a number of factors, such as its mass and structural properties or its environment. As argued by \citet{Peng+2010}, more massive galaxies are more likely to become quiescent regardless of what environment they reside in (`internal' or `mass quenching') and galaxies in denser environments are more likely to quench their star formation independent of their stellar mass (`external' or `environmental quenching'). Numerous observations reveal that the build up of the low-mass end of the red-sequence occurs at later times in the history of the Universe than that of the high-mass end (e.g. \citealt{Marchesini+2009, Moustakas+2013, Muzzin+2013}), which may simply be related to the fact that the star formation rate of low-mass galaxies declines more slowly than high-mass galaxies (e.g \citealt{Asari+2007}), or may indicate that quenching events are occurring at later cosmic times for low mass galaxies.

A popular candidate for the internal quenching mechanism in massive galaxies is feedback from the accreting supermassive black hole (active galactic nucleus, AGN) fueled by, e.g. disk instabilities \citep{Bournaud+2011} or a central bar \citep{Knapen+2000}, that can halt the star formation activity by modifying the interstellar gas conditions or expelling it in powerful galactic winds. External quenching may be driven by a variety of processes depending on the galaxy's immediate environment. In dense galaxy clusters these include: ram-pressure stripping of the cold interstellar medium \citep{GunnGott1972}, the removal of the hot gas halo or `strangulation'  \citep{Larson+1980, BaloghMorris2000} or fast encounters with other galaxies also known as `harassment' \citep{GallagherOstriker1972, Moore+1998}. In less dense environments it is more likely caused by galaxy mergers which can destroy disks in star-forming galaxies leading to morphologically and kinematically disturbed remnants that over time relax to a state characteristic of the red-sequence population (e.g \citealt{Barnes1992, NaabBurkert2003, Bournaud+2005}). Mergers of gas-rich galaxies can lead to powerful centralised starbursts followed by quenching of the star-formation, possibly also related to the AGN feedback on the interstellar medium (e.g. \citealt{Sanders+1988, Hopkins+2006, Johansson+2009a}). Alternatively quenching may occur without the presence of AGN feedback as once a galaxy acquires a spheroid-dominated morphology it can shut off its star-formation and turn red due to disk stabilisation against gas clouds fragmentation (`morphological quenching', \citealt{Martig+2009}). Discriminating between the different mechanisms driving the evolution of galaxies from blue to red is not trivial, especially since their relative importance is unlikely to have been constant over cosmic time. 

It is clear that a single class of galaxies will not hold the answers to all questions regarding the complex picture of galaxy evolution but building up our knowledge about the galaxies caught in transition between the main evolutionary stages, such as post-starburst galaxies, is a step in the right direction. 
One of the main challenges in studying post-starburst galaxies, perhaps apart from their scarcity, is the large diversity of selection criteria used in the literature. In what follows, we review the different selection methods, the corresponding post-starburst sample properties, and conclusions about their origin and importance for galaxy evolution.

\subsection{Quiescent post-starburst galaxies}

The first observation of post-starburst galaxies or `K+A' galaxies (here, \emph{quiescent} post-starburst galaxies) goes back to the early 1980s \citep{DresslerGunn1983}, when they were found in distant galaxy clusters ($0.3<z<0.6$). Further observations showed that at these intermediate redshifts `K+A' galaxies reside preferentially in such high density environments (see e.g. \citealt{Poggianti+2009} and the references therein). Morphological analysis of their optical images revealed that they are predominantly disk-dominated systems, some of which are interacting or obviously disturbed (e.g. \citealt{Couch+1994,Couch+1998, Dressler+1994, Oemler+1997, CaldwellRose1997, Dressler+1999}). Evidence of disk-like structures was also found in the kinematics of some cluster post-starburst galaxies, e.g. by \citet{Franx1993} or, more recently, in integral-field spectroscopic observations where kinematical configurations characteristic of fast rotators \citep{Emsellem+2007} were found in over $80\%$ of the studied cluster post-starburst galaxies (\citealt{Pracy+2009, Swinbank+2012, Pracy+2013}). The proposed mechanisms for the origin of `K+A' galaxies in dense environments include perturbations due to the cluster tidal field \citep{ByrdValtonen1990}, repeated encounters with other galaxies - `harassment' - \citep{Moore+1996, Moore+1998} which could induce  disturbance in galaxy morphology, or interactions with the intra-cluster medium of newly infalling galaxies to the cluster \citep{GunnGott1972}, leaving the stellar morphology undisturbed and possibly explaining the high incidence of disks in the galaxy samples. It has also been suggested (see e.g. \citealt{Poggianti+1999, Tran+2003}) that intermediate-redshift cluster post-starburst galaxies could be the progenitors of S0 galaxies that dominate the cores of present-day clusters, therefore playing a significant role in the evolution of the star-forming galaxy fraction in clusters over cosmic time (Butcher-Oemler effect, \citealt{ButcherOemler1984}). 

Other studies revealed that quiescent post-starburst galaxies are not exclusively related to clusters but can also be found in lower-density environments. In the local Universe, they are generally found in the field and loose groups, where dynamical conditions are more favourable for galaxy interactions and mergers (e.g \citealt{Zabludoff+1996, Blake+2004, Hogg+2006, Yang+2008, Yan+2009, Goto2007}). 
A connection to mergers is also revealed in the morphology and structural properties of many quiescent post-starburst galaxies, although the outcomes of different studies are quantitatively diverse. Morphological disturbance signifying an ongoing or past merger has been found in between $15\%$ and $70\%$ of cases, depending on selection criteria and image quality (e.g. \citealt{Zabludoff+1996, Goto2005, Yang+2008, Pracy+2009, Trouille+2013}). Many studies have reported a high incidence of bulge-dominated early-type morphologies and steep light profiles with high central concentration in post-starburst samples (e.g. \citealt{Goto2005, Quintero+2004, Tran+2004, Blake+2004, Poggianti+2009, Mendel+2013}), characteristics typical of red-sequence galaxies and also consistent with merger remnants seen in numerical simulations (e.g \citealt{ToomreToomre1972, Barnes1988, NaabBurkert2003}). Evidence for a merger origin has also been found in spatially-resolved studies of some post-starburst galaxies, which revealed centrally concentrated A/F stellar populations \citep{Pracy+2012, Swinbank+2011, Whitaker+2012}, in agreement with expectations of gas inflows to the centre of the merger seen in simulations.

Despite the diversity in the findings regarding the properties of the `K+A' galaxies, many studies agree that they are likely transitioning between the blue cloud and the red sequence, both in clusters and in the field. Many lie in the `green valley' of colour-magnitude diagrams (e.g. \citealt{Wong+2012}). \citet{Tran+2004} estimated that $\sim25\%$ of passive galaxies in the local field underwent a `K+A' phase at $z<1$ (increasing to $70\%$ if only early morphological types are considered) and \citet{Whitaker+2012} argued that their number density evolution of the `K+A' and red-sequence galaxies is consistent with all quiescent galaxies experiencing a `K+A' phase at $z>1$. However, other studies present a contrasting view of the role of post-starburst galaxies in the red-sequence growth. Reservoirs of both neutral and molecular gas have been found in over half of the investigated post-starburst galaxies \citep{Chang+2001, Buyle+2006, Zwaan+2013, French+2015}, meaning that these galaxies are not yet  devoid of fuel for star formation (although such conclusions are still limited to small samples). Furthermore, the low incidence of `K+A' galaxies in two clusters at $z\sim0.5$ found by \citet{DeLucia+2009} seems insufficient to represent a dominant channel for the formation of red sequence galaxies. A similar conclusion was arrived at by \citet{Dressler+2013}, who proposed that the majority of the `K+A' galaxies in both clusters and the field at $0.3<z<0.54$ represent a phase in an evolutionary cycle within the red sequence, where an already quiescent galaxy accretes a smaller gas-rich companion and passes through a brief post-starburst phase before returning to the red sequence. This was also supported by a morphological analysis of the sample by \citet{Abramson+2013}.
At somewhat higher redshifts ($0.47<z<1.2$) \citet{Vergani+2010} concluded that a variety of processes could lead to the post-starburst phase and that this channel provides a non-negligible contribution to the red sequence growth, although not higher than $\sim10\%$.

\subsection{Transitioning post-starburst galaxies}

All the studies mentioned above relate to post-starburst galaxies in which the star-formation has effectively been quenched, selected based on the lack of measurable nebular emission, usually the [OII] line (e.g., \citealt{DresslerGunn1983, Zabludoff+1996, Poggianti+1999}) or $H\alpha$ line (e.g., \citealt{Goto+2003, Quintero+2004, Balogh+2005}). However, a starburst is not an instantaneous event and, in fact, gas-rich merger simulations (which do not include significant AGN feedback) point to ongoing star-formation for several hundreds of Myr following the initial starburst (see e.g. \citealt{Hopkins+2006, Wild+2009}). It is therefore reasonable to expect some levels of star-formation to be visible for a while after the onset of the starburst. In light of this, one problem with the traditional definition of post-starburst galaxies is that \emph{the strict cut on emission lines excludes galaxies in the early transition stage between starburst and quiescence}. This early phase is important because the characteristics of the transitioning galaxies may contain information about the event that triggered the transformation and processes occurring during the transition, and these characteristics may fade by the time the galaxy enters the `K+A' phase. For example,  \citet{Tremonti+2006} measured high velocity outflows in very young (75-300\,Myr) post-starburst galaxies which appear to be caused by extreme starbursts rather than AGN as originally postulated \citet{SellTremonti2014}. Moreover, \emph{the strict cut on nebular emission lines in the traditional approach does not allow for selection of galaxies with ionisation mechanisms other than star-formation, e.g. AGN or shocks, leading to incomplete samples of post-starburst galaxies} (see also \citealt{Wilkinson+2017}). More recent studies turned their focus to an alternative broader definition of post-starburst galaxies, in which the condition of quiescence is relaxed. These studies, (examples discussed below), revealed the existence of galaxies whose optical spectra feature both strong Balmer absorption lines as well as nebular emission lines on a measurable level. 

However, it is important to note that the nature of galaxies with both strong Balmer lines in absorption and nebular emission lines is still under debate. One of the popular interpretations is that emission-line spectra with strong Balmer absorption, known as $e(a)$ spectra, indicate an \emph{ongoing} starburst with the youngest stellar populations obscured by dust \citep{Dressler+1999, Poggianti+1999}. This is supported by some observations which reveal that such spectral characteristics are more common among dusty starbursts and luminous infrared galaxies compared to normal star-forming galaxies \citep{LiuKennicutt1995, Smail+1999, PoggiantiWu2000}. Some studies have suggested that these ongoing dusty starbursts may be the progenitors of some quiescent post-starburst galaxies \citep{Poggianti+1999, Balogh+2005}. 

A second interpretation is that galaxies with strong Balmer lines in absorption and measurable emission lines are true post-starburst galaxies in which the star formation has not been fully quenched. Such decaying starbursts are on the way to becoming traditional `K+A' galaxies.  This evolutionary scenario was explored by \citet{Wild+2010} who used a Principal Component Analysis of galaxy spectra (PCA, \citealt{Wild+2007}), and quantified the shape of the continuum around 4000{\AA} and the relative strength of Balmer absorption lines to identify post-starburst galaxies \emph{without placing a cut on their emission-lines}. They selected a sample of 400 local galaxies whose spectral characteristics place them on an evolutionary sequence stretching over 600 Myr following the starburst and, from the decay in the $H\alpha$ emission, they found a characteristic star-formation decline timescale of $\sim300$ Myr in broad agreement with merger simulations. A morphological analysis of the images of these post-starburst galaxies revealed the presence of asymmetric faint tidal features in the outskirts of about half of the youngest subset ($t_{SB}<$100 Myr) and a clear decline in the incidence and asymmetry of such features with the starburst age over the following 500 Myr \citep{Pawlik+2016}. The same study found that the post-starburst galaxies have generally intermediate structural properties between those characteristic of normal star-forming and quiescent galaxies, with no significant structural evolution detected during the first 600 Myr following the starburst. \citet{Rowlands+2015} studied 11 galaxies at $z\sim0.03$ spanning the age sequence of $\sim1$ Gyr  from the onset of the starburst, finding a decrease in the molecular gas surface density and effective dust temperature with increasing starburst age. However, the gas and dust fractions were found to be higher than in red-sequence galaxies even 1 Gyr following the starburst. The monotonic trends in the star-formation rate, gas and dust conditions and visual morphology of the post-starburst galaxies with estimated starburst age speak in favour of an evolutionary link between the ongoing starbursts, transitioning post-starburst galaxies with measurable nebular emission, and quiescent `K+A' galaxies. 

A class of transitioning post-starburst galaxies was also studied by \citet{Yesuf+2014} who combined the traditional criterion of strong $H\delta$ absorption with a more relaxed criterion on the $H\alpha$ emission line, as well as galaxy colours and flux density ratios in the NUV-optical-IR regime to bridge the gap between starburst and quiescence at $z<0.1$. They found that at  $10.3 < \mbox{log}(\mbox{M}_{\star}/\mbox{M}_{\sun}) < 10.7$ the candidate transitioning post-starburst galaxies (with detectable emission lines) are five times as numerous as quiescent post-starburst galaxies and that their structure and kinematics are intermediate between those of blue cloud and red sequence galaxies. 


Transitioning post-starburst galaxies have also been found at higher redshifts where, similarly to `K+A' galaxies, they are more numerous compared with the local Universe. Using the above described PCA-based selection method, \citet{Wild+2009} reported an increase in the mass density of the post-starburst galaxies (both transitioning and `K+A') more massive than $\mbox{log}(\mbox{M}_{\star}/\mbox{M}_{\sun})=9.75$ by a factor of 200 between $z\sim0.07$ and $z\sim0.7$. They found that post-starbust galaxies selected with no emission-line cut are found across all environments with no significant difference in the distribution of local densities compared with control samples at $0.5<z<1.0$.
More recently, \citet{Wild+2016} used a Principle Component Analysis of the broad-band optical-NIR SED (supercolours, \citealt{Wild+2014}, see also \citealt{Maltby+2016}) to study the evolution of post-starburst galaxies from even earlier epochs and found that at $\mbox{log}(\mbox{M}_{\star}/\mbox{M}_{\sun})>10$ their fraction rises from $<1\%$ to $\sim5\%$ of the total galaxy population between $z\sim0.5$ and $z\sim2$. Based on the comparison of the mass functions of the post-starburst and red-sequence galaxies, they argue that rapid quenching of star formation can account for all of the quiescent galaxy population, in the case where the timescale for visibility of the post-starburst spectral features in broad band photometry is not longer than $\sim250$ Myr. 
A similar analysis by \citet{Rowlands+2017} using spectroscopic surveys at lower redshifts found that the importance of post-starburst galaxies (defined using spectral PCA, as in \citealt{Wild+2007}) in the build up of the quiescent galaxy population declines rapidly with decreasing redshift and may be insignificant by $z=0$.

\subsection{AGN and shocks in post-starburst galaxies}

Observationally, post-starburst galaxies with emission lines have also been linked with the presence of an AGN and shocks. This was not seen in the early works on post-starburst galaxies which employed emission-line cuts in their selection, as their samples were biased against objects with any kind of ionising sources, including AGN (particularly if the [OII]-line was used as the star formation indicator - see \citealt{Yan+2006}).
However, the connection was observed in numerous studies of AGN hosts. For example \citet{Kauffmann+2003a} argued that strong $H\delta$ lines in absorption are more common in luminous narrow-line AGN than in star-forming galaxies at $0.02<z<0.3$, and \citet{CidFernandes+2004a} found high-order Balmer absorption lines in the nuclear SED of nearly a third of their local low-luminosity AGN sample. Using a conservative $H\alpha$ emission-line cut to select quiescent post-starburst galaxies at $z<0.1$, \citet{Yan+2006} showed that most of them have AGN-like emission-line ratios. \citet{Goto2006} selected a sample of over 800 $H\delta$-strong AGN hosts and used resolved spectroscopy for three such objects at $z<0.1$ to reveal a spatial connection between the post-starburst region and the AGN. \citet{SellTremonti2014} find evidence for AGN activity in 50\% of their extreme post-starburst galaxies at $z\sim0.6$.

\citet{Wild+2007} used their PCA selection method to show that, at low redshift, AGN reside in over a half of the studied post-starburst galaxies and that, on average, they are the most luminous AGN within their samples.
AGN hosts with post-starburst characteristics may be essential to understanding the causal connection between starbursts and AGN activity, and consequently that between star-formation and black hole growth in galactic centres, as well as give a unique insight into the process of star formation quenching.
To that end \citet{Wild+2010} measured a delay between the starburst and AGN activity of about 250 Myr at $0.01<z<0.07$. A similar time delay was found following a different selection technique by \citet{Yesuf+2014} who concluded that AGN are not the primary source of quenching of starbursts, but may be responsible for quenching during the post-starburst phase (see also \citealt{Goto2006} and \citealt{Davies+2007} for time delays between peaks of starburst and AGN activity found in small AGN samples). 

Finally, it is important to note that, aside from star formation and AGN activity, the emission-line features in galaxy spectra may indicate other underlying processes, such as shocks. These are expected to be seen in transitioning galaxies, where the transition is attributed to violent dynamical mechanisms, likely to induce turbulence in the interstellar medium. \citet{Alatalo+2016} built a catalogue of `shocked' post-starburst galaxies or SPOGs, with emission line ratios indicative of the presence of shocks. Such shocks could be related to a number of physical mechanisms, including AGN-driven outflows, mergers or proximity to a cluster (for details see \citealt{Alatalo+2016} and references therein).

\subsection{Summary and goals of this work}

Many papers interpret post-starburst galaxies as a transition phase between the star-forming blue cloud and the quiescent red sequence blue cloud- the two major stages of galaxy evolution. However, their true importance for red sequence growth remains a matter of debate. The aim of this paper is to investigate the star-formation histories, visual morphologies, structural properties and environments of galaxies with strong Balmer absorption lines, and a range of emission line properties in order to determine their origins. We also aim to ascertain whether the different classes of Balmer strong galaxies are evolutionarily connected or following separate paths entirely.


This paper is organised as follows: Section 2 describes the samples and their selection criteria; Section 3 - the methodology used to obtain star formation histories, morphology, structure and environment; Section 4 - the results; Section 5 - a discussion, including the analysis of galaxy merger simulations, and focusing on the likely origin of the different post-starburst families and evolutionary pathways through the post-starburst phase; Section 6 - the summary of conclusions. We adopt a cosmology with $\Omega_{m}$ = 0.30, $\Omega_{\Lambda}$ = 0.70 and H$_{0}$ =70kms$^{-1}$Mpc$^{-1}$ and magnitudes are on the AB system. 



\section{Data and sample selection}

The spectroscopic catalogue of the SDSS (7th Data Release, SDSS DR7, \citealt{Abazajian+2009}), containing the optical spectral energy distributions (SED) of $\sim90,000$ galaxies, is a natural choice for selection of objects as rare as low redshift post-starburst galaxies.
In our study, we made use of both spectroscopic and imaging data provided by the survey. Additionally, we utilised the information regarding several spectral lines available in the SDSS-MPA/JHU value added catalogue\footnote{\url{http://www.mpa-garching.mpg.de/SDSS}}. 
The measurements of the Petrosian magnitudes and redshifts of the galaxies were taken directly from the SDSS catalogue and the stellar masses of the galaxies were measured from the five-band SDSS photometry  (J. Brinchmann, SDSS-MPA/JHU) using a Bayesian analysis similar to that described in \citet{Kauffmann+2003a}. Importantly, this method allows for bursty star formation, varying metallicity and 2-component dust attenuation.

The selection of the post-starburst galaxies as well as control galaxies with ordinary star formation histories was done based on their spectral characteristics. We note the relevance of the widely known aperture bias issue. At the low redshifts considered in this work the fixed 3\arcsec\ aperture diameter of the SDSS fibers probes only the central $\sim 0.6-3$ kpc of massive galaxies. This means that the resulting spectra and all derived quantities may be limited to the central regions of our galaxies and further investigation using spatially resolved spectroscopy is required to investigate the spectral characteristics of these galaxies on a global scale. However, the investigation of such ``central" post-starburst galaxies is still of significant interest, not least because in merger simulations the funnelling of gas to the central regions of the merger remnants leads to central starbursts which may be exactly the objects we are identifying in the observations. In what follows, we first introduce the basics of the technique adopted for the classification of the galaxy SED and then describe the criteria used to select the galaxy samples.

\subsection{Spectral analysis}\label{sec:spectralanalysis}

To distinguish post-starburst galaxies from those with other star-formation histories we used the Principal Component Analysis introduced by \citealt{Wild+2007}\footnote{\url{http://www-star.st-and.ac.uk/~vw8/downloads/DR7PCA.html}} - a multivariate analysis technique that combines features that vary together in a data set, in this case the optical spectra of galaxies. 
Regarding a single spectrum - traditionally a 1D array of $n$ flux values - as a single point in an $n$-dimensional space, we can visualise a collection of spectra as a cloud of points in $n$-dimensions. The principal components are the orthogonal vectors corresponding to the lines of greatest variance in the cloud of points, and they constitute the new basis onto which the galaxy spectra are projected upon. 
The components were calculated using a set of mock spectra created using the \citet[][BC03]{BruzualCharlot2003} spectral synthesis code, and therefore contain only stellar light.
As this work is focused on post-starburst galaxies, the spectral analysis was limited to the Balmer line region of the galaxy spectra, specifically 3175-4150{\AA}. Within that region the first two principal components contain information about: PC1 -  the 4000{\AA}-break strength, anti-correlated with Balmer absorption-line strength, which gradually increases with increasing mean stellar age; PC2 - the \emph{excess} Balmer absorption above that expected based on the measured 4000{\AA}-break strength, which identifies unusual `bursty' star formation histories.

The position in the PC1-PC2 parameter space depends on the stellar content and therefore the current and past star-formation activity of a galaxy. Galaxies with the highest specific star-formation rate are located on the left side of the distribution. Moving towards higher values of PC1 we find galaxies dominated by subsequently older stellar populations and lower specific star-formation rates.
PC2 traces the recent star-formation history of the galaxies. Due to the short lifetimes of the most massive O/B stars, the stellar content of galaxies changes rapidly after a starburst, and after about 1 Gyr the galaxy enters an evolutionary sweet spot where A/F stars dominate its energy output. This evolution is imprinted on the galaxy SED as the A/F stars are characterised by the strongest Balmer lines among all stellar types. 
Therefore we can select robust samples of $\sim$ 1Gyr-old post-starburst galaxies from the `bump' at the top of the distribution in PC1-PC2.

This selection method does not require any emission-line cut and therefore is suitable for the selection of both the traditional post-starburst galaxies in which the star-formation has been quenched, as well as those with detectable emission caused by either ongoing star-formation or other ionisation mechanisms, such as AGN or shocks.

\begin{table*}
 \caption{Balmer-strong galaxy counts in the final samples and a summary of the sample selection criteria (see Section \ref{sec:spectralanalysis} for details). Cuts that were omitted during selection of particular samples are marked as not applicable (N/A) - note this does not mean that a given feature is not present in the resulting sample. The numbers in brackets in columns 2 and 3 correspond to galaxy counts in the reduced `clean' samples, i.e. they exclude galaxies for which the SDSS images are contaminated sue to the presence of a nearby bright source (as explained in Section \ref{sec:res_morph}). These samples are of particular importance for the interpretation of the results of automated image analysis of the galaxies.}
 \label{tab:galcounts}
 \begin{tabular}{|c||c|c|cccc|}
 \hline
Sample & Counts & Counts &  Balmer-strong &  H$\alpha$ & Dusty & AGN  \\
& $\mbox{M}_{\star}/\mbox{M}_{\sun}< 3\times10^{10}$ & $\mbox{M}_{\star}/\mbox{M}_{\sun}> 3\times10^{10}$ & (PC1-PC2) & emission & (Balmer decrement) & (\citealt{Kewley+2001}) \\
\hline
\hline
qPSB & 36 (24) & 5 (5) & Yes & No & N/A & No \\
agnPSB & 33 (26) & 5 (5) & Yes & N/A & N/A & Yes \\
ePSB & 57(43) & 10 (6) & Yes & Yes & No & No \\
dPSB & 31(23) & 12 (10) & Yes & Yes & Yes & No \\
\hline
 \end{tabular}
\end{table*}

\begin{figure*}
\centering
  \centering
  \includegraphics[scale=0.54]{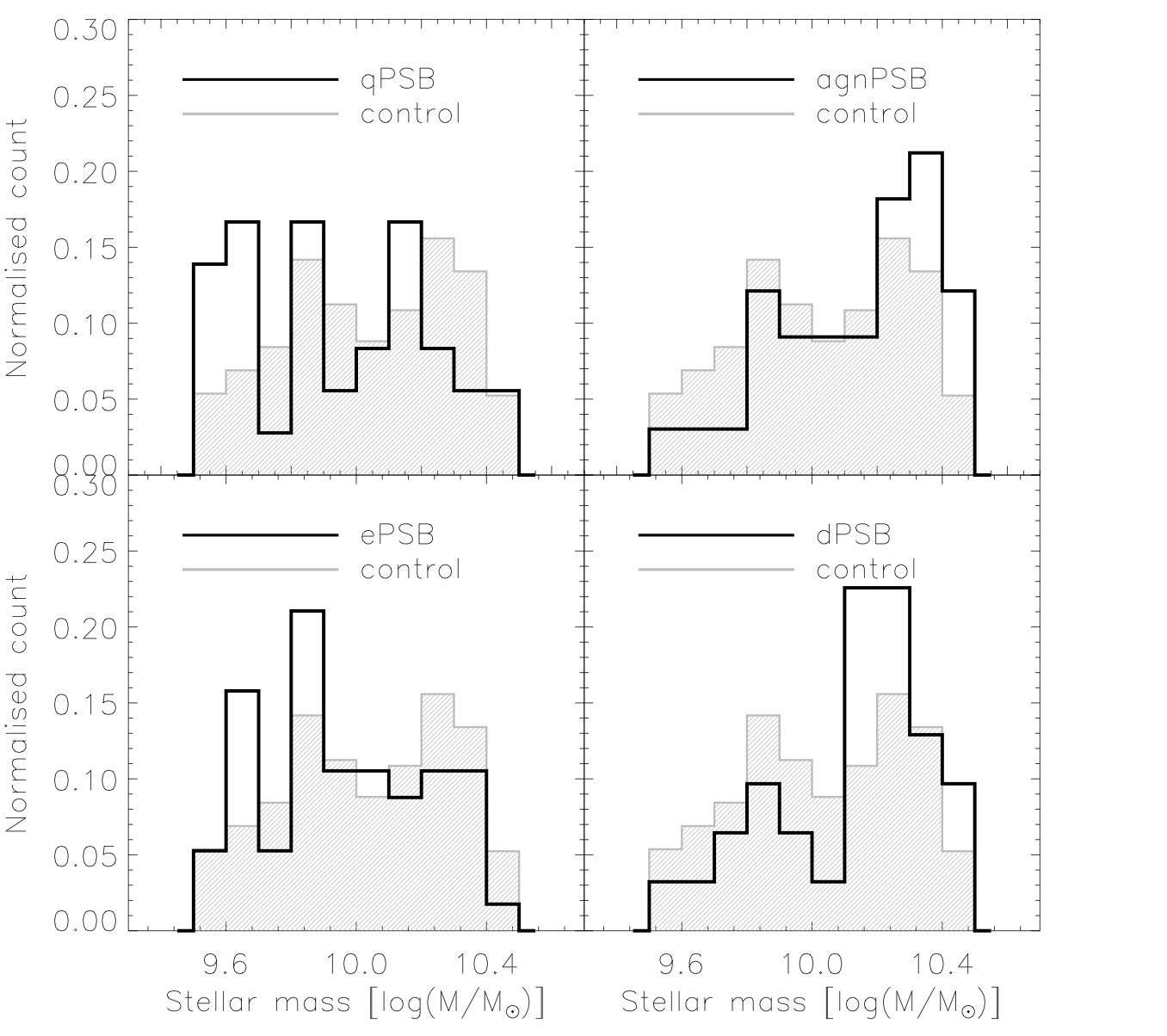}
    \includegraphics[scale=0.54]{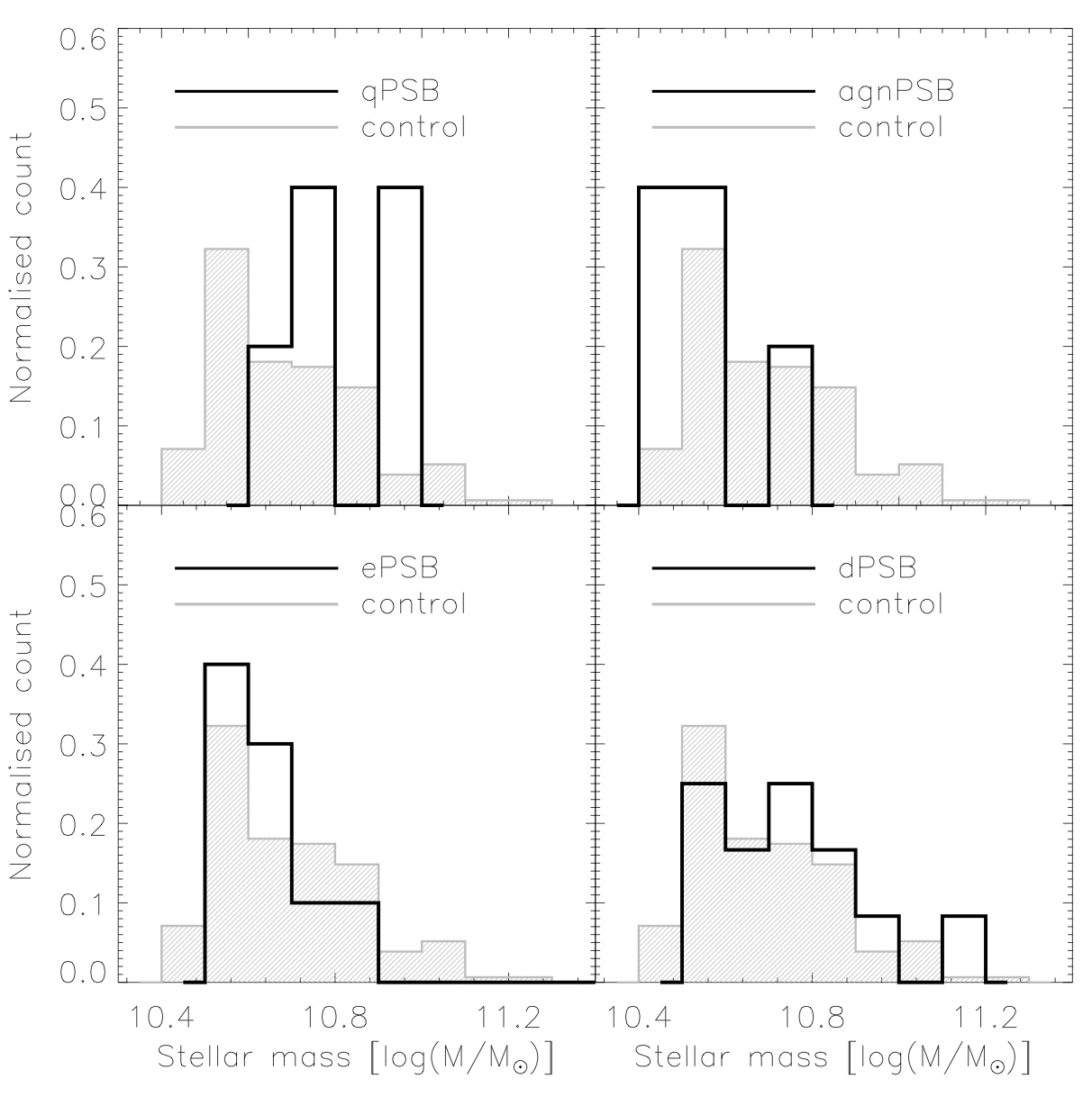}
\caption{Stellar mass distributions of the Balmer-strong galaxy and control samples. As a consequence of pair-matching the control  galaxies onto the Balmer-strong samples (see Section \ref{sec:controlselection}), the distributions are the same for the star-forming and quiescent control samples and they match the combined distribution for all Balmer-strong galaxies, in both low-mass and high-mass regime.}
\label{fig:mass_hist}
\end{figure*}

\subsection{Sample selection criteria}\label{sec:sampleselection}

We began our selection with a sample containing 83634 spectroscopically confirmed galaxies with Petrosian $r$-band magnitudes $14.5<\mbox{m}_{r}<17.7$ at $0.01<\mbox{z}<0.05$. In this redshift range selecting galaxies with stellar masses above $\mbox{M}_{\star}/\mbox{M}_{\odot}=10^{9.5}$ yields samples that are statistically complete in red-sequence galaxies, which are defined as galaxies for which the spectral indices PC1 and PC2 (see Section \ref{sec:spectralanalysis}) satisfy the relation PC2 $\leq$ PC1 + 0.5. Applying this mass criterion reduced the number of galaxies in the sample to 49148. We then applied a cut on spectral per-pixel signal-to-noise ratio: SNR $>$ 8 in the $g$-band, to ensure accurate measurements of the spectral indices and emission line properties. This removed further $11\%$ of the sample, leaving us with 43811 objects. 

Motivated by the bimodal nature of several galaxy properties in the local Universe separating the majority of local galaxies into two distinct families at  $\mbox{M}_{\star}/\mbox{M}_{\odot}=3\times10^{10}$ (e.g. \citealt{Kauffmann+2003a}), we split our sample into these two different mass regimes and refer to the resulting subsets as \emph{low-mass} and \emph{high-mass} galaxies.
The low- and high-mass parent samples contain 33438 and 10373 objects, respectively.
Before selection of the Balmer-strong galaxies and control samples we applied a further restriction by removing from the parent samples all galaxies observed `edge-on', with projected axis ratio\footnote{We defined the projected axis ratio using two SDSS parameters: \emph{expAB} and \emph{devAB} (axis ratios from exponential and deVaucouleurs fits, respecively), measured in the $r$-band. Through careful visual inspection of the galaxy images, we found that the value of $0.32$ works well for isolating the `edge-on' objects and, therefore, we required both parameters to have values above that limit. Assuming that the galaxies have a characteristic intrinsic axis ratio of 0.2, the measured value 0.32 corresponds to 75$^{o}$ inclination.} greater than 0.32. The purpose of this cut was to minimise potential biases in our measurements due to strong attenuation of the stellar light by dust in inclined galactic disks and it resulted in the reduction of the low- and high-mass samples by removing $17\%$ and $13\%$ of the galaxies, respectively. Our final samples from which the Balmer-strong and control samples were drawn contain 27901 and 9001 galaxies in the low- and high-mass regimes, respectively.

\subsubsection{Balmer-strong/post-starburst galaxies}\label{sec:psbselection}

We found that placing a cut at $\mbox{PC2}=0.0$ at 1$\sigma$ works well for selecting galaxies with prominent Balmer absorption lines, yielding 157 and 32 galaxies in the low- and high mass regimes, respectively. 
Based on the models of top-hat starbursts superimposed on an old stellar population, investigated by \citet{Wild+2007}, the galaxies selected from this extremum of the PC1-PC2 parameter space are expected to have starburst ages greater than $\sim0.6$ Gyr (see also the measured starburst ages in \citealt{Wild+2010}).
 We then classified these Balmer-strong galaxies based on their emission-line measurements.
We used a cut on the $H\alpha$ emission line equivalent width (EQW) to determine whether the galaxies have ongoing star formation at a measurable level (indicated by $\mbox{EQW}>3${\AA} with $\mbox{SNR}>3$). We further used the BPT diagnostics \citep{Baldwin+1981} to identify potential AGN-host candidates. For this purpose we used the condition introduced by \citet{Kewley+2001}, again, requiring that the emission lines have $\mbox{SNR}>3$.
We chose this criterion over that introduced by \citet{Kauffmann+2003a} to ensure the selection of galaxies with AGN-dominated emission only, excluding those in which the contributions from the AGN and star formation are comparable. 

A non-negligible number of the emission-line Balmer-strong galaxies were found to have high values of the Balmer decrement, i.e. the flux ratio of $H\alpha$ to $H\beta$ emission lines measured with respect to the intrinsic ratio of 2.87, indicating considerable dust content. We separate out the `dustiest' galaxies in both mass regimes, in order to test whether they are a separate class of dust-obscured starburst galaxies (\citealt{Dressler+1999, Poggianti+1999}). A cut on the Balmer decrement was placed to identify the top 10\% of dusty galaxies in the parent samples, corresponding to $H\alpha$/$H\beta>5.2$ and $H\alpha$/$H\beta>6.6$ in the low- and high-mass regime, respectively. Although arbitrary, this provides a good base for determining whether the dust-obscured Balmer-strong galaxies are truly different from the dust-unobscured ones. We find $35\%$ and $42\%$ of the Balmer-strong galaxies with emission lines have Balmer decrements above these cuts in the low- and high-mass samples, respectively. 
Considering the above criteria, we distinguish between four types of Balmer-strong galaxies:
\begin{itemize}
\item {\bf `Quiescent' Balmer-strong galaxies (qPSB)} - with no measurable $H\alpha$ emission, equivalent to the traditional definition of post-starburst (or `K+A') galaxies. 
\item {\bf Balmer-strong AGN host galaxies (agnPSB)} - located above the \citet{Kewley+2001} demarcation line in the BPT diagram.
\item {\bf Emission-line Balmer-strong galaxies (ePSB)} - galaxies with measurable $H\alpha$ emission line, not classified as dusty or AGN-host candidates. The normal dust content suggests that these are unlikely to be contaminating dust-obscured starbursts. This will be assessed in the paper. 
\item {\bf Dusty Balmer-strong galaxies (dPSB)} - with measurable $H\alpha$ emission, classified as dusty but not as AGN host candidates; the subset of ePSB with the highest dust content as indicated by the Balmer decrement. These may be dust obscured starbursts, and are the most likely contaminants of post-starburst samples defined without an emission line cut.
\end{itemize}
In all cases, we use the label `PSB' for conciseness. 

\subsubsection{Control galaxies}\label{sec:controlselection}
Within both mass regimes, we selected control samples of quiescent and star-forming galaxies pair-matched with the post-starburst galaxies in stellar mass, within $\Delta\mbox{M}_{\star}/\mbox{M}\odot=10^{0.1}$. We randomly selected 5 star-forming and 5 quiescent control galaxies per Balmer-strong galaxy, from the highest-density regions of PC1-PC2 space coinciding with the blue cloud and the red sequence, respectively. As shown in the top left panel of Figure \ref{fig:pc12} the star-forming galaxies were selected from regions defined by $-1.0<\mbox{PC2}<-0.5$ with the PC1 criterion depending on the mass regime: $-4.5<\mbox{PC1}<-3.4$ (low-mass), $-3.5<\mbox{PC1}<-2.4$ (high-mass), and the quiescent galaxies from regions defined by $\mbox{PC2}>0.8\times\mbox{PC1}-0.2$ and $\mbox{PC2}<0.8\times\mbox{PC1}-0.6$, and by $-1.0<\mbox{PC1}<-0.2$ (low-mass) and
$-0.8<\mbox{PC1}<0.0$ (high-mass).
As there are no clear boundaries between the different classes of galaxies in the PC1-PC2 space, we chose to select samples from the regions of highest number density within the blue cloud and the red sequence in order to create clean samples of control galaxies with ``typical'' properties, and avoid selecting objects with either extreme or intermediate properties. 

Additionally, we build a dusty star-forming control sample using the same regions of PC1-PC2 space as the star-forming control samples, with an additional constraint on the Balmer decrement to match the limits used to select the dPSB galaxies. 

\subsection{Summary of sample properties}

\begin{figure*}
\centering
  \includegraphics[scale=0.53]{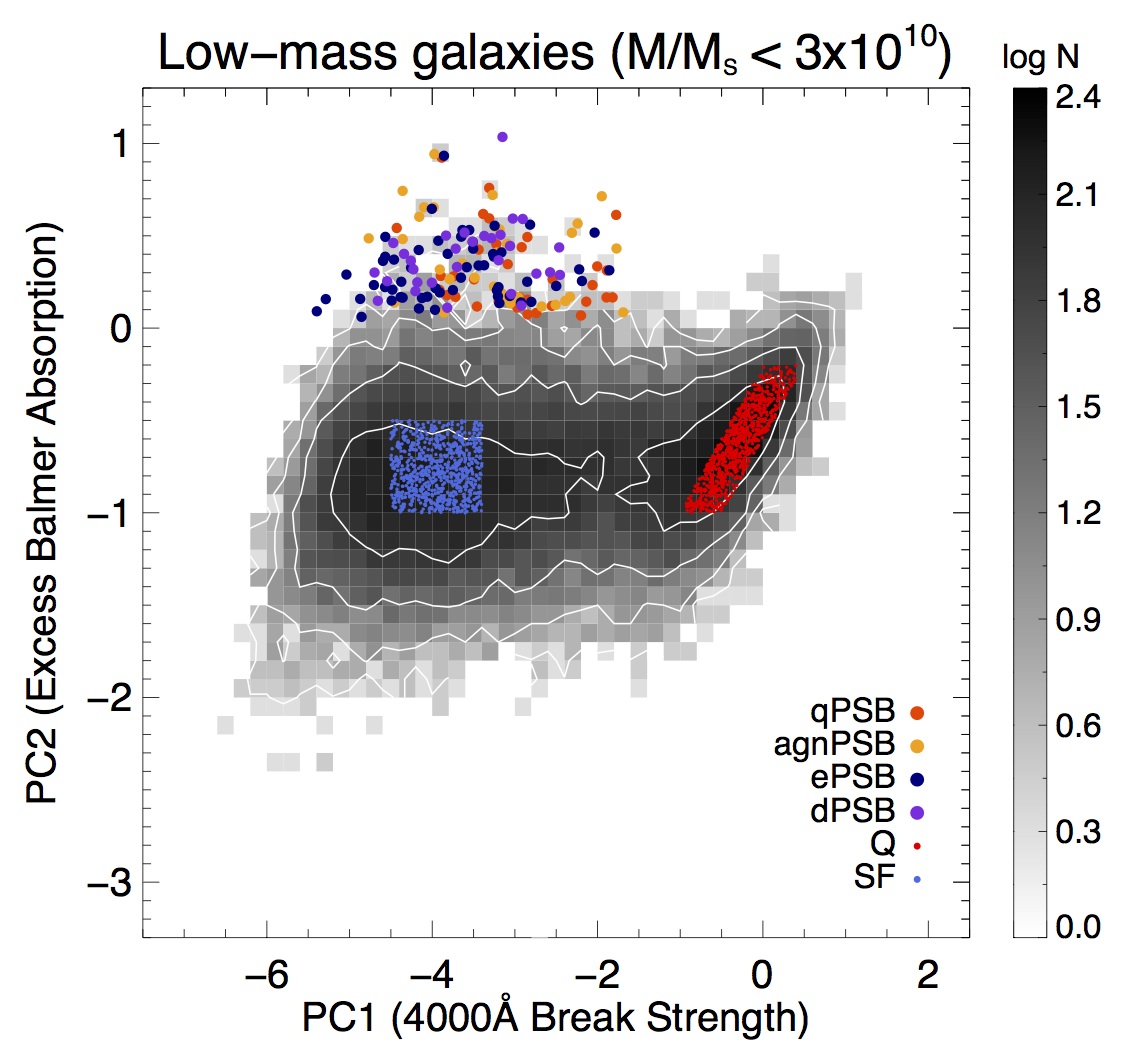}
  \includegraphics[scale=0.53]{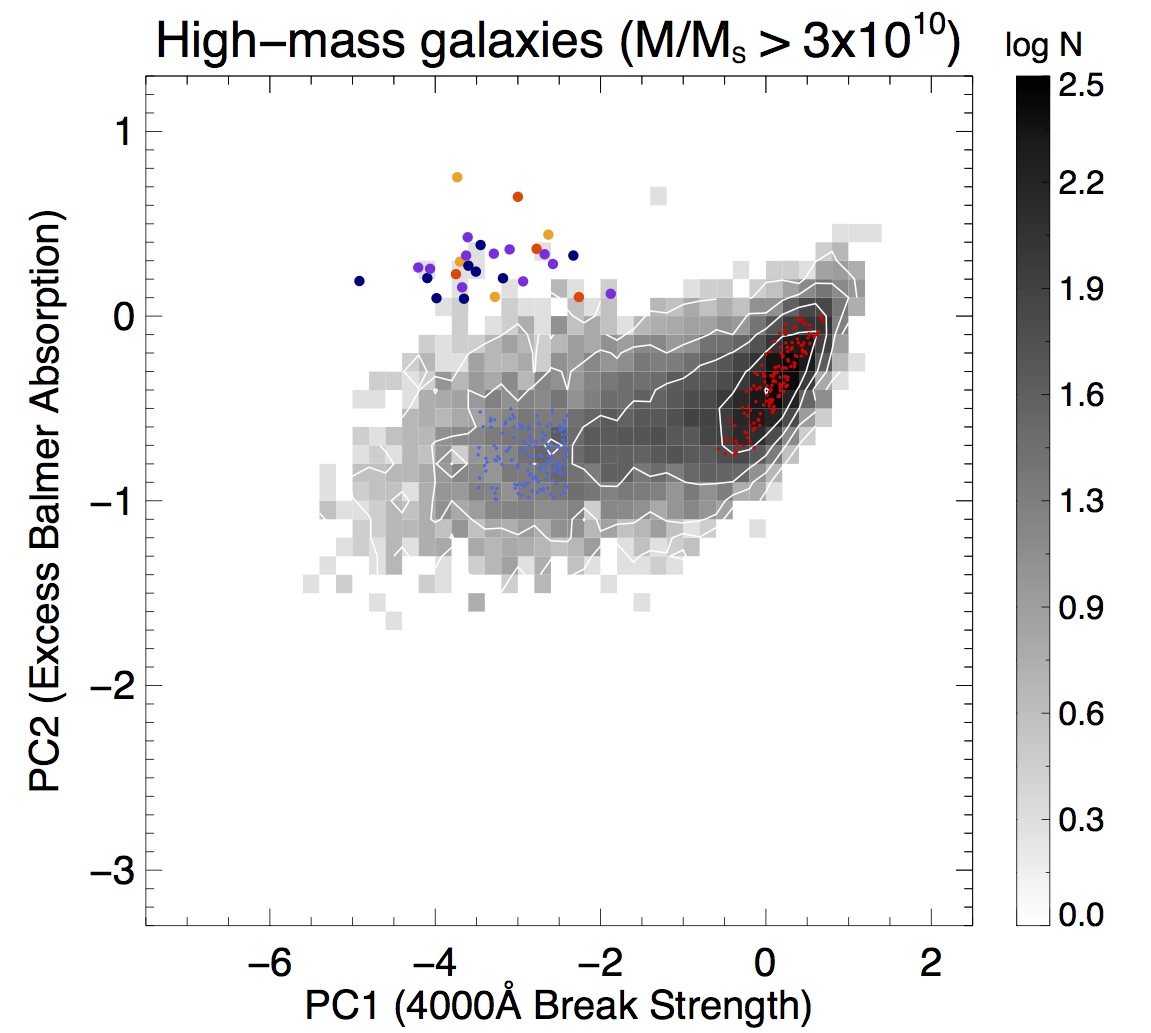}
  \includegraphics[scale=0.53]{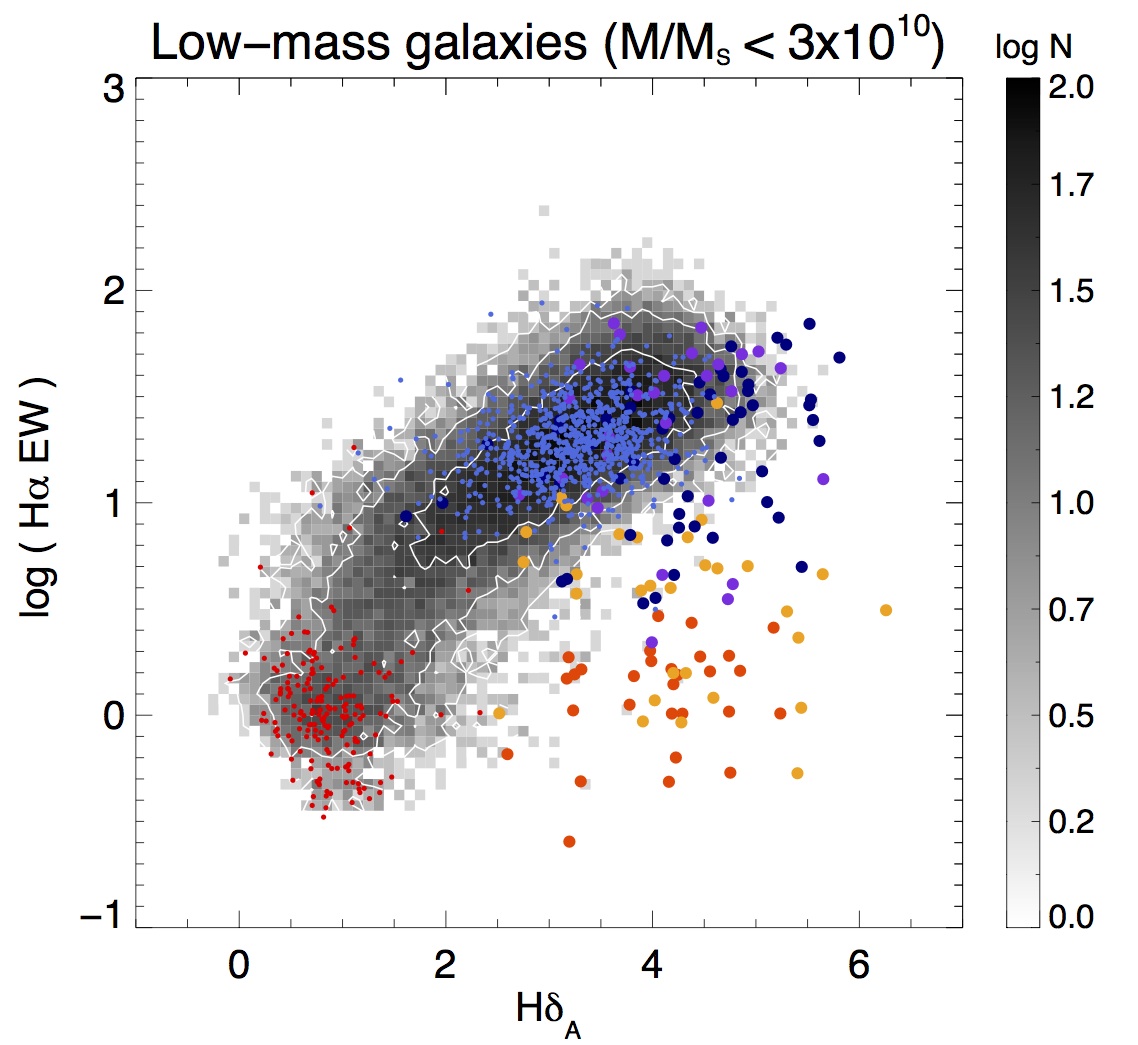}
  \includegraphics[scale=0.53]{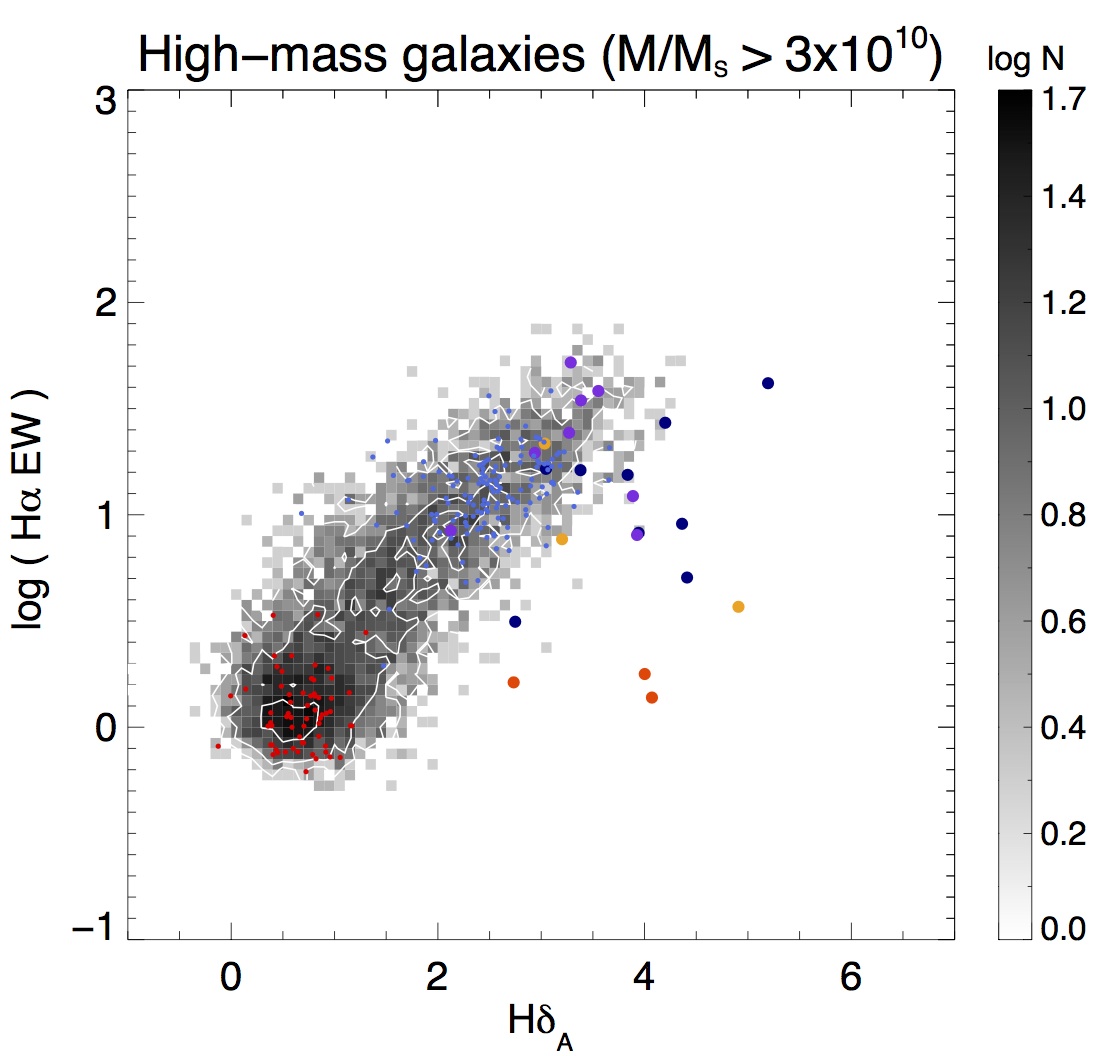}
  \includegraphics[scale=0.53]{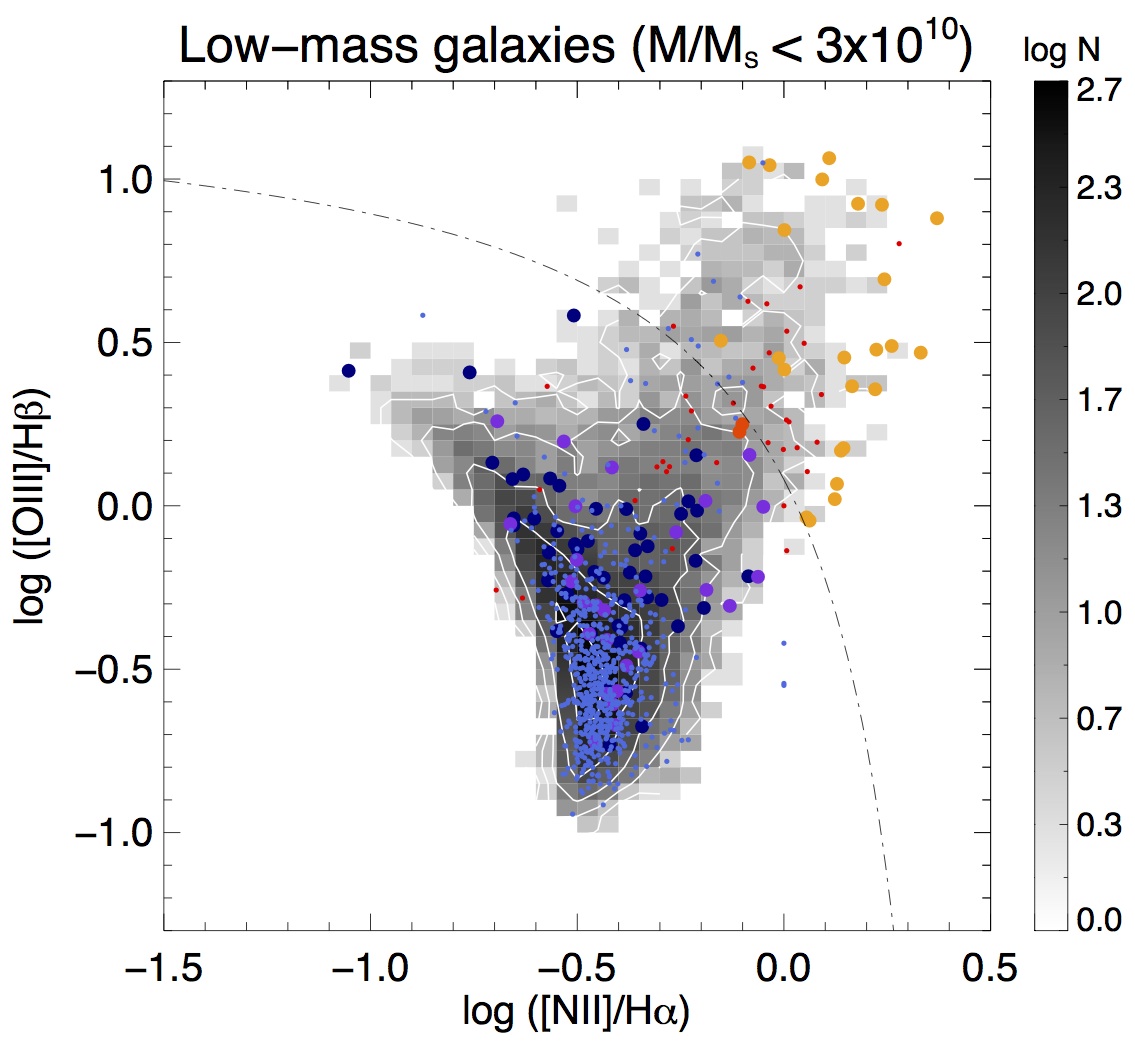}
  \includegraphics[scale=0.53]{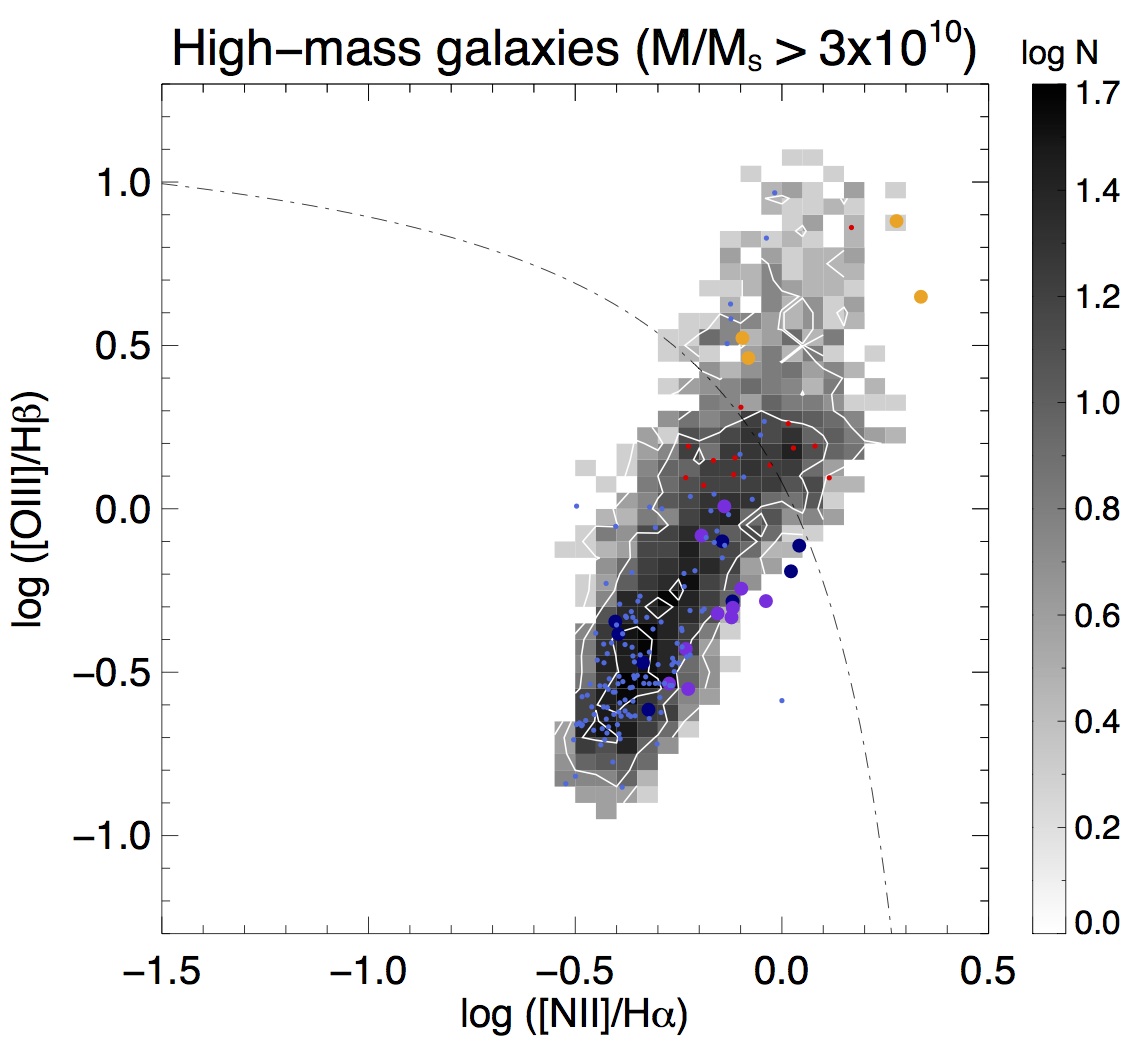}
\caption{The distribution of various spectral indicators for all galaxies in the parent samples (grayscale) with the final Balmer-strong galaxies (see Section \ref{sec:psbselection} for selection details) overplotted in orange (qPSB), yellow (agnPSB), dark blue (ePSB) and purple (dPSB). The control samples of star-forming (SF) and quiescent (Q) galaxies are plotted in light blue and red, respectively.
\emph{Top row} - principal component amplitudes, PC1 and PC2, used for the  selection of Balmer-strong galaxies; \emph{Middle row} - the traditionally used $H{\delta}_A$ Lick absorption line index and the equivalent width of the $H_{\alpha}$ emission line; \emph{Bottom row} - the BPT diagram \citep{Baldwin+1981}, with the dashed-dotted line showing the \citet{Kewley+2001} criterion for distinguishing between systems in which a significant contribution to the emission comes from an AGN.}
\label{fig:pc12}
\end{figure*}

\begin{figure*}
\centering
  \centering
  \includegraphics[scale=0.83]{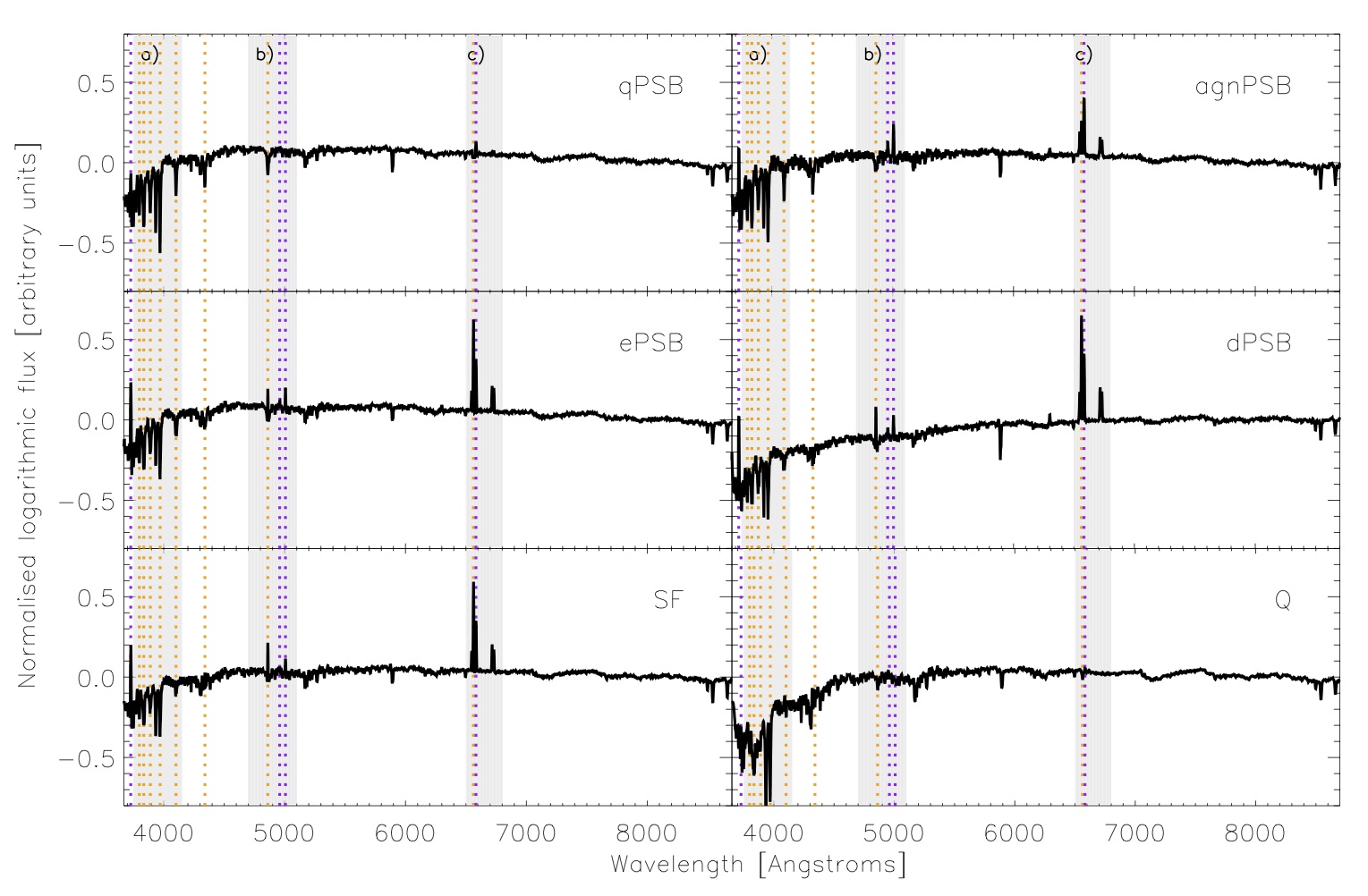}
      \includegraphics[scale=0.83]{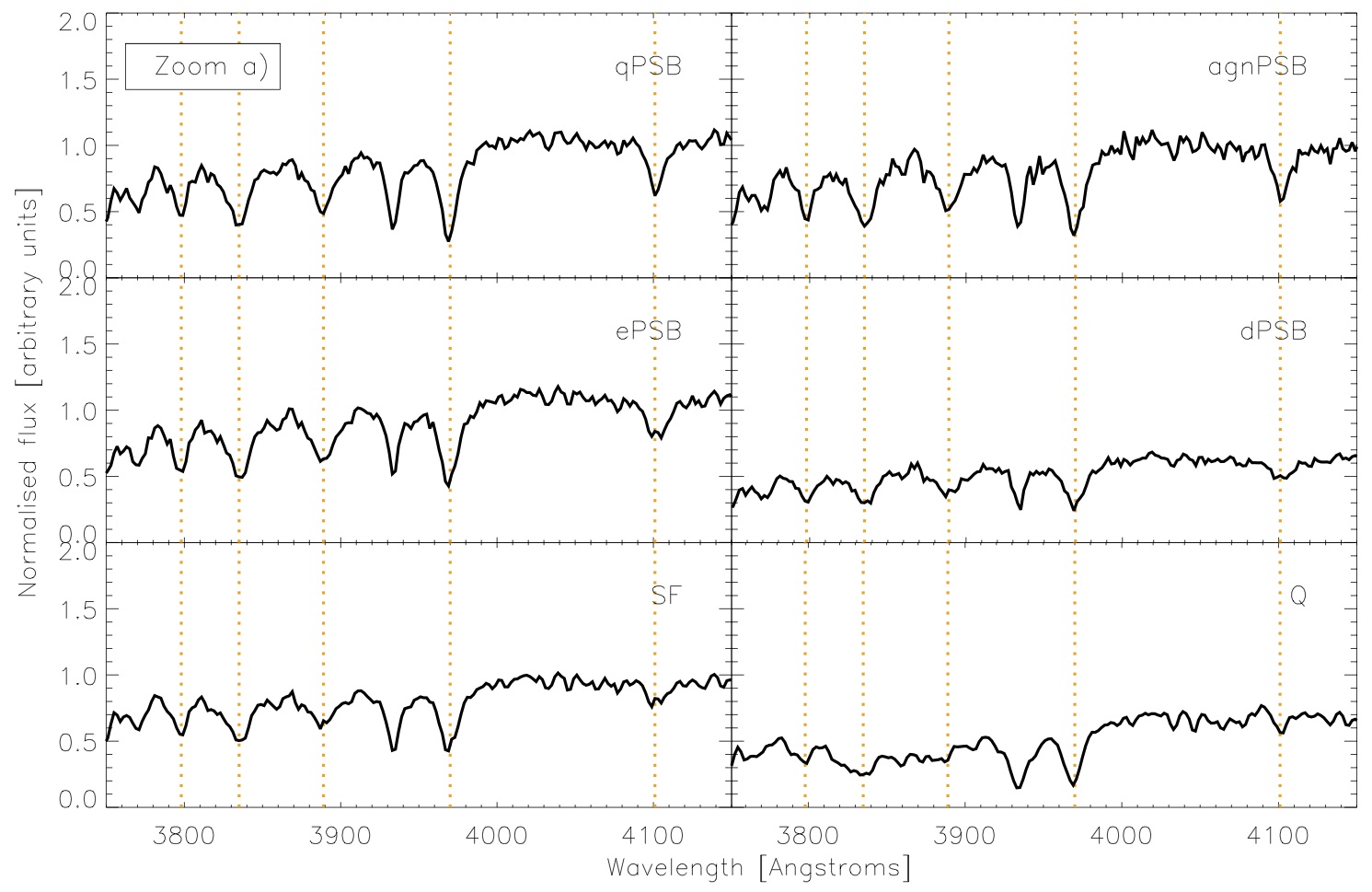}
\caption{Top panel: the stacked SED of the different galaxy samples, normalised at 8000{\AA}, with the flux plotted in logarithmic units so that both emission lines and the shape of the continuum can be seen in all samples. The orange lines show the location of the Hydrogen Balmer series including, from right to left, H$\alpha$6563, H$\beta$4861, H$\gamma$4341, H$\delta$4102, H$\epsilon$3970, H$\zeta$3889, H$\eta$3835 and H$\theta$3798, and the purple lines that of [NII]6583{\AA}, [OIII]4959,5007{\AA} and [OII]3727{\AA}. The grey shaded areas show the locations of the regions, a, b and c, which are zoomed-in on in the bottom panel and in Figure \ref{fig:SED_stack2}. Note that in all three zoomed-in regions the flux is plotted in linear units.
Bottom panel: zoom a) - the region in which the PCA indices are calculated containing the 4000{\AA}-break and higher order Balmer absorption lines. The orange lines include, from right to left, H$\delta$4102, H$\epsilon$3970, H$\zeta$3889, H$\eta$3835 and H$\theta$3798. Note that H$\epsilon$ coincides with CaH at 3969{\AA}.}
    \vspace{30pt}
\label{fig:SED_stack1}
\end{figure*}

\begin{figure*}
\centering
  \centering
    \includegraphics[scale=0.8]{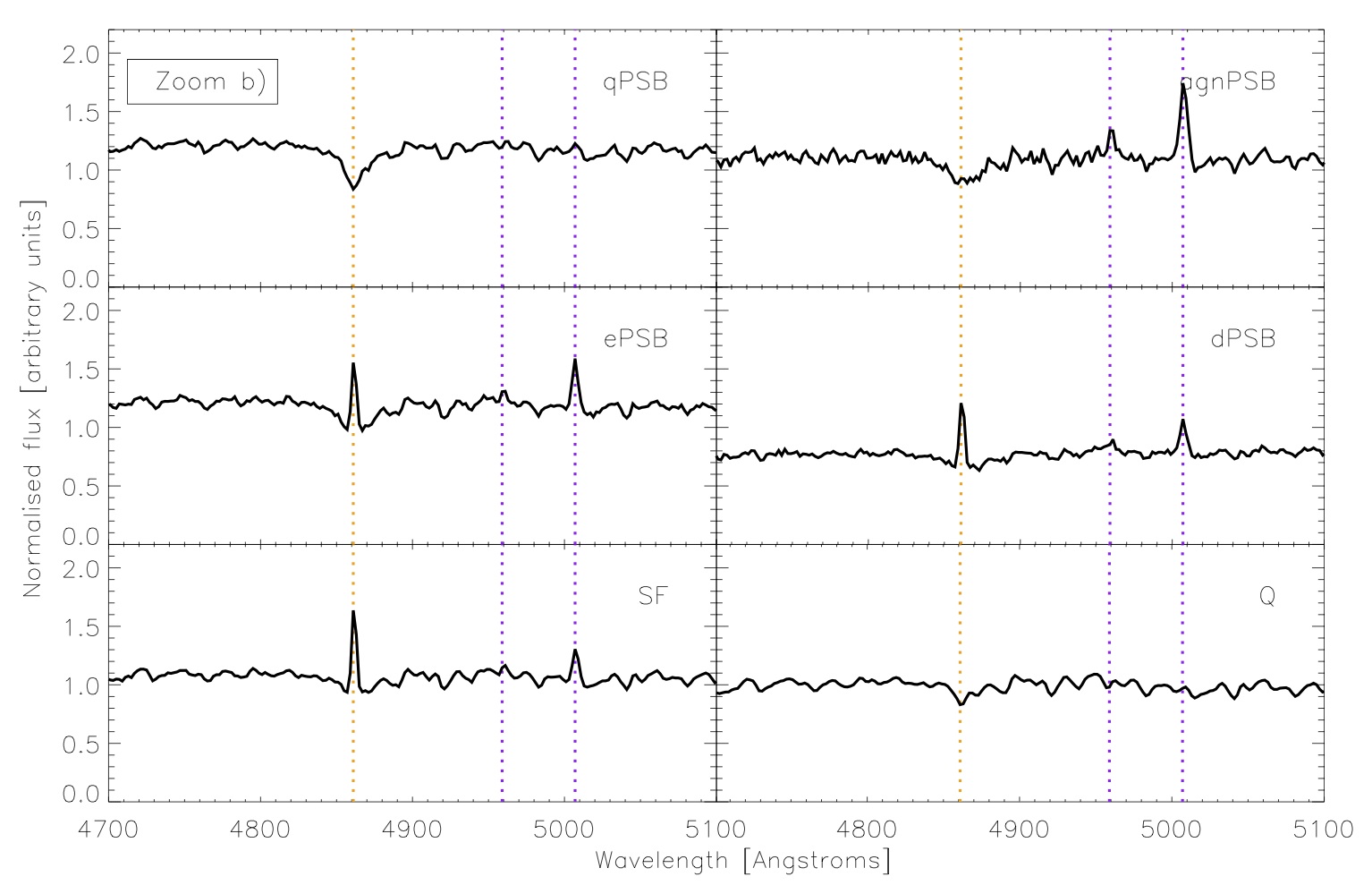}
    \includegraphics[scale=0.8]{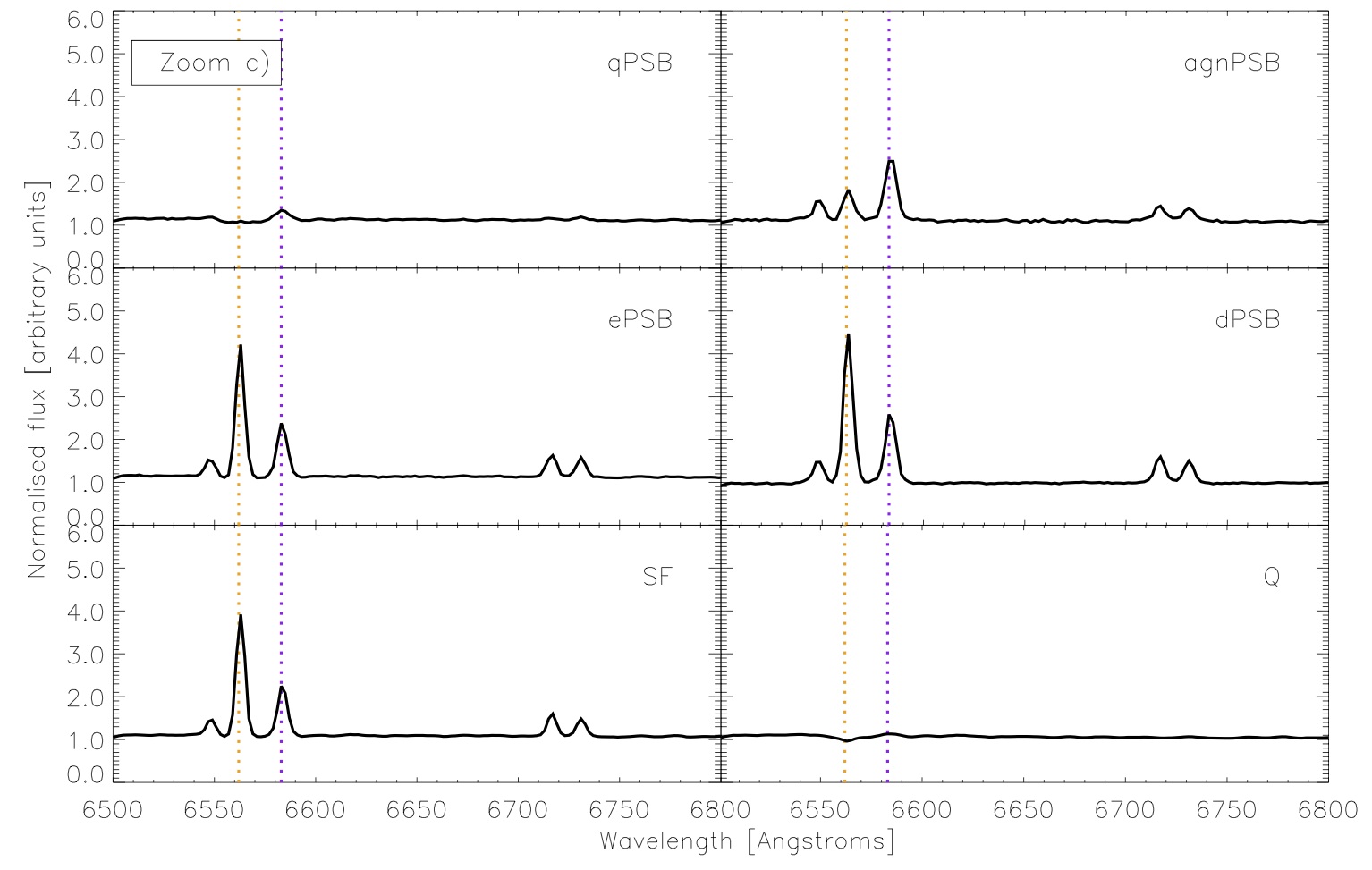}
\caption{Portions of the stacked SED presented in the top panel of Figure \ref{fig:SED_stack1} containing particular regions of interest. Top panel: zoom b) - region containing the H$\beta$4861 line (orange) and the [OIII]4959,5007{\AA} lines (purple). Bottom panel: zoom c) - region surrounding the  H$\alpha$6563 and [NII]6583{\AA} lines (marked in orange and purple, respectively).}
    \vspace{30pt}
\label{fig:SED_stack2}
\end{figure*}

The galaxy counts of all the Balmer-strong samples along with their selection criteria are summarised in Table \ref{tab:galcounts} and, in Figure \ref{fig:mass_hist}, we show the stellar mass distributions of the Balmer-strong samples and compare them with the mass distribution of the control samples. 
In Figure \ref{fig:pc12} we present the key spectral indices and line ratios used during  the sample selection, measured for all galaxies in Balmer-strong galaxies as well as control samples.
Finally, in Figures \ref{fig:SED_stack1} and \ref{fig:SED_stack2} we show the stacked spectra of the different samples, both across the full optical wavelength range and focusing on the individual regions: 1) 3750-4150\AA\, over which the PCA indices are calculated; 2) 4700-5100\AA\, containing the $H\beta$ and OIII  lines; 3) 6500-6800\AA\, where the $H\alpha$ and NII spectral lines are located.

\section{Methodology}


\subsection{Star formation histories}

The galaxy SED in the UV-to-IR regime is generally dominated by the light emitted by its stellar components, reprocessed by the surrounding reservoirs of the interstellar medium (ISM). Therefore, it contains information about the galaxy's star-formation rate and its star formation history, as well as its dust content.
The star formation history can be extracted from the SED through the process of spectral synthesis which essentially breaks down the galaxy SED into its base components. In practice, this is done by fitting the SED with a range of models - here we use an unparameterised approach, meaning that the star formation history is not constrained to be a particular form. We fit a linear combination of starbursts, called simple stellar populations (SSP), spanning a wide range of ages and metallicities.

\subsubsection{Spectral synthesis}\label{sec:meth_specsynth}

In this work we utilised the SED fitting code STARLIGHT \citep{CidFernandes+2005}, to fit an updated version of the BC03 evolutionary synthesis models, with dust attenuation modelled as a single foreground dust screen. The fitting procedure in STARLIGHT is carried out using a sophisticated multi-stage $\chi^{2}$-minimisation algorithm combining annealing, Metropolis and Markov Chain Monte Carlo techniques. Prior to the fitting, the galaxy spectra were sky-residual subtracted longward of $6700${\AA}, in order to correct the effects of the incomplete subtraction of the OH emission lines by the SDSS spectroscopic reduction pipeline \citep{Wild+2006}\footnote{\url{http://www.sdss.jhu.edu/skypca/spSpec/}}. All fluxes were corrected for Galactic extinction using the extinction values provided in the SDSS catalogue which are based upon the \citet{Schlegel98} dust emission maps and the Milky Way extinction curve of \citet{CCM98}. The spectra were moved onto air wavelengths to match the models and nebular emission lines of the star-forming, ePSB and dPSB galaxies were masked using a bespoke mask defined from the stacked star-forming galaxy spectrum. 

The processed spectra were fit with a linear combination of 300 SSPs spanning 60 stellar ages, that range from 1\,Myr to 14\,Gyr, and 5 metallicities: Z/Z$_\odot$ = 0.02, 0.2, 0.4, 1 and 2.5, where we have assumed Z$_\odot=0.02$. The ages were chosen to cover the whole of cosmic time, roughly linearly spaced in log age between $10^8$ and $10^{10}$ years, and with slightly sparser sampling for models younger than $10^8$ years. 
Although 60 ages bins are far more than can be constrained from a single optical spectrum, sufficient coverage is required across the main sequence lifetime of A/F stars (between 100\,Myr and 2\,Gyr) where the strength of the Balmer lines and distinctive shape of post-starburst galaxy spectra change rapidly with time, in principle allowing accurate age dating of the population. We used our star-forming galaxy control sample to test both the standard BC03 population synthesis models which use the Stelib stellar spectral library \citep{LeBorgne2003}, and a new set of models which combine both the Stelib and MILES libraries \citep{miles2006} to give a total wavelength coverage of 3540.5-8750{\AA}\footnote{XMILESS 2016, available from G. Bruzual on request}. Redwards of 8750{\AA} both models are based on the theoretical BaSeL 3.1 spectral library \citep{Westera2002}. Additional changes to the model atmospheres used to model the UV are not relevant to this work. Both models are based on the \emph{Padova 1994} set of stellar evolution tracks. 

The traditional BC03 models caused an artificial peak in the star formation history (SFH) of star-forming control galaxies between 1.6-1.9\,Gyr, as well as a smaller peak at 1.3\,Gyr; these peaks in star formation histories are a known problem with the models and visible in previous analyses where SFHs from unparameterised fits are presented in a relatively unsmoothed format (e.g. mass-assembly in Fig. 8 of \citealt{Asari+2007}, \citealt{Delgado2017}). The spikes are particularly visible in our Balmer-strong samples, which have a significant fraction of their mass formed in this time interval. The updated models provide a more continuous stacked star formation history for the control star-forming galaxies around the critical $0.5-2\times10^9$year timescale which is relevant for quantifying the burst strength and age in post-starburst galaxies. However, a smaller artificial bump is still evident at $\sim$1.3\,Gyr. Unfortunately this feature in the stellar population models limits the accuracy with which we can age-date the starburst in older post-starburst galaxies. Further investigation as to how these problems can be mitigated when fitting post-starburst galaxy spectra is on-going, but for the purposes of this paper we limit our analysis to parameters that are robust to changes in the library i.e. the total fraction of stars formed in the last 1 and 1.5\,Gyr, and the star formation history prior to 2\,Gyr. Additionally, we use our control sample of star-forming galaxies to ensure that bursts are detected above any artificial signals in the star-forming sample. 

During fitting, we adopted the attenuation law for starburst galaxies of \citet{Calzetti+2001}. Repeating the analysis with the Milky Way extinction law of \citet{Cardelli+1989} produced quantitatively slightly different SFHs, but did not alter our conclusions when comparing SFHs between samples.

\subsection{Morphology and structure}\label{sec:method_morph}

To characterise the morphology and structure of our galaxies we applied a range of automated measures to their sky-subtracted images in the $g$, $r$ and $i$ bands of the SDSS. The analysis was performed on 1 arc minute cutouts of the SDSS images, centred on the galaxy of interest as defined by the coordinates provided in the SDSS database. 
Prior to the analysis all images were inspected visually to identify bright light sources in close proximity to the galaxies of interest. Bright nearby sources have the potential to severely contaminate the measures of morphology and structure and, therefore, all images in which such sources had been identified were excluded from the image analysis. This resulted in the reduction of the sample sizes as summarised in Table \ref{tab:galcounts}. 

Our methods follow those presented in \citet{Pawlik+2016}; here we outline only the most important details. First, we created a binary pixel map which identifies pixels associated with the galaxy, as opposed to the surrounding sky. The algorithm loops around the SDSS position pixel, searching for connected pixels above a given threshold (1$\sigma$ above the median sky background level). Combined with a running average smoothing filter, the algorithm picks out contiguous features in an image down to a low surface brightness ($\sim$24.7 mag/arcsec$^{2}$). 

The binary pixel maps were used to estimate the galaxy radius, $R_{max}$, as the distance between the centre (brightest galaxy pixel) and the most distant pixel from that centre within the map. Generally, this definition of galaxy radius agrees well with the commonly used Petrosian radius \citep{Petrosian1976, Blanton+2001,Yasuda+2001}; however, it provides an advantage in the case of galaxies with extended faint outskirts, like tidal tails. 

We then followed standard procedures to measure the S{\'e}rsic index ($n$)\footnote{Computed using the 1D surface-brightness profiles defined by circular apertures.}, the concentration index ($C$)\footnote{Here we use the growth curve radii enclosing 20$\%$ and 80$\%$ of the total light.}, the light-weighted asymmetry ($A$), the Gini index ($G$) and the moment of light ($M_{20}$). We additionally measured the new shape asymmetry ($A_{S}$), presented in \citet{Pawlik+2016}, which quantifies the disturbance in the faint galaxy outskirts. The shape asymmetry is computed using the same expression as the light-weighted asymmetry parameter, under a 180-degree rotation, but with the measurement performed using the binary pixel maps rather than the galaxy images. This approach allows for equal weighting of all galaxy parts during the measurement, regardless of their relative brightness. Finally, we computed $A_{S90}$ to quantify the shape asymmetry under a 90-degree rotation. This can be used in conjunction with $A_{S}$ to indicate whether the features in galaxy outskirts are elongated (e.g. tidal tails) or circular (e.g. shells). Further details of the methods used to measure each of these parameters are given in \citet{Pawlik+2016} and references therein.

\subsection{Environment}\label{sec:method_env}

We adopted the projected number density of galaxies in the vicinity of the target galaxy as a measure of the environment. The number density was calculated following the method described in \citet{Aguerri+2009}, using the projected comoving distance of the target galaxy, $d_{N}$, to its $N$th nearest neighbour:
\begin{equation}
\Sigma_{N}=\frac{N}{\pi d_{N}^2}.
\end{equation}
 
The nearest neighbours were defined in two ways. The first definition includes all galaxies with spectroscopic redshifts, $z_{s}$, within $\pm$1000km/s of the target galaxy, and with an absolute magnitude difference of not more than $\pm$2. These criteria are similar to those used by \citet{Balogh+2004a} and are designed to limit the contamination by background/foreground galaxies even when using projected distances. The second definition uses the SDSS photometric redshift measurements and selects galaxies within $\Delta z_{p}$=0.1 from the target galaxy (see \citealt{Baldry+2006} for a similar approach). This does not suffer the same incompleteness of the spectroscopic samples, but has higher contamination due to the less accurate photometric redshifts. 

It is important to realise that the values of $\Sigma_{5}$ are approximate estimates of the local number density, with the spectroscopic and photometric measurements representing the lower and upper boundaries\footnote{The spectroscopic measure is taken as the lower limit because it is expected to underestimate the galaxy number density due to the incompleteness in the spectroscopic redshift data, while the photometric measure will overestimate the true value due to projection effects and is therefore taken as the upper limiting value.}, and the uncertainties associated with $\Sigma_{5}$ are expected to be large. To examine the accuracy of the density measurements we also considered the 3rd, 8th and 10th nearest neighbours and found a good agreement with $\Sigma_{5}$. Furthermore, we flagged all galaxies with $d_{N}$ greater than the distance to the edge of the survey, as for such locations the density measurements may be unreliable. 

To provide one single estimate of local environment, we took the mean $\Sigma_{5}$ of the spectroscopic and photometric measurements. We stress that the purpose of the mean is merely to provide a single value which is likely to be closer to the true value than the individual measurements, rather than to serve as any statistical measure. 
We find that the qualitative results and conclusions remain generally unchanged whether we use the mean $\Sigma_{5}$ or the individual spectroscopic/photometric measurements, except for the ePSB sample. For this case we comment on the discrepancies when discussing the results in Section \ref{sec:results_env}.

To relate the values of $\Sigma_{5}$ to the different types of environment, we follow the definitions used by previous studies (e.g. \citealt{Aguerri+2009}, \citealt{Walcher+2014}):
\begin{itemize}
\item $\Sigma_{5}<1\mbox{Mpc}^{-2}$ - very low-density environments, 
\item $1 \mbox{Mpc}^{-2}<\Sigma_{5}<10 \mbox{Mpc}^{-2}$ - loose groups,
\item $\Sigma_{5}>10 \mbox{Mpc}^{-2}$ - compact groups and clusters.
\end{itemize}




\section{Results}

In Table \ref{tab:results} we present the recently formed mass fractions estimated by STARLIGHT ($f_{M1}$, $f_{M15}$ - within the last 1Gyr and 1.5 Gyr, respectively), the projected galaxy number density ($\Sigma_{5}$) and the $r$-band measurements of several structural and morphological parameters ($n$, $C$, $A$, $A_{S}$, $A_{S90}$, $G$, $M_{20}$) for all Balmer-strong galaxies studied in this work. Here we present only the top ten rows (galaxies ordered by their \emph{specobjid} number); the full table is available online. 

\begin{table*}
\centering
 \caption{Analysis results for the post-starburst galaxies (full table available online as supplementary material). The columns contain: the SDSS $specobjid$; the post-starburst type (Section \ref{sec:psbselection}); the logarithmic stellar mass in units of $M\sun$; the spectral indices - PC1 and PC2; the line ratios from the BPT diagram; the equivalent width of the $H\alpha$ emission line; the Balmer decrement ($H\alpha/H\beta)$; the fractions of recently formed stellar mass - $f_{M1}$ and $f_{M15}$ (Section \ref{sec:meth_specsynth}); the indicators of structure and morphology measured in the $r$-band - $n$, $C$, $A$, $A_{S}$, $A_{S90}$, $G$ and $M_{20}$ (Section \ref{sec:method_morph}) and the corresponding $r$-band image contamination flag; the local galaxy density measurements (photometric and spectroscopic) used to compute $\Sigma_{5}$ (Section \ref{sec:method_env}) and the corresponding measurement flags related to the proximity to survey boarders.}
 \vspace{0.25cm}
 
 \label{tab:results}
 \begin{tabular}{|c|ccccccccc|}
 \hline
ID & SDSS specobjid & PSB type & $log(M)$ & $PC1$ & $PC2$ & $log([NII]/H\alpha)$ & $log([OIII]/H\beta)$ & $H\alpha\,EQW$ & $H\alpha/H\beta$ \\ 
\hline
1 & 75657056748568576 & agnPSB & 10.06 & -4.348 & -0.399 & 0.049 & 0.268 & 4.58 & 2.17 \\
2 & 78754789168513024 & qPSB & 9.64 & -2.641 & -0.408 & -- & -- & -- & -- \\
3 & 78754790712016896 & ePSB & 10.13 & -3.048 & -0.903 & 0.081 & -- & 3.97 & -- \\
4 & 78754790720405504 & qPSB & 9.59 & -3.311 & -0.759 & -- & -- & 0.54 & -- \\
5 & 79034666610327552 & qPSB & 10.11 & -1.892 & -0.166 & -- & -- & 0.66 & -- \\
6 & 82695481800523776 & qPSB & 10.27 & -2.200 & -0.068 & 0.046 & -- & 1.49 & -- \\
7 & 83539864024252416 & dPSB & 10.67 & -2.082 & -0.825 & -0.099 & -0.326 & 23.78 & 4.07 \\
8 & 86071546359054336 & agnPSB & 10.15 & -3.908 & -0.317 & 0.165 & 0.366 & 4.91 & 1.37 \\
9 & 94798927357804544 & ePSB & 10.76 & -3.490 & -0.309 & -0.385 & -0.515 & 15.80 & 1.81 \\
10 & 100146046004363264 & ePSB & 9.62 & -4.596 & -0.364 & -0.543 & 0.061 & 59.87 & 1.72 \\
... & ... & ... & ... & ... & ...  & ...  & ... & ... & ... \\
 \end{tabular}

\vspace{0.5cm}

 \begin{tabular}{|c|cccccccccccccc|}
 \hline
ID & $f_{M1}$ & $f_{M15}$ & $n$ & $C$ & $A$ & $A_{S}$ & $A_{S90}$ & $G$ & $M_{20}$ & IMG FLAG & $\Sigma_{5p}$ & $\Sigma_{5s}$ & $\Sigma_{5p}$-flag & $\Sigma_{5s}$-flag \\
\hline
1 & 6.39 & 8.64 & 2.58 & 3.12 & 0.076 & 0.121 & 0.228 & 0.720 & -1.76 & 0 & 2.87 & 1.28 & 0 & 0 \\
2 & 12.63 & 48.52 & 2.82 & 2.88 & 0.077 & 0.076 & 0.331 & 0.606 & -1.80 & 0 & 31.91 & 0.36 & 0 & 0 \\
3 & 1.84 & 19.35 & 1.97 & 3.08 & 0.044 & 0.132 & 0.189 & 0.638 & -2.06 & 1 & 10.41 & 2.20 & 0 & 0 \\
4 & 22.10 & 63.59 & 5.48 & 6.09 & 0.283 & -- & 0.555 & 0.657 & -3.24 & 1 & 25.25 & 25.25 & 0 & 0 \\
5 & 3.81 & 44.30 & 3.26 & 3.22 & 0.073 & 0.146 & 0.433 & 0.741 & -2.05 & 0 & 3.11 & 0.39 & 0 & 0 \\
6 & 4.16 & 19.01 & -- & 2.77 & 0.015 & 0.158 & 0.248 & 0.761 & -1.67 & 0 & 5.33 & 0.61 & 0 & 0 \\
7 & 0.02 & 0.04 & 2.96 & 3.21 & 0.034 & 0.157 & 0.175 & 0.605 & -1.78 & 0 & 6.25 & 0.47 & 0 & 0 \\
8 & 14.48 & 18.40 & 5.53 & 3.73 & 0.198 & 0.222 & 0.173 & 0.748 & -0.37 & 0 & 0.66 & 0.20 & 0 & 0 \\
9 & 4.09 & 7.54 & 3.49& 3.67 & 0.059 & 0.170 & 0.442 & 0.652 & -2.39 & 0 & 0.26 & 0.18 & 0 & 0 \\
10 & 6.01 & 10.26 & 3.07 & 3.29 & 0.067 & 0.134 & 0.446 & 0.579 & -2.07 & 0 & 17.77 & 2.81 & 0 & 0 \\
... & ... & ... & ... & ... & ... & ... & ... & ... & ... & ... & ... & ... & ... & ... \\
 \end{tabular}
\end{table*}

\subsection{Star formation history}\label{sec:res_sfh}

Using the output of the STARLIGHT code, we explored both recent and earlier (pre-burst) star-formation histories of the Balmer-strong galaxies and compared them with the control samples. The star-formation histories of the dPSB galaxies and associated dusty star-forming control sample output by the code showed almost $100\%$ of their mass assembled at very early cosmic times - inconsistent with their current star forming properties as evident from their nebular emission lines. We therefore believe that the spectral fitting is likely impacted by the high dust contents of these galaxies, and the fact that STARLIGHT can only fit a single-component dust screen, so we exclude these galaxies from this part of the analysis. Further details on the colours and spectral fits of this sample are given in Appendix \ref{app:dpsb}.


\subsubsection{Recent star formation}

\begin{figure*}
  \centering
  \includegraphics[scale=0.85]{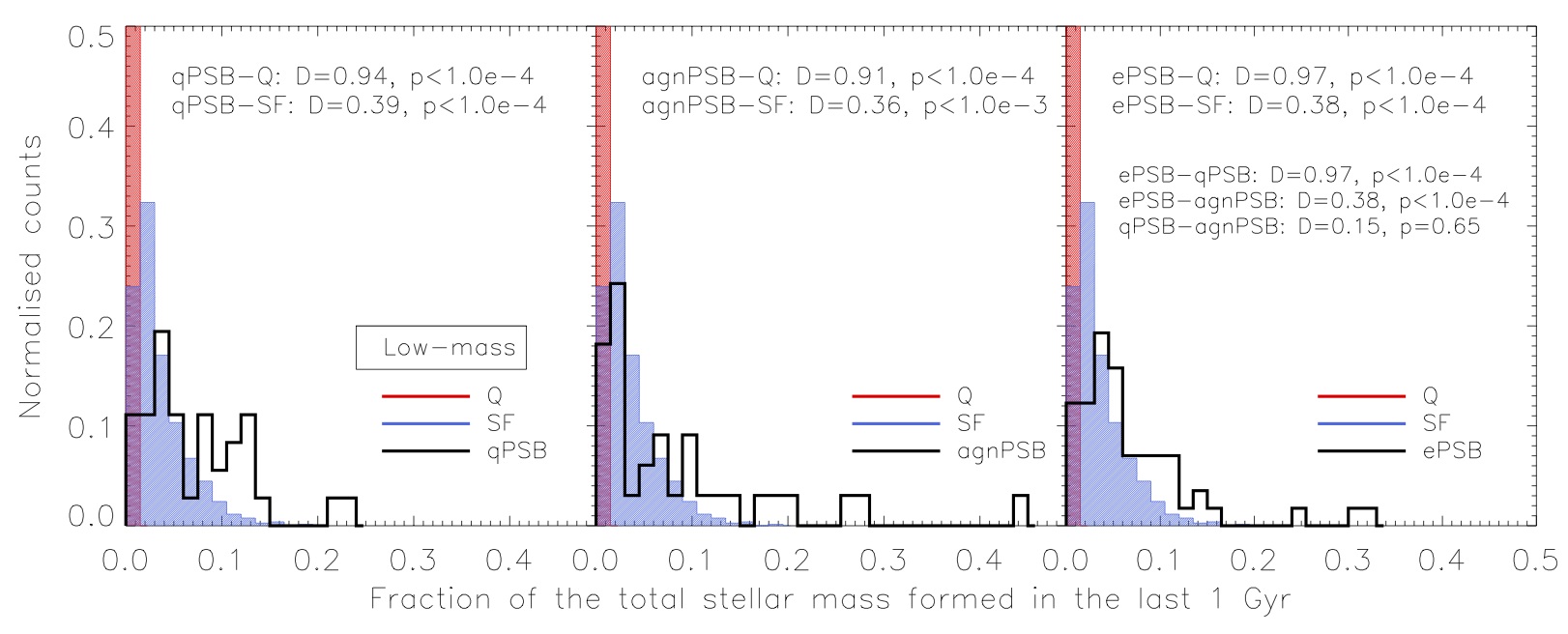}
   \includegraphics[scale=0.85]{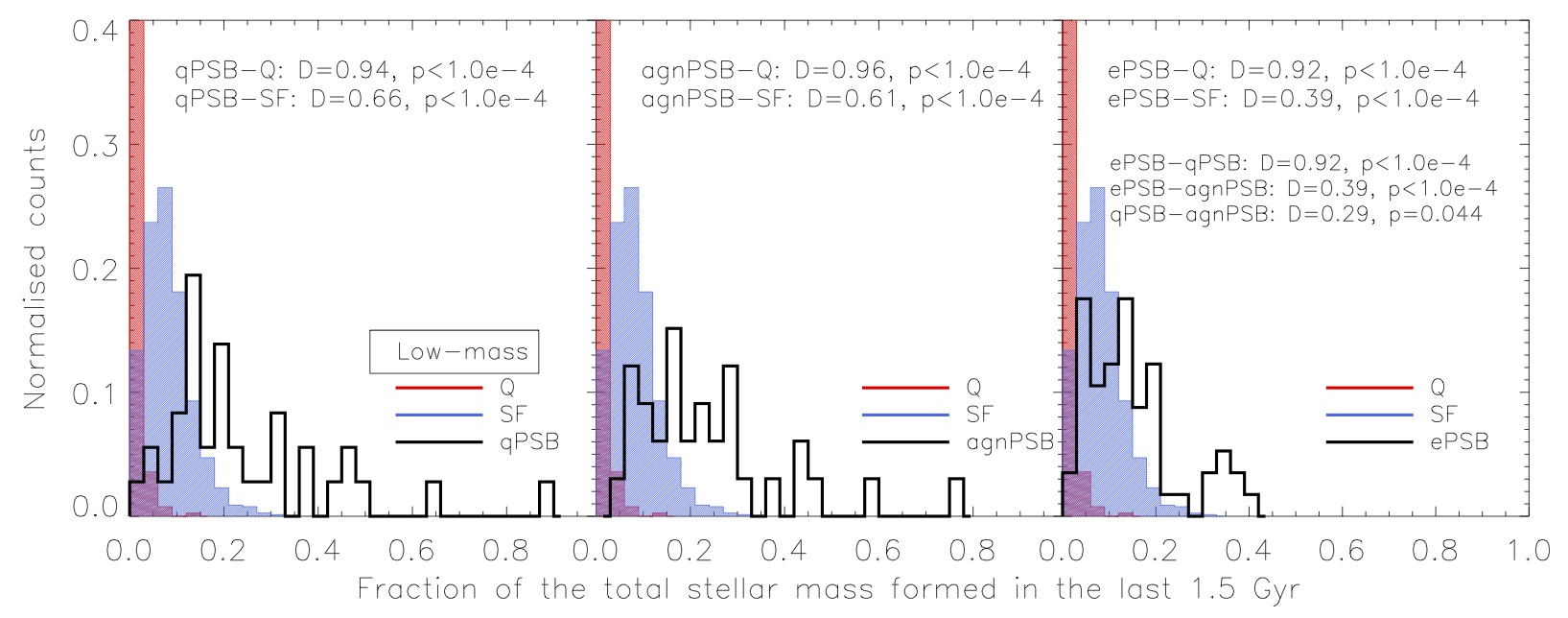}
\caption{The fraction of total stellar mass of the low-mass galaxies formed in the last 1 and 1.5\,Gyr (top and bottom, respectively), as estimated by STARLIGHT. The legend shows the results of the K-S test for comparison of the distributions in the given panel. As a value of zero is measured for most of the quiescent galaxies, the y-axis has been shortened to better show the data for the other samples. Note the difference between the x-axis ranges in the upper and lower panel.}
\label{fig:massfrac_lowM}
\end{figure*}

\begin{figure*}
  \centering
  \includegraphics[scale=0.85]{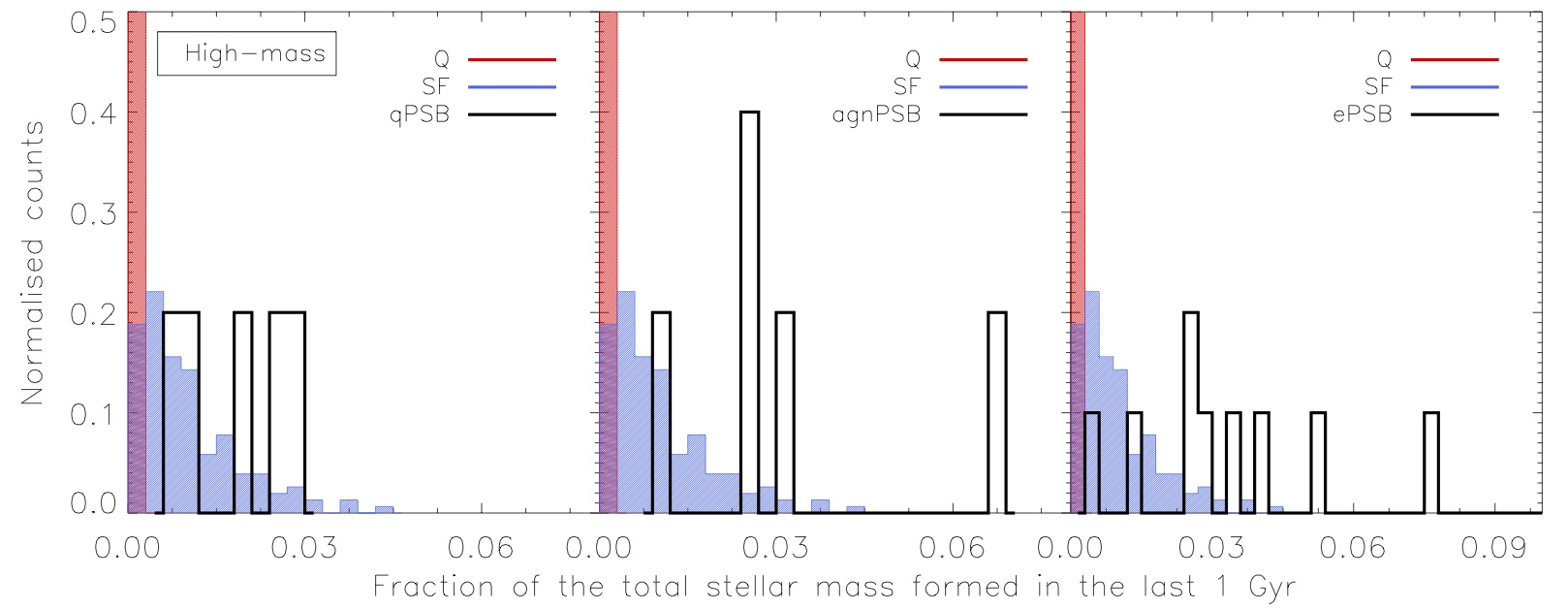}
   \includegraphics[scale=0.85]{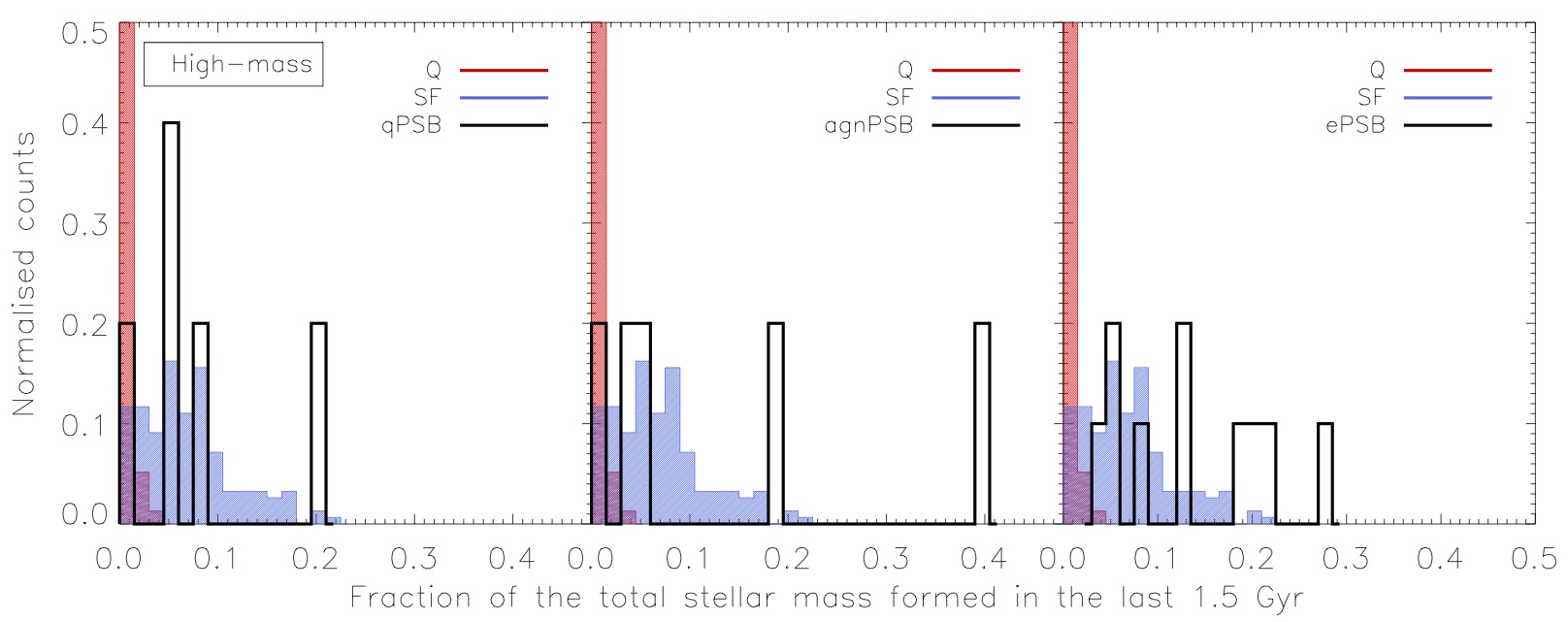}
   \vspace{-5pt}
\caption{Fractions of recently formed mass for the high-mass samples (see the caption of Figure \ref{fig:massfrac_lowM}) for more information. Note the difference between the x-axis ranges in the upper and lower panel.}
\label{fig:massfrac_highM}
\end{figure*}

The histograms in Figures \ref{fig:massfrac_lowM} and \ref{fig:massfrac_highM} show the distributions of the fractions of recently formed mass for the low-mass and high-mass samples, respectively. The quantities $f_{M1}$ and $f_{M15}$ correspond to the fractions of mass formed in the last 1\,Gyr and 1.5\,Gyr, respectively. Each panel shows one Balmer-strong sample (qPSB, agnPSB, ePSB), compared to the control samples of quiescent and star-forming galaxies. Note the change in $x$-axis range between each row. In both samples, we can immediately see that the distributions of $f_{M1}$ and $f_{M15}$ for the Balmer-strong galaxies are skewed towards higher values compared with the control distributions. For the low-mass samples, we show K-S test results comparing the different distributions. The small numbers of objects in the high-mass samples make such tests less useful so are not shown. Table \ref{tab:massfrac} presents the percentage of galaxies with very high and very low $f_{M1}$ and $f_{M15}$ in each PSB sample and the control star-forming sample. 

In both mass bins, the values of $f_{M1}$ and $f_{M15}$ for the quiescent control samples are consistent with zero and the K-S results clearly show that none of the low-mass Balmer-strong samples are consistent with matching the quiescent control sample. For the low-mass star-forming control galaxies we find the majority $(\sim80\%)$ have $f_{M1}<0.05$ and $f_{M15}<0.12$, and a very small fraction (less than $1\%$) have $f_{M1}>0.10$  and $f_{M15}>0.20$. Looking at Table \ref{tab:massfrac}, compared to the star-forming galaxies, (1) a much lower fraction of low-mass Balmer-strong galaxies  ($14\%-54\%$) formed less than 5$\%$ and $12\%$ of their stellar mass in the last 1\,Gyr and 1.5\,Gyr, respectively; 2) a considerably higher fraction of low-mass Balmer-strong galaxies ($21\%-48\%$) have more than $10\%$ and $20\%$ of there stellar mass formed in the last 1\,Gyr and 1.5\,Gyr, respectively. This effect is particularly pronounced in the qPSB and agnPSB samples, pointing to a stronger starburst compared with the ePSB galaxies. The K-S tests additionally show that the distributions of $f_{M1}$ and $f_{M15}$ for the agnPSB and qPSB samples are statistically identical, with all other distributions being different from one another. 

In the high-mass samples the values of $f_{M1}$ and $f_{M15}$ are generally lower than in the low-mass samples, in agreement with previous studies which found that at low redshifts low-mass galaxies tend to have younger stellar ages and higher specific star-formation rates than the high-mass galaxies (e.g. \citealt{Kauffmann+2003b, Asari+2007}). Unfortunately the small sample sizes do not allow for a meaningful statistical analysis, however both Figure \ref{fig:massfrac_highM} and the results in Table \ref{tab:massfrac} suggest a similar picture to the low-mass sample. About 20-50$\%$ of high-mass Balmer-strong galaxies have $f_{M1}>0.03$ and 20-30$\%$ have $f_{M15}>0.2$, compared with $\sim3\%$ and $\sim2\%$ of the star-forming galaxies. In contrast with the low-mass bin, the fractions of recently formed mass are highest for the ePSB sample. 

It is clear that a notable number of the Balmer-strong galaxies, particularly in the low-mass ePSB sample, have a fraction of recently formed mass that is consistent with that found in star-forming galaxies. It is of course entirely possible that these are true post-starburst galaxies with weaker bursts than the others, as it is actually the rapid decline in star formation that leads to the distinctive spectral shape of post-starburst galaxies picked up by the PCA selection method, and this is not exactly what is measured by $f_{M1}$ and $f_{M15}$.  We may expect weaker bursts to fail to use up the entirety of the available gas, thereby accounting for the ongoing star formation. However, it does raise the question of whether they are true post-starburst galaxies or interlopers with strong Balmer absorption lines caused by something other than their star formation history. A careful investigation showed that the values of $f_{M1}$ and $f_{M15}$ are independent of stellar mass, structure and the environment of the galaxies; however, we found some dependence on the signal-to-noise ratio ($SNR$) and the dust content. Out of the ePSB galaxies with $SNR<15$ (20/57), $70\%$ and $55\%$ have the lowest measured fractions of recently formed stellar mass ($f_{M1}<0.05$ and $f_{M15}<0.1$), compared with $46\%$ and $24\%$, respectively, of those with $SNR\geq15$. This suggests that higher SNR spectra than than the typical in SDSS-DR7 are required to reliably identify recent bursts weaker than $\sim$10\% by mass. Furthermore, $76\%$ and $53\%$ of ePSB galaxies with the largest dust content ($H_{\alpha}/H_{\beta} > 4.6$) coincide with $f_{M1}<0.05$ and $f_{M15}<0.1$, respectively, compared with $39\%$ and $25\%$ of the less-dusty ePSB galaxies. This could imply either: (a) the assumption of a single dust screen prevents STARLIGHT from recovering a recent burst in the dustier ePSB galaxies, but they fundamentally do not differ from the rest of the ePSB sample other than by their dust content, or (b) the stronger than average Balmer absorption lines do not reflect a decaying starburst but actually arise from a dust-star geometry such that the O/B stars being obscured behind more dust than average, i.e. these galaxies are not simply more dusty ePSB galaxies but less extreme versions of the ``dusty starburst" galaxies \citep{PoggiantiWu2000}. Untangling these two possibilities is very tricky, pushing us to the limits of spectral fitting techniques, and will be the subject of a future study. We conclude that, while a significant fraction of the ePSB galaxies have had a recent burst of star formation in the past in which typically $\gtrsim10\%$ of the stellar mass was formed, the effects of noise and dust on the galaxy spectra may cause some level of contamination of post-starburst samples in which only weak bursts are identified by spectral fitting. Higher SNR spectra will be needed in order to understand the cause of strong Balmer absorption lines in the majority of ePSB galaxies from their spectra alone. In the following subsections we turn to other properties to further constrain their origins.

\begin{table}
\centering
 \caption{The percentage of galaxies in the Balmer-strong samples and control star-forming galaxies that formed a given portion of their stellar mass in the last 1\,Gyr ($f_{M1}$) and 1.5\,Gyr ($f_{M15}$), as estimated by STARLIGHT. }
 \label{tab:massfrac}
 \begin{tabular}{|c||ccc|c|}
  \hline
Low-mass galaxies & qPSB &  agnPSB & ePSB & SF \\
 \hline
 \hline
  $f_{M1}<0.05$ & 44$\%$ &$45\%$ & $54\%$ & $78\%$  \\
 $f_{M1}>0.1$ &$28\%$ &$30\%$ &$21\%$& $3\%$ \\
 $f_{M15}<0.12$ & $14\%$&$18\%$&$44\%$ & $82\%$\\
 $f_{M15}>0.2$ &$46\%$& $48\%$ &$23\%$&$3\%$ \\
 \hline
 \hline
 High-mass galaxies & qPSB &  agnPSB & ePSB & SF \\
 \hline
 $f_{M1}<0.015$ & 40$\%$&  20$\%$ & 20$\%$ & 77$\%$  \\
$f_{M1}>0.03$ &$20\%$& $40\%$ &$50\%$ & 3$\%$  \\
 $f_{M15}<0.1$ & $80\%$ & $60\%$ &$40\%$ &  $81\%$\\
 $f_{M15}>0.2$ &$20\%$& $20\%$&$30\%$& $2\%$ \\
\hline
 \end{tabular}
\end{table}


\begin{figure*}
  \centering
  \includegraphics[scale=0.90]{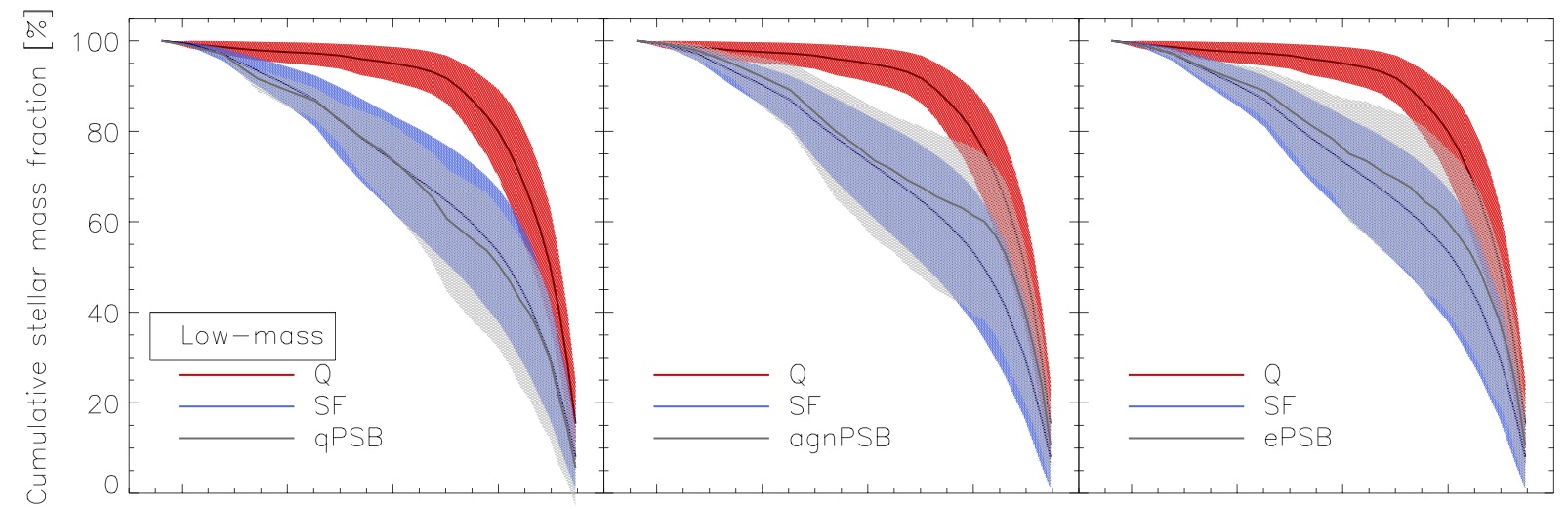}
    \includegraphics[scale=0.90]{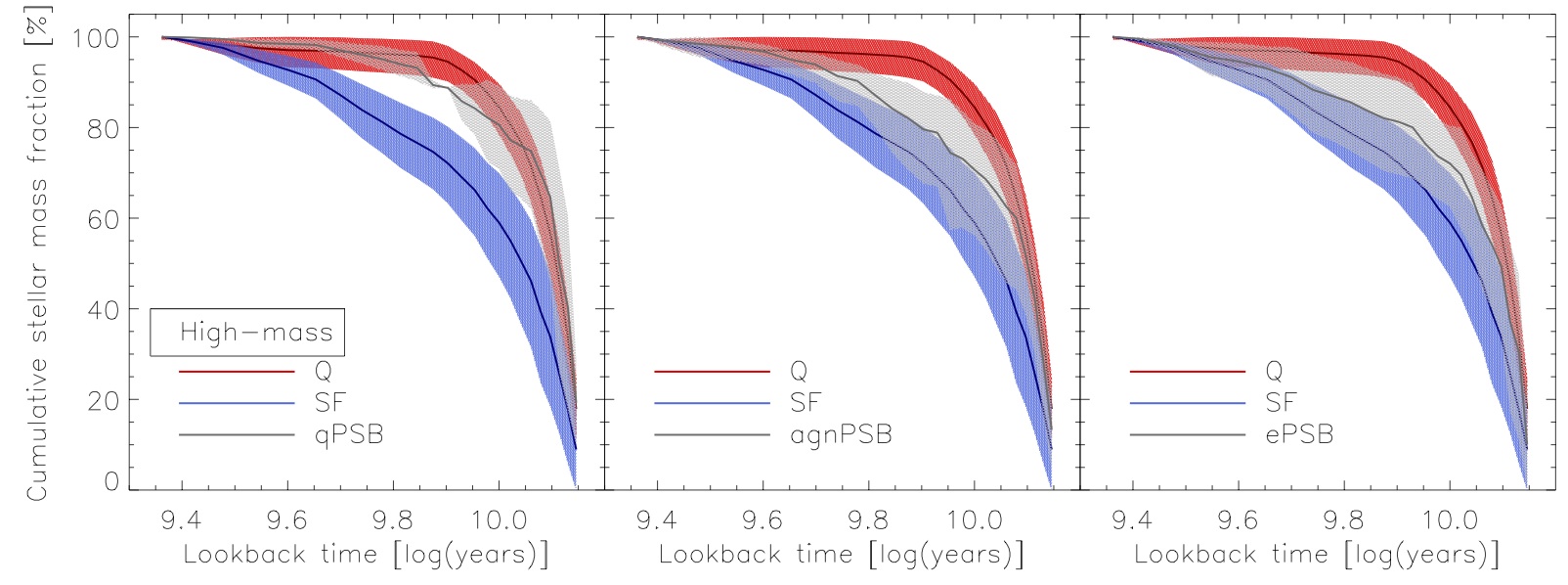}
\caption{The pre-burst star formation histories of the Balmer-strong galaxies, compared with control galaxy samples. Plotted are the stacked cumulative pre-burst star-formation histories, normalised to unity at 2\,Gyr lookback time, obtained for the Balmer-strong galaxies (grey) and of the control star-forming and quiescent galaxies (blue and red, respectively). The top and bottom rows show results for the low-mass and high-mass samples, respectively. For each sample, the solid lines represent the mean values and the shaded regions illustrate the spread of values within the sample (measured by the standard deviation from the mean).
}
\label{fig:sfh}
\end{figure*}

\subsubsection{Star formation prior to the starburst}

Figure \ref{fig:sfh} shows the stacked time evolution of the cumulative fraction of the total stellar mass of the galaxies, with the total mass calculated at 2\,Gyr in lookback time. This allows us to investigate the star formation history of the galaxies prior to the starburst. For each sample, the solid lines represent the mean values and the shaded regions illustrate the spread of values within the sample, quantified by the  standard deviation from the mean. As for the recent star formation history, the difference between the quiescent and star-forming galaxies is clear and in agreement with expectations: the quiescent galaxies clearly build a higher fraction of their stellar mass at earlier times.

In the low-mass regime, the pre-burst star formation histories of all the three Balmer-strong samples, qPSB, agnPSB and ePSB, are almost indistinguishable from the histories of the star-forming galaxies and clearly distinct from the quiescent galaxies. This is consistent with the low-mass Balmer-strong galaxies originating from gas-rich star-forming, rather than quiescent, progenitors. Interestingly, the same is not true at high-mass. In particular, the pre-burst star-formation histories of the massive qPSB galaxies are distinct from the star-forming control, and overlap with those of the quiescent galaxies. This points to red-sequence progenitors, perhaps rejuvenating through minor mergers with gas-rich dwarfs. For the agnPSB and ePSB samples, the stellar mass build-up prior to the burst falls between the two control samples. More detailed inspection revealed that the star-formation histories of both ePSB and agnPSB split roughly equally between those that resemble the quiescent population and those that look more like the star-forming galaxies.

\subsection{Morphology and structural properties}\label{sec:res_morph}

Using the output from the image analysis code described in Section \ref{sec:method_morph} we investigated the morphology and structural properties of all galaxies without nearby stars or other image contaminants (`clean' samples, Table \ref{tab:galcounts}). In Appendix \ref{appendix:AGN} we investigated whether emission from narrow-line AGN affects the measurements of galaxy structure and morphology in our samples. We found no significant effect on any of the light-weighted parameters ($n$, $C$, $G$, $M_{20}$, $A$) measured in the $r$-band and conclude that we can use these measurements to meaningfully compare between galaxies with and without narrow-line AGN. Additionally, in Appendix \ref{appendix:struct} we present relations between selected parameters that may be of interest to some readers. These include $A-C$, $G-M_{20}$ and $n-log(\Sigma_{5})$.

Here we present the results of the analysis of the $r$-band images (we found that the analysis of the $g$- and $i$-band images led to the same conclusions). We additionally visually inspected the 3-colour images of the galaxies for signs of past mergers, which can be difficult to identify with automated measurements. These include tidal features that do not form an asymmetric pattern when observed from a given direction and are therefore not detectable with the shape asymmetry ($A_{S}$). The images were inspected by only one reviewer as the aim of the visual classification was merely to provide subsidiary information to that inferred from the automated proxies - the main component of our analysis.

\begin{figure*}
  \centering
  \includegraphics[scale=0.76]{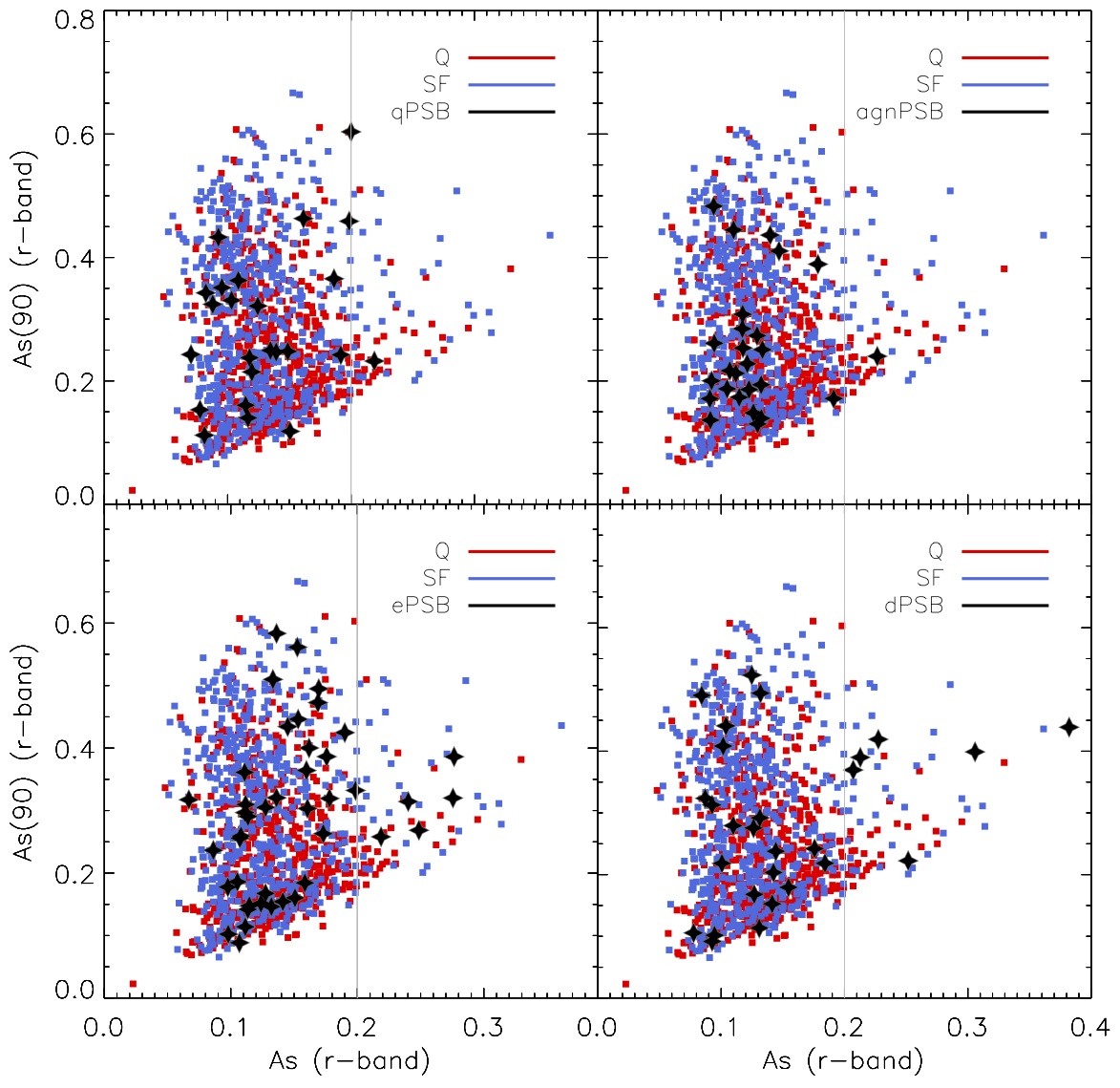}
    \includegraphics[scale=0.76]{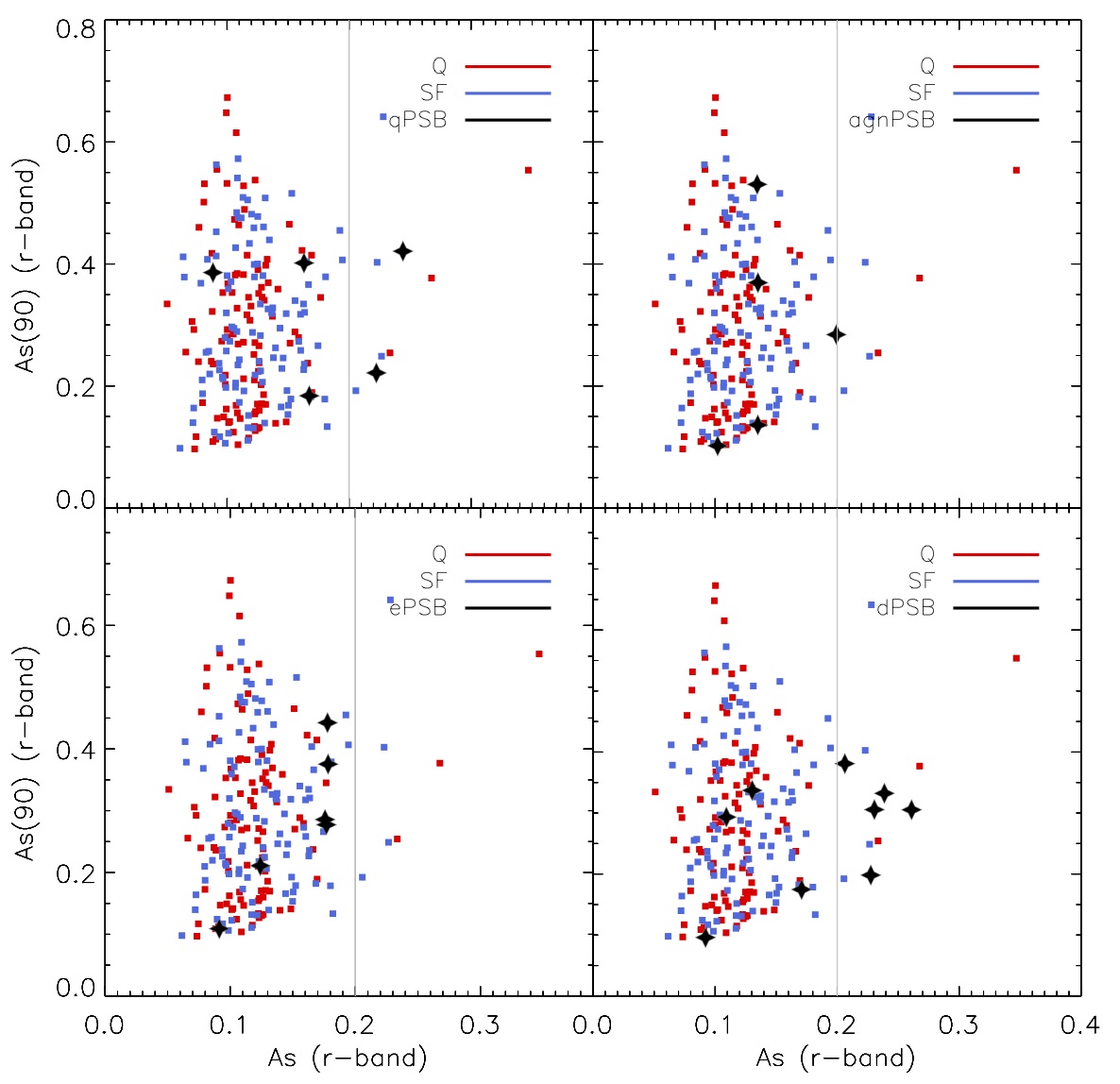}
\caption{The relation between the shape asymmetry parameters, $A_{S}$ and $A_{S90}$ (measured under 180$^{o}$- and 90$^{0}$-rotation, respectively) for the Balmer-strong galaxies (qPSB, agnPSB, ePSB, dPSB) and the control samples of star-forming (SF) and quiescent (Q) galaxies. Both parameters were measured in the $r$-band. The top and bottom panels show results for the low-mass and high-mass samples, respectively. The vertical line marks the value of $A_{S}=0.2$ separating the galaxies with asymmetric projected shapes from those with normal morphologies.}
\label{fig:morph_AsAs90}
\end{figure*}

\begin{table*}
\setlength{\tabcolsep}{10pt}
 \caption{The percentage of galaxies in the Balmer-strong samples with given morphology/structure. {\it Top:} presence of asymmetric tidal features implied by $A_{S}\geq0.2$. {\it Middle:} visually identified morphological disturbance signifying a recent interaction. {\it Bottom:} steep light profiles, characterised by  $n\geq2.0$. The numbers in brackets show the total number of galaxies used to calculate the percentage values in each case. 
 For visual image analysis this equals to the total numbers of galaxies found in our samples but in the case of automated analysis the numbers are lower as they exclude galaxies for which the code failed to obtain a measurement of $A_{S}$ or $n$).}
 \label{tab:morph}
 \begin{tabular}{|c||cccc|cc|}
  \hline
 & qPSB &  agnPSB &  ePSB & dPSB & Q & SF \\
 \hline
\hline
Low-mass galaxies with $A_{S}\geq0.2$  & $8\%$ (24) & $4\%$ (26) & $12\%$ (43) & $17\%$ (23) & $6\%$ (602) & $6\%$ (592)\\
High-mass galaxies with $A_{S}\geq0.2$ & $40\%$ (5) & $0\%$ (5) & $0\%$ (6) & $60\%$ (10) & $3\%$ (101) & $4\%$ (108) \\
\hline

\hline
Low-mass post-mergers & $11\%$ (36) & $3\%$ (33) & $21\%$ (57) & $16\%$ (31) & $2\%$ (783) & $8\%$ (785) \\
High-mass post-mergers  & $80\%$ (5) & $80\%$ (5) & $40\%$ (10) & $67\%$ (12) & $0\%$ (155) & $0\%$ (154) \\
\hline

\hline
Low-mass galaxies with $n\geq2.0$  &  $82\%$ (22) & $84\%$ (25) & $53\%$ (40) & $22\%$ (23) &$92\%$ (537) & $26\%$ (549)  \\
High-mass galaxies with $n\geq2.0$  &  $100\%$ (5) & $80\%$ (5) &  $100\%$ (5) & $63\%$ (8)  & $98\%$ (94) & $55\%$ (103) \\
\hline
 \end{tabular}
\end{table*}

\subsubsection{Asymmetries and signs of interaction}\label{sec:results_morph_A}

Both the visual inspection and automated measurements agree that the Balmer-strong galaxies in our samples are not ongoing mergers. In both low- and high-mass samples, the majority have low light-weighted asymmetry values $A<0.2$, characteristic of normal galaxy types and none have $A>0.35$ commonly found in ongoing mergers. Furthermore, they occupy a similar region of the $G-M_{20}$ parameter space as the control galaxies, with only a few `outliers' in the merger region. The light-weighted asymmetry vs. concentration index and Gini index vs. $M_{20}$ are presented in Figures \ref{fig:morph_CA} and \ref{fig:morph_GM20} respectively. Given the short visibility timescales for merger signatures (0.2-0.4\,Gyr), peaking before coalescence \citep{Lotz+2008}, and the estimated ages of the starburst ($>0.6$\,Gyr), it is not surprising to see few ongoing mergers and this does not rule out a merger origin for the Balmer-strong galaxies. 

As a merger-induced starburst is believed to occur at coalescence of the progenitor galaxies (except bulgeless galaxies, in which case it may occur earlier, see e.g. \citealt{MihosHernquist1996}), it is more likely to observe \emph{post-merger} signatures, such as tidal features, in post-starburst galaxy samples (see \citealt{Pawlik+2016}). In Figure \ref{fig:morph_AsAs90} we show the shape asymmetry measured under 90$^{o}$ and 180$^{o}$ rotation;  generally the values of $A_{S}$ fall below 0.2, meaning that the galaxies do not have visible asymmetric post-merger signatures, such as tidal tails. This is consistent with the results of \citet{Pawlik+2016} who found that by 600\,Myr following the starburst, the shape asymmetry had largely returned to levels similar to control samples. In the top rows of Table \ref{tab:morph} we present the fraction of galaxies in each sample with $A_{S}>0.2$. At low-mass, the ePSB and dPSB samples contain the highest fractions of objects with $A_{S}\geq0.2$ ($12\%$ and $17\%$, respectively), which is a little higher than found in the control samples ($6\%$ for both quiescent and star-forming galaxies). The proportions of low-mass qPSB and agnPSB galaxies with $A_{S}\geq0.2$ are low, consistent with those found in the control samples. At high mass, the qPSB and dPSB galaxies have much higher fractions of post-merger candidates than the control samples ($40\%$ [2/5] and $60\%$ [6/10], respectively), but the other two samples have no positive detections (0/5).  

In the middle rows of  Table \ref{tab:morph} we present the fraction of galaxies in each sample identified as post-merger candidates by visual inspection. We stress that the two post-merger definitions are not equivalent, as the visual classification does not rely on a high degree of asymmetry in the morphological disturbance and is therefore more inclusive. At low-mass, the fraction of ePSB galaxies visually classified as post-mergers is significantly higher than measured with $A_{S}$, and at high mass the same is true for qPSB, agnPSB and ePSB samples.  At high mass, the measured fractions reach $80\%$ (4/5) in both qPSB and agnPSB. This increase in post-merger fractions is due to features which are not asymmetric enough to be detected by $A_{S}$. Interestingly, those Balmer-strong galaxies with $A_{S} \geq 0.2$ tend to have low/moderate values of $A_{S90}$, which also points to tidal features with little azimuthal asymmetry.
Given that some simulations have shown that symmetric tidal feature patterns such as shells may be formed not only through satellite accretion but also in late stages of major mergers (see e.g. \citealt{HernquistSpergel1992}, Pop et~al. submitted\nocite{Pop+2017subm}), it is interesting to speculate that symmetric tidal features are more common in more evolved systems, consistent with the starburst ages of $\sim1-1.5$\,Gyr in these samples. However, as minor mergers may lead to similar signatures, further analysis of simulations would be required to confirm this.

\subsubsection{The profile and central concentration of light} 

\begin{figure*}
  \centering
  \includegraphics[scale=0.57]{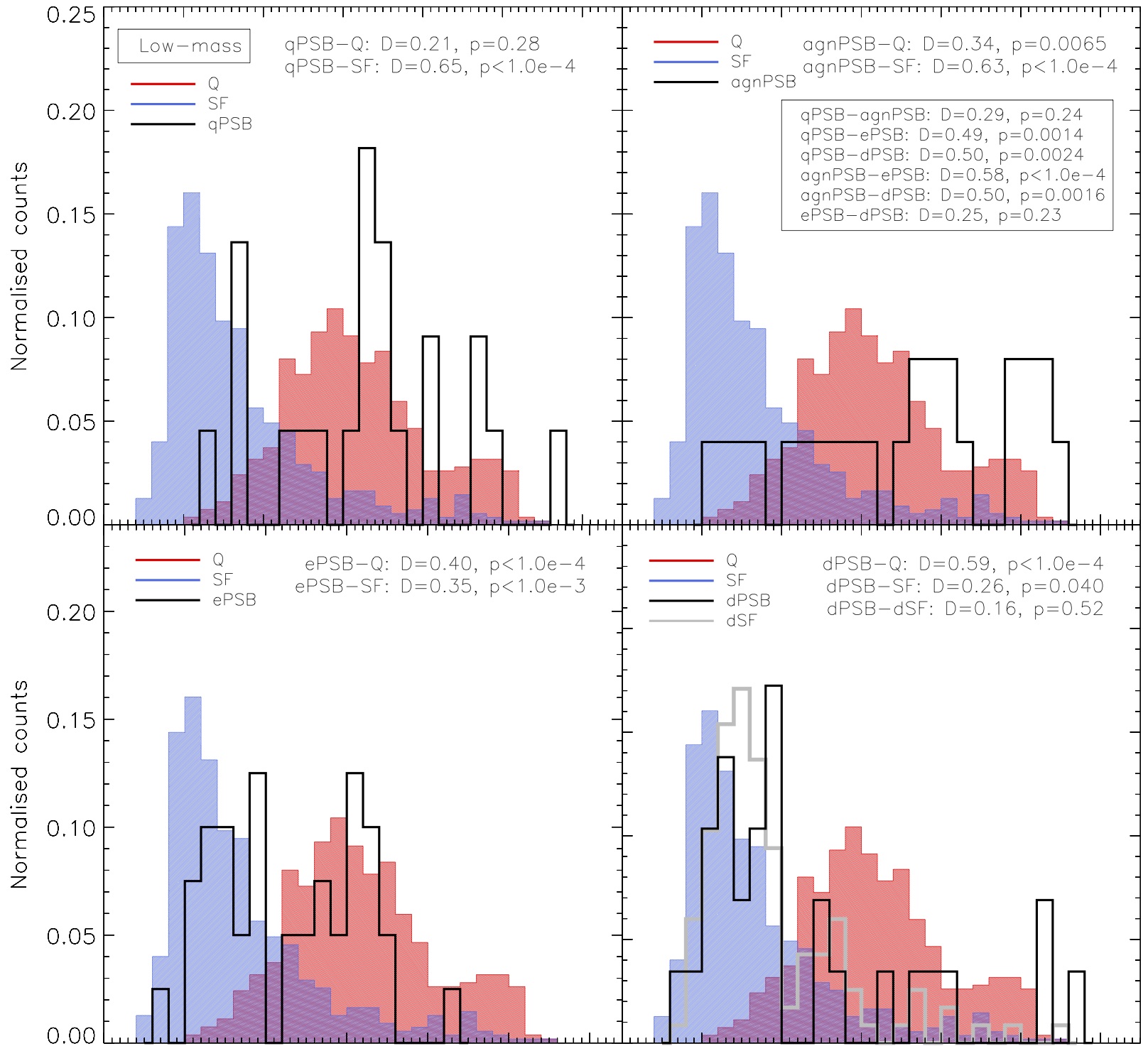}
      \includegraphics[scale=0.57]{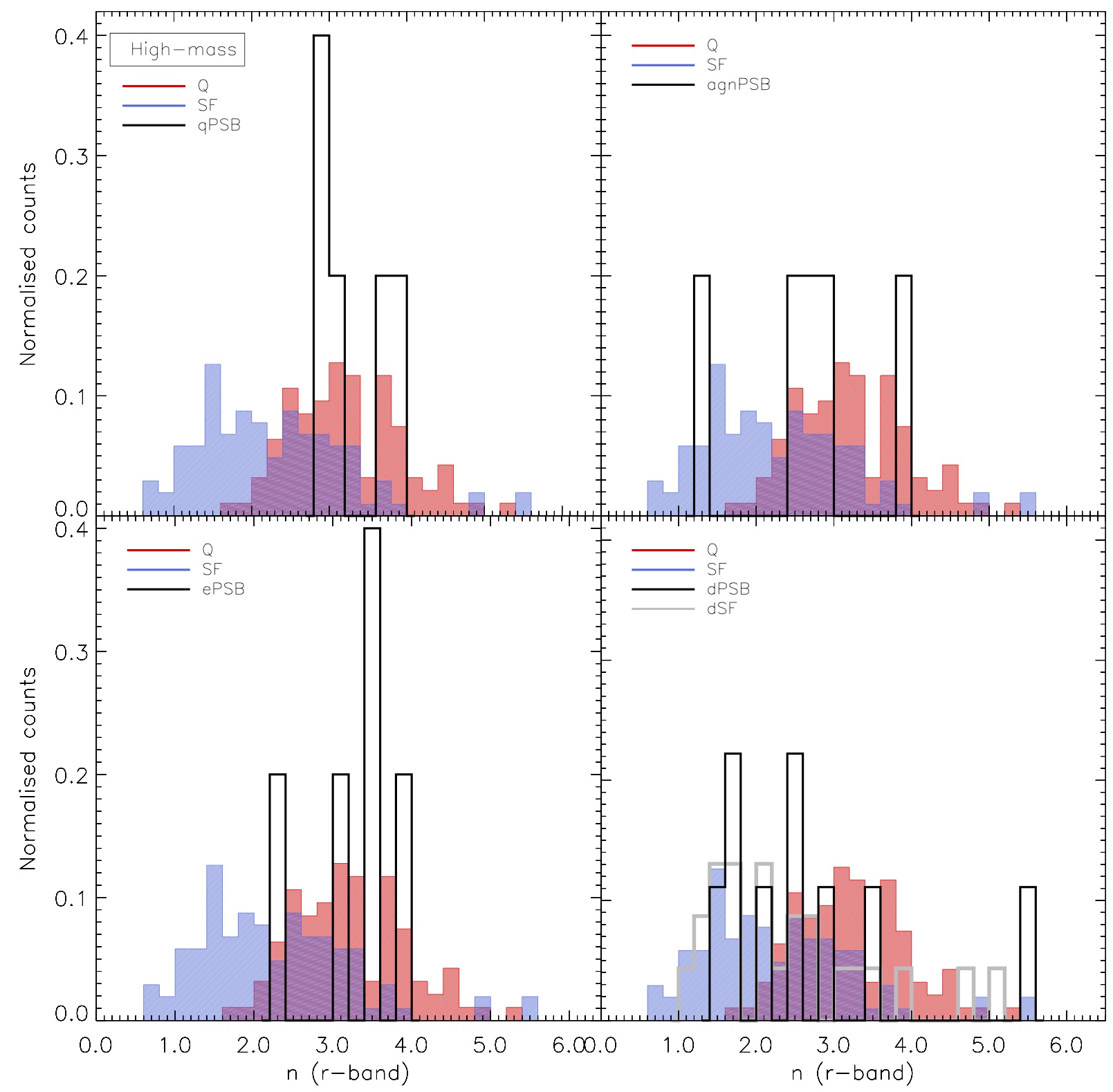}
\caption{The values of the $r$-band S{\'e}rsic index measured for the low-mass (top) and high-mass (bottom) galaxies in the `clean' samples (Table \ref{tab:galcounts}). The legend shows the results of the KS-test for comparison of the different distributions. }
\label{fig:Sersic}
\end{figure*}

\begin{figure*}
  \centering
  \includegraphics[scale=0.56]{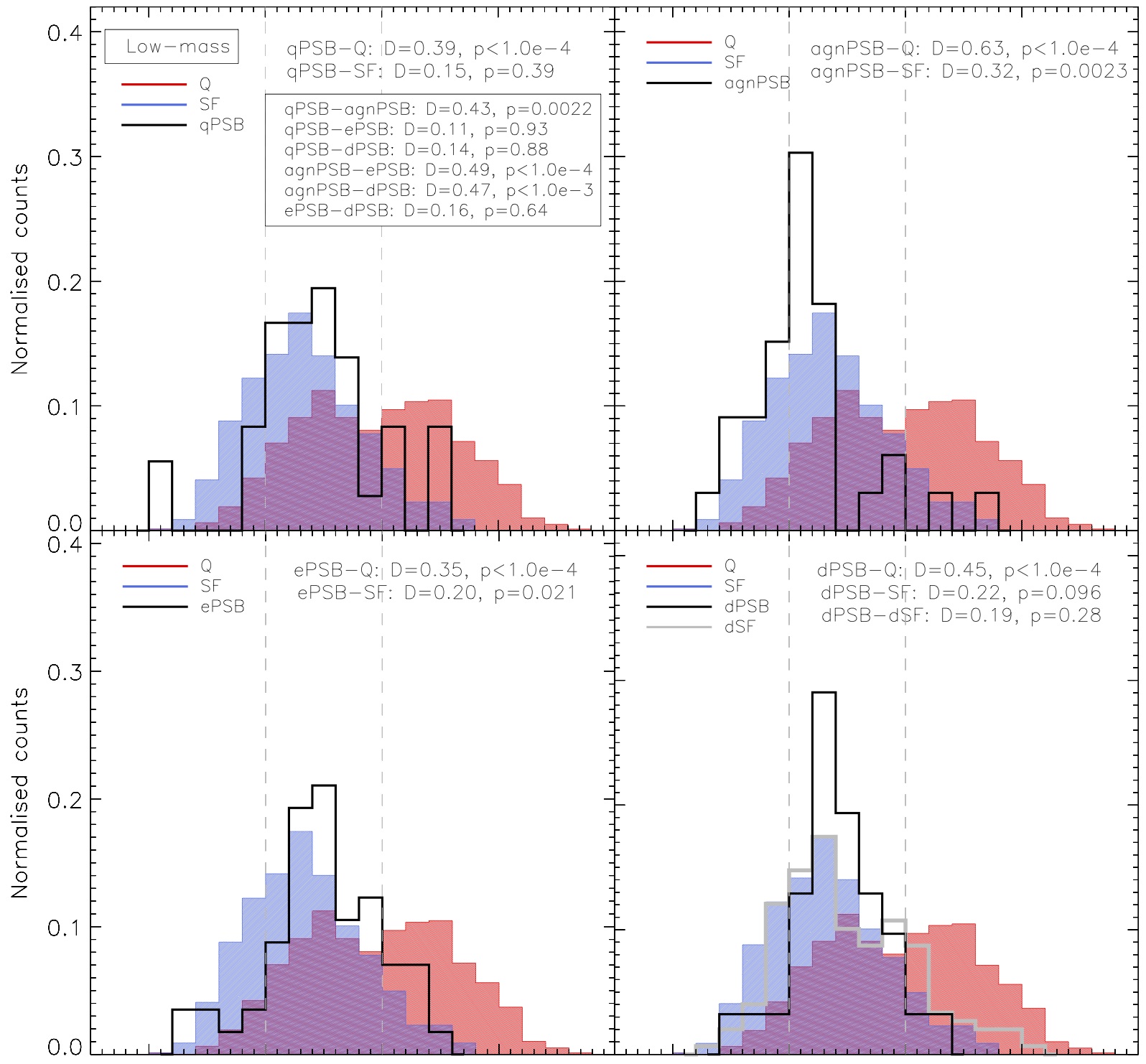}
    \includegraphics[scale=0.56]{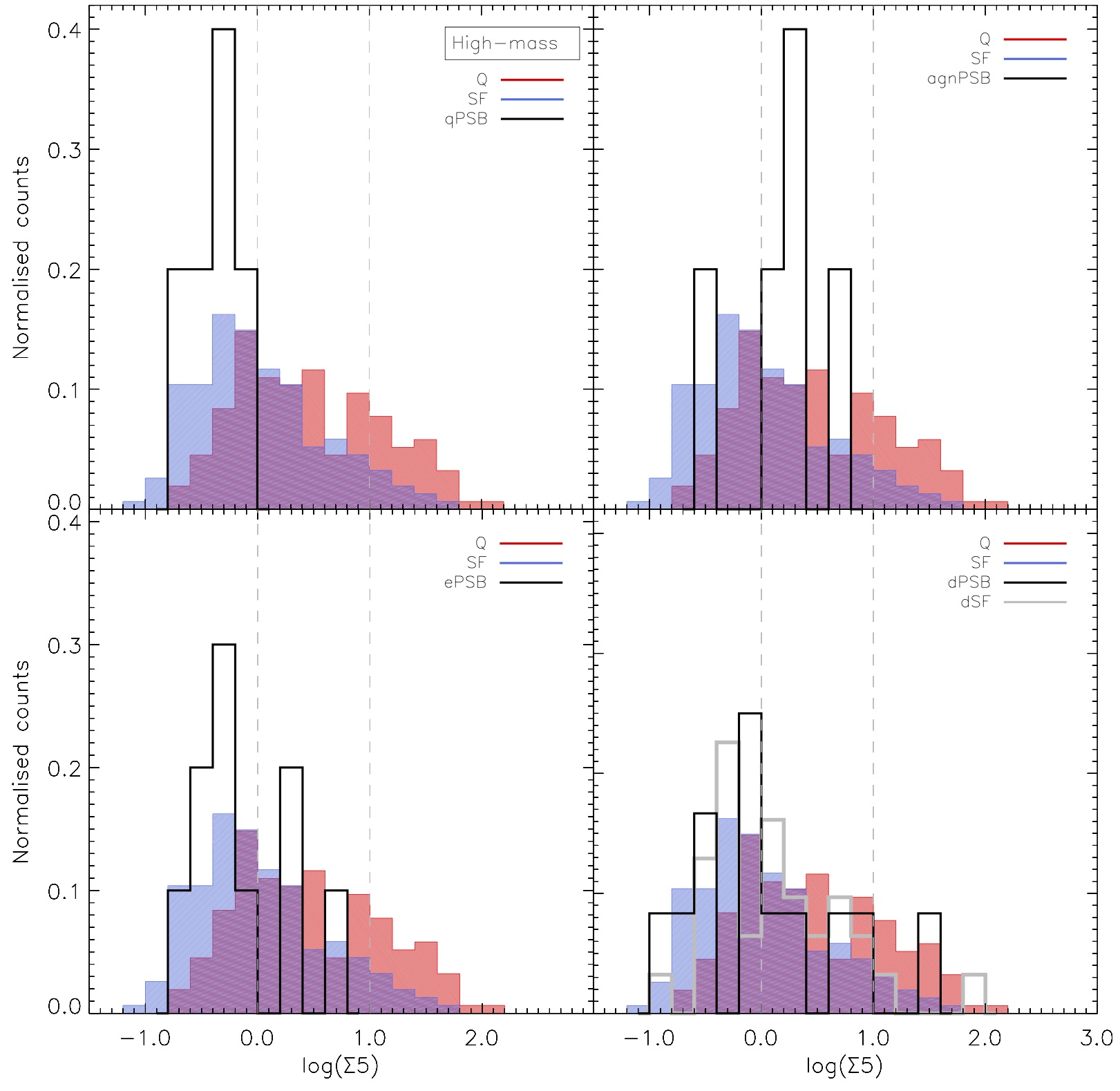}
\caption{The values of the projected density measured to the 5th nearest neighbour ($\Sigma_{5}$) measured for the low-mass (top) and high-mass (bottom) samples (showing only those galaxies that are not close to a survey boundary). The legend shows the results of the KS-test for comparison of the different distributions for the low-mass samples. The dashed lines show the demarcation between very low-density environments, loose groups, and compact groups and clusters used in the text.}
\label{fig:env}
\end{figure*}
\begin{table*}
 \caption{The percentage of low-mass galaxies with values of $\Sigma_{5}$ pointing to different environment types. The numbers in brackets correspond to the total number of galaxies in each sample and exclude those galaxies for which the measurements of $\Sigma_{5}$ are unreliable due to close proximity to the survey edge.}
 \label{tab:env}
 \begin{tabular}{|c|c||cccc|cc|}
  \hline
& & qPSB & agnPSB & ePSB & dPSB & Q & SF \\
\hline
\hline
Low-mass sample &  $\Sigma_{5}<0.0$ & $9\%$ (35) & $36\%$ (33) & $4\%$ (55) & $7\%$ (30) & $3\%$ (775) & $19\%$ (770)  \\
 & $0.0\leq\Sigma_{5}<1.0$  & $74\%$ (35) & $58\%$ (33) & $82\%$ (55) & $87\%$ (30) & $49\%$ (775) & $74\%$ (770) \\
 & $\Sigma_{5}\geq1.0$ & $17\%$ (35) & $6\%$ (33) & $15\%$ (55) & $7\%$ (30) &  $48\%$ (775) & $7\%$ (770) \\
\hline
High-mass sample &  $\Sigma_{5}<0.0$ & $100\%$ (5) & $20\%$ (5) & $70\%$ (10) & $58\%$ (12)  & $21\%$ (153) & $52\%$ (151)  \\
 &  $0.0\leq\Sigma_{5}<1.0$   & $0\%$ (5) & $80\%$ (5) & $30\%$ (10) & $33\%$ (12) & $64\%$ (153) & $45\%$ (151) \\
 &  $\Sigma_{5}\geq1.0$ & $0\%$ (5) & $0\%$ (5) & $0\%$ (10) & $8\%$ (12)  & $15\%$ (153) & $3\%$ (151) \\
\hline
 \end{tabular}
\end{table*}

In Figure \ref{fig:Sersic} we show the distribution of S{\'e}rsic indices, with K-S test results comparing distributions to the control samples and each other. The lower rows of Table \ref{tab:morph} show the fraction of each sample with steep light profiles, characterised by  $n\geq2.0$. 

The low-mass Balmer-strong galaxies span the whole dynamic range in S{\'e}rsic index, with $0.5\le n \le 5.5$, pointing to a range of structural properties, from highly concentrated single component spheroids to disk-dominated systems. The distributions found for qPSB and agnPSB are comparable to the quiescent control sample ($D\sim0.2$, $p=0.28$ and  $D\sim0.3$, $p=0.0065$), indicating high central concentration characteristic of massive spheroids ($82\%$ and $84\%$ with $n\geq2$, respectively, compared with $92\%$ of the quiescent galaxies). The ePSB sample has typically lower values of $n$ than both the qPSB and agnPSB samples, with K-S statistics showing the distribution is distinct from the quiescent control sample. The dPSB sample has the lowest values of $n$ for all Balmer-strong samples, with the distribution comparable with the control sample of star-forming galaxies ($D\sim0.3$, $p=0.04$ and $22\%$ with $n\geq0.2$, compared with $26\%$ of the star-forming galaxies). The slight shift towards higher values of $n$ could be due to the high dust content of the dPSB galaxies as a similar shift is observed when comparing the distribution of $n$ between the star-forming galaxies with the control sample of dusty star-forming galaxies. The distributions of $n$ for the dPSB and dusty star-forming control are consistent with being drawn from the same underlying distributions ($D\sim0.2$, $p=0.5$).

The picture is very different at high-mass, where we found high values for the  S{\'e}rsic index comparable with the quiescent control sample for the qPSB, agnPSB and ePSB samples ($100\%$, $80\%$ and $100\%$ with $n>2.0$). As for the low-mass sample, the distribution of $n$ for the dPSB galaxies resembles the dusty star-forming control sample. 

Similar conclusions are reached from the concentration and Gini indices, however, the separation between samples is less apparent than in the values of $n$ (see Figures \ref{fig:morph_CA} and \ref{fig:morph_GM20}).

\subsection{Environment}\label{sec:results_env}


In Figure \ref{fig:env} we present the distributions of the projected number density ($\Sigma_{5}$), and the results are summarised in Table \ref{tab:env}. For the low-mass galaxies we show the K-S test results comparing the different distributions. We tested two different cuts on the samples, firstly just removing those galaxies that fell near the edge of the survey, and secondly also removing objects for which the difference between the spectroscopic and photometric value for $\log(\Sigma_{5})$ was greater than 0.4\,dex \citep{Baldry+2006}. The results using the additional cuts were not significantly different. 

Overall, we see that both low-mass and high-mass Balmer-strong galaxies tend to occupy the low/medium-density environments log($\Sigma_{5})<1.0$, similar to the star-forming control samples. This is evident in the K-S test statistics for the low-mass sample. The only possible difference is for the low-mass agnPSB, where the distribution shifts towards lower values ($36\%$ have log($\Sigma_{5})<0.0$ compared to $19\%$ of the star-forming control), although the K-S test shows that any difference is not formally significant ($D=0.32$, $p=0.0023$).
At high-mass there is a possible indication that the qPSB and ePSB galaxies are preferentially found in lower density environments than the star-forming control sample. However, the small numbers prevent any firm conclusions to be drawn. 

Qualitatively the above results are generally unaffected by the choice of measurement of the number density, i.e. the distributions of the mean $\Sigma_{5}$ shown in Figure \ref{fig:env} point to the same local environments of the post-starburst galaxies relative to the control samples, as the distributions of the individual photometric/spectroscopic measurements. The one exception is the ePSB sample, where $\Sigma_{5}$ derived solely from the photometric redshifts suggests slightly higher-density environments relative to the control samples than the mean values.

We verified that there were no trends in the structural properties of the low-mass Balmer-strong galaxies as a function of their local environment (Figure \ref{fig:morph_nE5}), although the few morphologically disturbed qPSB and agnPSB do tend to reside in under-dense environments ($\Sigma_{5}\lesssim0.0$). 

The fact that the environments of the low-mass Balmer-strong galaxies are broadly consistent with the star-forming control sample is in agreement with the results presented in Section \ref{sec:res_sfh} showing that the pre-burst star formation histories are also characteristic of star-forming galaxies. The possible tendency of the high-mass Balmer-strong galaxies to be found in lower density environments than the control samples, while their pre-burst star formation histories suggest they are originate from a mix of star-forming and quiescent galaxies, might indicate that environment is the most important factor driving the occurrence of starbursts at high mass. However, larger samples would be needed to verify this conclusion.




\section{Discussion}

The wide range of properties of the Balmer-strong galaxies revealed during our analysis implies that there is no unique pathway that leads to their formation, and that the different conclusions drawn about their origins in the literature may all be valid in certain circumstances. Previous observations of Balmer-strong or post-starburst galaxies have suggested they are present in a range of different environments, with varying incidence depending on the cosmic epoch, stellar mass and environment (see e.g. \citealt{Poggianti+2009}). As such, post-starburst galaxies may represent a phase that is common to a variety of different mechanisms driving galaxy evolution. 

A commonly suggested mechanism for the formation of post-starburst galaxies is through mergers of gas-rich galaxies, which can leave faint visual signatures in the morphology of the merger remnant. We therefore begin our discussion by investigating the timescale of visibility of morphological signatures of a past merger using mock galaxy images from hydrodynamical merger simulations (Section \ref{disc:mergers}). We then bring together the results of our analysis of the star-formation histories, morphologies, structural properties and the environments of the local Balmer-strong galaxies to discuss which of our samples are true post-starburst galaxies and which are more likely to be interlopers (Section \ref{disc:psb}), and to investigate the evolutionary pathways that lead to their formation (Section \ref{disc:pathways}). 

\subsection{Timing the visibility of post-merger features}\label{disc:mergers}

Although the structural evolution of galaxies in merger simulations has already been studied, using various measures of galaxy structure and morphology, the previous studies have not included the new merger-remnant sensitive shape asymmetry. Using hydrodynamical simulations we created mock images of galaxies undergoing a merger, recorded at 20-Myr time intervals, in order to study the evolution of the merger morphology as measured by $A_{S}$. 

The simulations, described in detail in Appendix \ref{sec:mergers}, were performed using the entropy conserving smoothed particle hydrodynamics code \gadget\ \citep{Springel2005}, with improved SPH implementation - SPHGal \citep{Hu+2014,Eisenreich+2017} and include radiative cooling, star formation and feedback from stars and supernovae.
In this work we focus mainly on equal-mass mergers of gas-rich galaxies with three different initial morphologies (Sa, Sc and Sd), and in three different dynamical configurations (prograde-prograde 00, prograde-retrograde 07 and retrograde-retrograde 13; see \citealt{NaabBurkert2003}). We also analyse three simulations with a mass ratio of 1:3, involving galaxies with the same morphologies (Sc) but different orbital configurations (prograde-prograde 00, prograde-retrograde 07 and retrograde-retrograde 13).
We limit ourselves to a small number of simulations only to illustrate how the timescale of visibility of the tidal features varies with the conditions of the interaction. A more detailed analysis of a full suite of simulations is left for future work. The mock images were created with the noise properties of the SDSS imaging data (see Appendix \ref{appendix:mergers_img}, and analysed with the same code as the real data (see Section \ref{sec:method_morph}), to ensure a truly meaningful comparison with the results presented in this paper.

\begin{figure}
\centering
  \includegraphics[width=\columnwidth]{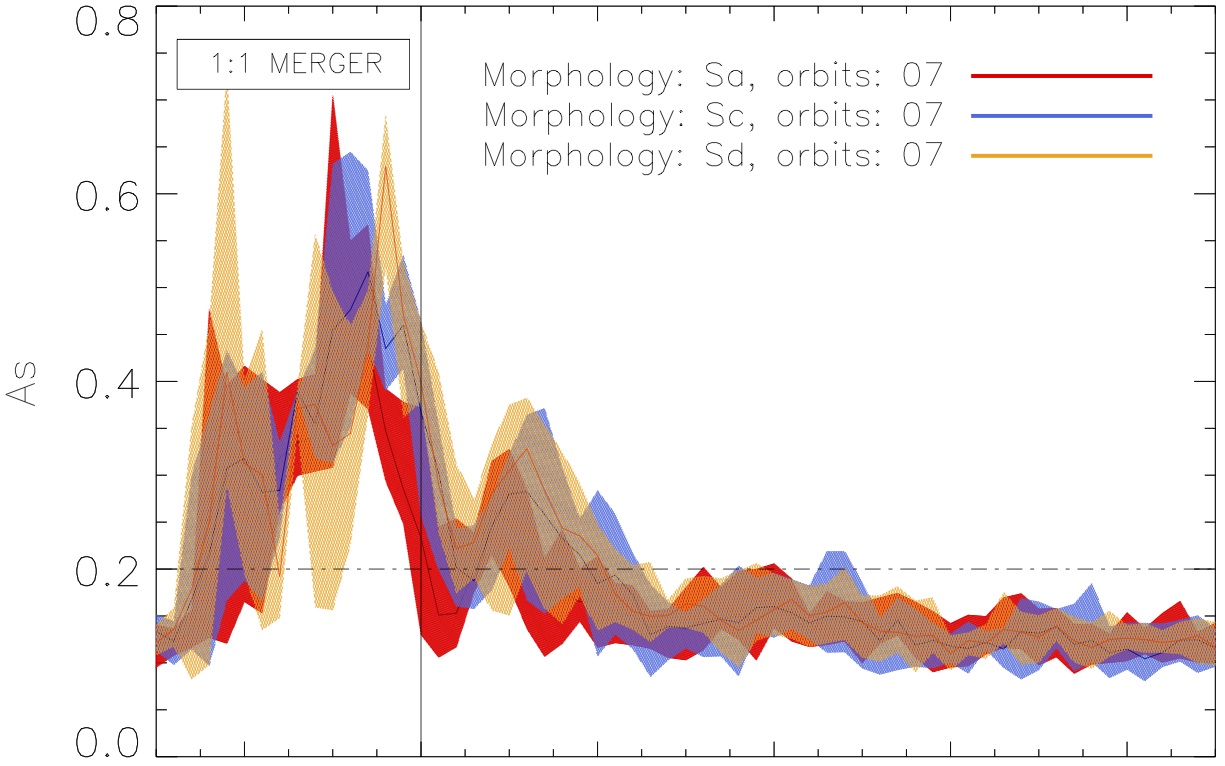}
  \includegraphics[width=\columnwidth]{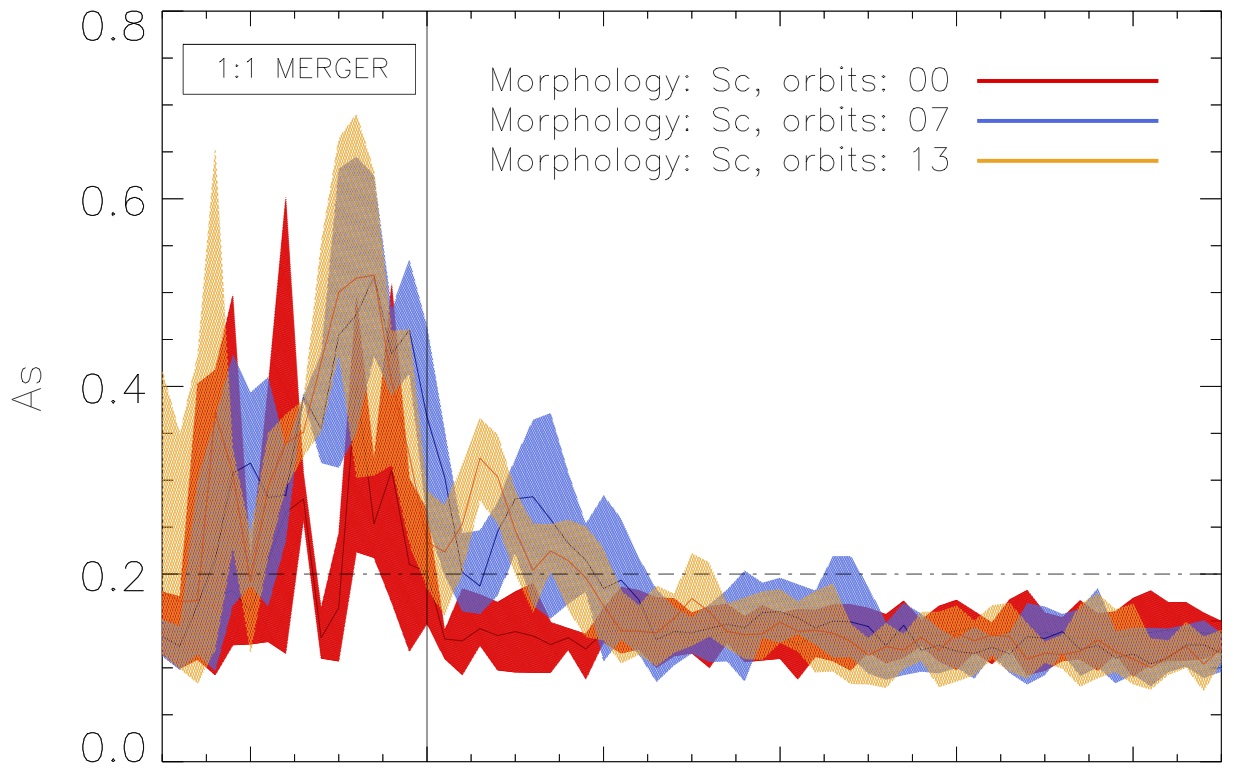}
   \includegraphics[width=\columnwidth]{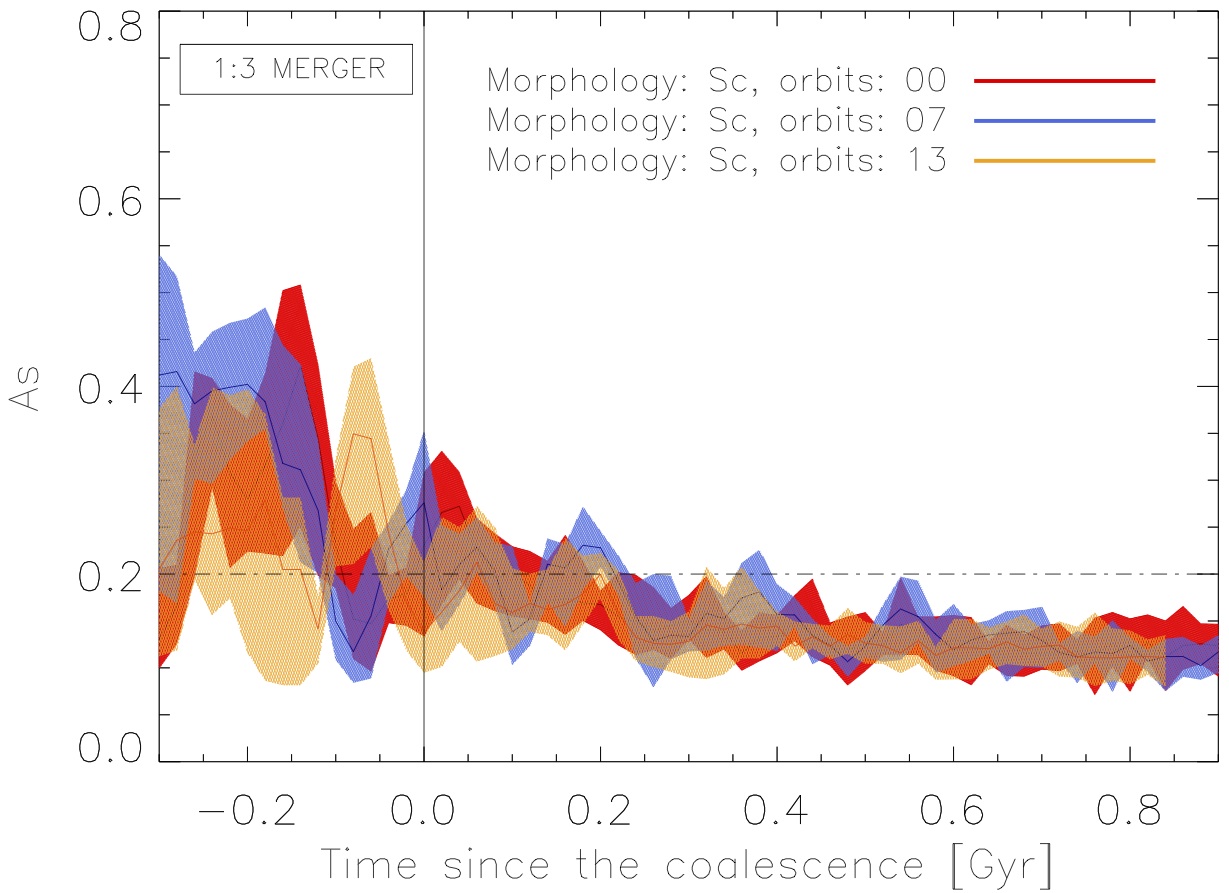}
\caption{The time-evolution of the shape asymmetry (see Section \ref{sec:method_morph}) as measured in the mock images of the simulated galaxy mergers. The spread in the data represents the minimum and maximum values calculated per given simulation, in images synthesised using six different viewing angles. Equal-mass mergers are shown in the top and middle panels and those with mass ratio 1:3, can be found in the bottom panel. In the top panel we vary the progenitor morphology, while in the middle and bottom panels the initial orbital parameters, as indicated in the legend and described in Appendix \ref{sec:mergers}. }
\label{fig:mergers}
\end{figure}

In Figure \ref{fig:mergers} we show the evolution of the shape asymmetry, $A_{S}$, with time since the coalescence of the two galaxies.
As described in Section \ref{sec:method_morph}, $A_{S}$ can be used to identify asymmetric tidal features in galaxies, signifying a recent morphologically disruptive event, such as a merger. Using SDSS images, \citet{Pawlik+2016} concluded that galaxies displaying asymmetric features are characterised by $A_{S}\geq0.2$, while values of $A_{S}\sim 0.1$ correspond to regular, undisturbed morphologies. This is in good agreement with the end point of the simulations. In the top and middle panel we show equal-mass mergers, and in the bottom panel we show 1:3 mass ratio mergers. The different data sets correspond to different simulations, with varying progenitor morphology (upper panel), and initial orbital configuration of the progenitors (middle and bottom panels). 

The timescale of visibility of the tidal features, measurable by $A_{S}\geq0.2$ varies with morphology, orbital configuration and the mass ratio of the progenitors. For the equal-mass mergers, the shortest timescales, measured from the time of coalescence, are observed for the early-type Sa morphology ($<200$ Myr) and the coplanar prograde-prograde orbital configuration (00), in which case the tidal features vanish immediately after coalescence. In the cases where the galaxies have smaller bulge components (Sc and Sd morphologies) and where they collide in more asymmetric dynamical configurations (07 and 13) the asymmetric tidal features induced by the interaction prevail for longer. This is consistent with expectations, as in major merger simulations retrograde configurations produce the most violent effects due to the anti-alignment of the galactic versus orbital angular momenta. In the case of unequal-mass mergers the asymmetry of the tidal features tends to be lower and vanish more rapidly following the coalescence of the progenitors, compared with the 1:1 mergers with the same morphology and orbital parameters.

Depending on the progenitor morphology and initial configuration of orbits, a post-merger with the merger age of $\sim100$ Myr can have a a wide range of values of the shape asymmetry ($\sim0.1$ -- $\sim0.4$). Regardless of the initial conditions, the tidal features generally fade away after $\sim500$\,Myr from the coalescence in 1:1 mergers, and after $\sim400$\,Myr in those with a mass ratio of 1:3.

The above results could explain the lack of visible post-merger features among our Balmer-strong samples (Section \ref{sec:results_morph_A}), given that their estimated starburst ages are greater than $\sim0.6$\,Gyr. The lack of such features is therefore not sufficient to rule out a merger origin of the Balmer-strong galaxies.

\subsection{The different families of Balmer-strong galaxies}\label{disc:psb}


As described in Section \ref{sec:psbselection} we separated the Balmer-strong galaxies at $0.01<z<0.05$ based on their emission line properties into quiescent `K+A' galaxies (qPSB) with no/weak emission lines, those with a measurable level of nebular emission from ongoing star-formation (ePSB and `dusty' dPSB) or AGN/shock activity (agnPSB). We obtained samples of 36, 33, 57 and 31 qPSB, agnPSN, ePSB and dPSB, respectively, in the low-mass regime ($10^{9.5}<\mbox{M}_{\star}/\mbox{M}_{\sun}<3\times10^{10}$) and 5, 5, 10 and 12 at higher masses ($\mbox{M}_{\star}/\mbox{M}_{\sun}>3\times10^{10}$). The lower mass limit was set to ensure that all samples were complete, including the quiescent control sample. In this section we look at the similarities and differences between the different samples, assessing the likelihood of them being true ``post-starburst" galaxies, before progressing onto their likely origins and fate in the following subsection. 

\subsubsection{The quiescent Balmer-strong galaxies and AGN/shocks hosts (qPSB and agnPSB)}

As the origin of post-starburst galaxies is often linked with violent dynamical processes, we might expect their spectra to show evidence of AGN and shocks. Previous results have shown that strong Balmer absorption lines are common in the spectra of narrow-line AGN samples (e.g. \citealt{Kauffmann+2003c,CidFernandes+2004a}) and narrow-line AGN are common in samples of galaxies with strong Balmer absorption lines (e.g. \citealt{Yan+2006, Wild+2007}). This motivated \citet{Wild+2009} to discard the emission line cut when selecting post-starburst galaxies. \citet{Tremonti+2006} and \citet{Alatalo+2016} also found evidence of galactic winds and shocks in post-starburst galaxies, respectively. Furthermore, AGN have often been invoked to aid the quenching of star-formation following simulated galaxy mergers, causing the galaxies to become quiescent (e.g. \citealt{Hopkins+2006}).
In this section we compare the properties of the ``classical" quiescent Balmer-strong galaxies (qPSB), with the sample that have emission line ratios that lie above the \citet{Kewley+2001} demarcation line (agnPSB). While we have focussed on an AGN as the most likely origin of the high ionisation emission lines, we remind the reader that our selection does not entirely rule out shocks as an alternative origin. However, the requirement for the equivalent width of H$\alpha$ to be larger than 3\AA\ does rule out weak shocks, as well as the class of ``retired" galaxies where the high ionisation lines are caused by evolved stellar populations \citep{CidFernandes2011}.  

At both high and low masses, the close similarities between the structural properties, star-formation histories and environments of the qPSB and agnPSB suggest that they are of the same physical origin. The distributions of their structural and morphological parameters are indistinguishable from one another, and closely resemble those of the quiescent control sample (Figures \ref{fig:morph_AsAs90}, \ref{fig:Sersic}, \ref{fig:morph_CA}, \ref{fig:morph_GM20}) apart from a possible enhancement in visually identified post-merger features (Table \ref{tab:morph}). Both are found in intermediate-density environments, matching the star-forming control sample (Figure \ref{fig:env}), although at low-mass there is tentative evidence that the agnPSB galaxies are found in slightly lower density environments than the qPSB galaxies on average. In the low mass sample, the mean and distribution of pre-burst star formation histories are barely distinguishable, matching the star-forming control sample (Figure \ref{fig:sfh}), and the distributions of recently formed stellar mass are again very similar with $\sim50\%$ indicating a recent starburst in which $>$20\% of the stellar mass was formed (Figure \ref{fig:massfrac_lowM} and Table \ref{tab:massfrac}). At high mass the pre-burst star formation histories of the qPSB and agnPSB galaxies are mixed, although the majority (4/5 and 3/5) closely resemble the quiescent control sample. 

The STARLIGHT spectral analysis does not identify a high fraction of mass formed in the last 1.5\,Gyr in 14\% (18\%) of the low-mass qPSB (agnPSB) galaxies, and in 80\% (60\%) of the high-mass qPSB (agnPSB) galaxies. Either the fraction of mass formed in the starbursts was $\lesssim10\%$, and therefore difficult to identify given the limited SNR of the data and limitations of the models, or some Balmer-strong galaxies with no ongoing star-formation are not actually post-starburst. While the latter seems unlikely, further detailed spectral analysis would be required to confirm that these Balmer-strong galaxies are post-starburst, including two-component dust, covariances between fitted parameters, and an assessment of the limitations (and means of improvement) of the stellar population models used. 

\emph{From the samples and observations available at the present time, there is no evidence that low-redshift galaxies with strong Balmer absorption lines and high ionisation emission lines do not have the same origin as traditional ``K+A" galaxies with strong Balmer absorption lines and weak emission lines.} It would appear that the AGN activity switches on and off on timescales shorter than the post-starburst phase. Because our samples are complete in stellar mass, we can use the relative number of qPSB and agnPSB galaxies to constrain the duty cycle of the AGN during the post-starburst phase (ignoring the possible contribution of shocks to the agnPSB samples). In our low (high) mass sample we find 36 (5) qPSB and 33 (5) agnPSB i.e. the AGN must be ``on" for $\sim$50\% of the time. 
This indicates that previous work that selected post-starburst galaxies with stellar masses $>10^{9.5}\mbox{M}_\odot$ using cuts on the emission lines will have underestimated their number density by a factor of two. 
A similarly high fraction  of AGN ($\sim40\%$) was recently found in a sample of bright compact star-forming galaxies at $z=2$, which are likely progenitors of compact quiescent galaxies found at that epoch \citep{Kocevski+2017}. While the masses and effective radii of our post-starburst galaxies do not imply significant compactness, it is encouraging to find similarly high incidence of AGN among potential progenitors of quiescent galaxies at both low and high redshift.

\subsubsection{The star-forming and dusty Balmer-strong galaxies (ePSB and dPSB)}\label{sec:discussion_edpsb}

As the transition between the starburst and quiescent post-starburst phases is not instantaneous, declining levels of star-formation should be observed following the starburst. Balmer-strong galaxies with emission indicative of ongoing star-formation are possible candidates for the transitioning phase (e.g. \citealt{Wild+2010,Pawlik+2016,Rowlands+2017}); however their true nature is still under debate as some studies regard them as ongoing dust-obscured starbursts (e.g. \citealt{Dressler+1999, Poggianti+1999}), and they may also be caused by an event that was not sufficiently disruptive to permanently halt the star-formation \citep{Rowlands+2015}. In this paper we deliberately included all Balmer-strong objects, regardless of their emission line characteristics, in order to make an objective assessment of their likely origins. They account for $\sim60\%$ of the total number of Balmer-strong galaxies (Table \ref{tab:galcounts}). Here we compare the properties of the Balmer-strong galaxies with emission-line galaxies indicating star-formation (ePSB) with those in which the Balmer decrement (ratio of $H\alpha$ and $H\beta$ emission lines) points to significant dust contents (dPSB).


The difference in structure between the ePSB and dPSB samples clearly suggests that the two classes represent objects of different physical nature. The ePSB are split roughly equally between low and high values of $n$ and at low-mass the distribution resembles neither the star-forming nor quiescent control samples (Figure \ref{fig:Sersic}). On the other hand, the distribution of $n$ measured for the dPSB class is unimodal and resembles that found for the star-forming control sample, in particular those that are highly dust-obscured. In fact, there is no measurement that we were able to make in which the dPSB sample differs significantly from the star-forming control sample, except in the fraction of galaxies with post-merger signatures. Given that their dust content prevents us from analysing their star formation histories, \emph{we are unable to find any evidence that the dPSB galaxies are truly ``post-starburst" and we concur with previous literature that they may be dust-enshrouded star-forming galaxies, in which the strong Balmer absorption lines are caused by the preferential obscuration of O/B stars.} It remains possible that the dPSB are progenitors of (some of) the post-starburst galaxies \citep{Yesuf+2014}, however, we are unable to make such a link with the data currently available.

The ePSB galaxies are the most difficult class in which to understand the origin of the strong Balmer lines. In \citet{WildGroves2011} we obtained Spitzer IRS spectra for 11 dusty Balmer-strong galaxies and found that dust-correcting the optical emission lines using the measured $H\alpha/H\beta$ Balmer decrement and a standard attenuation law led to emission line strengths that were entirely consistent with those measured in the mid-IR. We concluded that the Balmer decrement is a reliable estimate of the dust attenuation in Balmer-strong galaxy spectra when it is measurable. The ``normal" Balmer decrement distribution matching the star-forming control sample, and normal continuum colours of the ePSB class compared to the very red continuum of the dPSB class both suggest that only a minority of the ePSB galaxies can be dust-enshrouded star-forming galaxies. This can be seen in the stacked spectra presented in Figure \ref{fig:SED_stack1},  but to better illustrate the difference in the continuum colors of the ePSB and dPSB galaxies we present their optical and mid-IR colours in Figure \ref{fig:gi_w1w2} as well as some individual examples of spectra in Figure \ref{fig:sedfits_dpsb}. The mid-infrared colours of the dPSB galaxies clearly indicate that they have higher dust content compared with the other samples. However, the dPSB sample is just an arbitrary cut to identify the extreme dustiest of the ePSB class (with the adopted cut dPSB make up $35\%$ and $55\%$ of the ePSB galaxies, respectively). It is possible that this cut should be placed lower, and some of the ePSB class are dusty star-forming galaxies rather than true post-starburst galaxies. Indeed, we find that over half of the dustiest ePSB galaxies coincide with the lowest measured fractions of recently formed stellar mass in our spectral analysis, although this may be indicative of the limitations of spectral fitting models.

As well as their very broad distribution of $n$, covering both star-forming and quiescent control samples, the ePSB galaxies also show on average a higher fraction of stellar mass formed in the recent past than the star-forming control (Figures \ref{fig:massfrac_lowM} and \ref{fig:massfrac_highM}, Table \ref{tab:massfrac}), and a higher fraction have post-merger features (Figure \ref{fig:morph_AsAs90} and Table \ref{tab:morph}). Again, this suggests that at least a fraction are truly post-starburst, but exactly what fraction is difficult to determine. Conservatively taking only the tail of objects for which our spectral analysis identifies a high fraction of stellar mass formed in the last 1-1.5\,Gyr compared to the star-forming control, we conclude that $\gtrsim20\%$ of the low-mass ePSBs and $\gtrsim50\%$ of the high-mass PSBs are truly post-starburst galaxies. This will exclude objects with smaller starbursts that we are unable to constrain with the spectral analysis. To obtain a less conservative estimate, we can combine the spectral and structural information. While almost all of the ePSBs have pre-burst star formation histories that are consistent with the star-forming control sample, $53\%$ of the low mass and $100\%$ of the high-mass ePSBs have $n>2$ indicative of a spheroid. We can speculate that these galaxies have undergone a violent event leading to a burst of star formation and increase in light concentration through growth of a bulge, or the dominant progenitor was already spheroidal. This is also supported by the higher fraction with post-merger features than seen in the star-forming control sample.\emph{Overall, we find  only $20\%$ of the low-mass ePSB sample and none of the high-mass ePSB sample where there is no evidence of either a structural change, recent significant enhancement in star-formation or asymmetric tidal features (i.e. $f_{M1}<0.1$ and $f_{M15}<0.2$ and $n<2$ and $A_{S}<0.2$), compared with $61\%$ and $28\%$ of the low- and high-mass star-forming control samples, respectively.} 

Finally, using the merger simulations described in Section \ref{disc:mergers} and Appendix \ref{sec:mergers}, we find further evidence supporting the notion that the ePSB class are truly transitioning post-starburst galaxies with declining levels of star-formation. We created mock spectra of the merging galaxies and combined this with the ongoing SFR to estimate the equivalent width of the $H\alpha$ emission line. We found that the timescale of visibility of the ePSB phase reaches at least $0.7-1.3$\,Gyr following the starburst\footnote{In 8 out of 9 cases this estimate is limited by the length of the simulation and it is likely that the ePSB phase will prevail for longer. In the single case where the ePSB phase clearly ends before the end of the simulation, the timescale of visibility is  $\sim$0.8\,Gyr.}, which is roughly the characteristic age of our qPSB galaxies. Naturally, this is a very crude estimate and investigating a larger suite of simulations run for a longer time, that include a wide range of progenitor characteristics and orbital configurations is required to constrain this quantity further. Nevertheless, it is encouraging that all simulations considered in this work show that a merger-induced starburst is followed by the ePSB transition phase during which the star-formation rate in the merger remnant declines over $\sim1$\,Gyr - a timescale long enough to be observed at the characteristic age of the traditional quiescent post-starburst galaxies.

Determining how many ePSB galaxies are truly transitioning to the red-sequence is also difficult. One possibility is that the ePSB galaxies have experienced a more recent burst than the qPSB/agnPSB galaxies, and will subsequently become qPSBs before entering the red-sequence. The morphologies of 53\% of the low-mass and 100\% of the high mass ePSB galaxies are consistent with this scenario. The STARLIGHT results do indicate that the ePSBs have younger bursts than the qPSB/agnPSB galaxies on average (Table \ref{tab:massfrac}): $f_{M1}$ and $f_{M15}$ are high in $\sim20\%$ of low-mass ePSBs, compared to $f_{M15}$ being high in $\sim50\%$ of qPSB/agnPSBs and $f_{M1}$ high in only $\sim30\%$. However, the aliasing seen in the star formation histories of the star-forming control close to the expected age of the starburst prevents strong conclusions from being drawn. Probably a more profitable approach would be through the measurement of the gas contents of the different classes of Balmer-strong galaxies, to investigate the amount of fuel remaining for future star formation (e.g. \citealt{Rowlands+2015, French+2015}).



\subsection{Pathways through the post-starburst phase}\label{disc:pathways}

\begin{figure*}
\centering
  \includegraphics[scale=0.45]{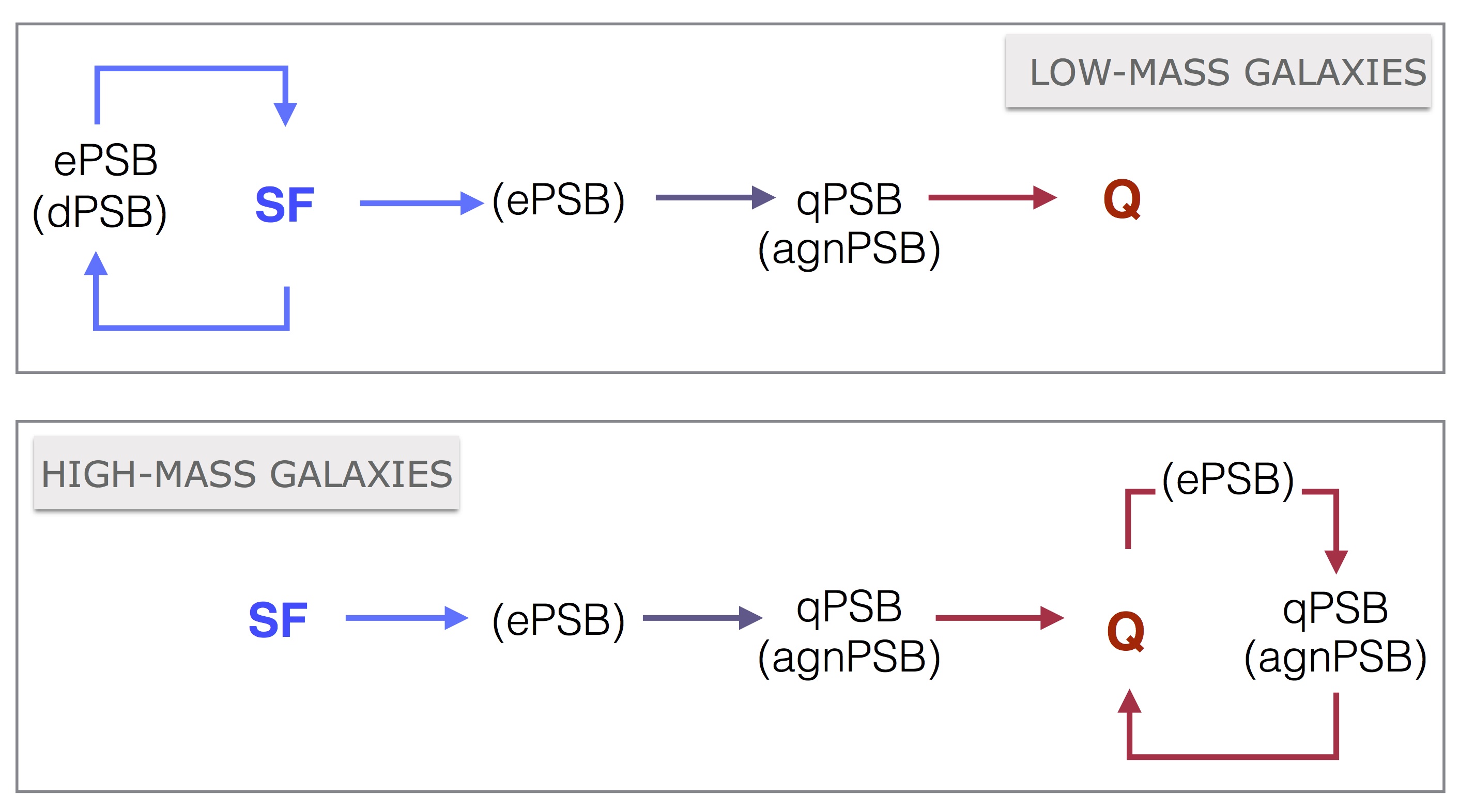}
\caption{A schematic representation of the pathways to the post-starburst phase in the low- and high-mass regimes. Post-starburst galaxies are often a transition phase between the blue cloud and the red sequence. Additionally, the low-mass and high-mass post-starburst galaxies can represent a phase in a cyclic evolution within the blue cloud and the red sequence, respectively. See Sections \ref{sec:disc_pathways_lowM} and \ref{sec:disc_pathways_highM} for more details.}
\label{fig:psbevol}
\end{figure*}

With an insight to the physical characteristics of the different spectral classes of Balmer-strong galaxies we can build a picture of the different evolutionary pathways of galaxies that can lead to a post-starburst phase. We summarise our conclusions in Figure \ref{fig:psbevol}. In this section we do not discuss the dPSB class further, as we concluded in the previous sub-section that the strong Balmer lines were likely due to dust/star geometry rather than recent star formation history. As we are unable to separate the ePSB galaxies that are most likely post-starburst from those that could be interlopers due to dust distribution, we include them here together, but note that there may be some contaminants in this class. 

 \subsubsection{Low-mass regime}\label{sec:disc_pathways_lowM}

 
{\it The pre-burst star formation histories and environments of the low-mass qPSB, agnPSB and ePSB galaxies are entirely consistent with those of the star-forming control galaxies, implying that they all originated from star-forming progenitor galaxies.} The structure of $83\%$ qPSB/agnPSB and $53\%$ of the ePSB galaxies is consistent with the quiescent control sample, supporting the scenario that $\sim70$\% of low-mass post-starburst galaxies are a transition phase between star-forming spirals and quiescent spheroids. We cannot directly conclude from this study that the starburst and morphological transition were concurrent, although it does not seem plausible that only galaxies that are already spheroidal undergo starbursts strong enough to produce strong Balmer lines. A more plausible scenario is that the same event that triggers the starburst causes the morphological transition.  Overall, a higher fraction of the post-starburst galaxies show post-merger signatures than the control samples, although our merger simulations concur that the merger likely occurred too far in the past for signatures to be visible in the majority of the SDSS images. Most of the Balmer-strong galaxies occupy intermediate-density environments (i.e. loose groups), where the dynamical conditions are favourable for galaxy mergers to occur. The remaining ePSB galaxies have a low central concentration of light, characteristic of disks, which suggests they may be returning onto the blue-cloud to continue forming stars in an ordinary fashion. 


Based on the above characteristics of the different samples, we propose the following two evolutionary scenarios that lead to the post-starburst phase in the low-mass regime:
\begin{itemize}
\item Due to some violent triggering event, a gas-rich star-forming disk galaxy experiences a strong starburst and morphological transition, likely followed by strong AGN activity as the galaxy evolves to become a quiescent spheroid. Whether a galaxy passes through an ePSB phase before reaching the qPSB/agnPSB phase may be determined by the conditions of the triggering event (e.g. orbital configurations in the case of a merger), as well as the properties of the progenitors, such as their structure or their gas content. The limited set of merger simulations considered in this work suggest that the ePSB phase should be visible for $\gtrsim$0.7\,Gyr following the initial starburst, irrespective of the progenitor morphology and orbital configurations, but a larger variety of simulations is required to better constrain this value.
\item A less violent mechanism leads to a starburst that fades more gradually through the ePSB phase, and may or may not result in morphological transition. Depending on the remaining gas reservoir, the galaxy will either return to the star-forming blue cloud, or become a qPSB/agnPSB and subsequently join the red sequence.
 \end{itemize}
If the triggering of the starburst occurs via galaxy mergers, then the strength of the starburst and the associated post-starburst evolutionary pathway may plausibly be determined by the conditions of the interaction (e.g. more major mergers leading to stronger starburst, more rapid shut down in star formation and a more significant morphological transformation). 
 

\subsubsection{High-mass regime}\label{sec:disc_pathways_highM}

Due to the limited sample sizes, the origins of the high-mass post-starburst galaxies are harder to constrain; however, as our analysis points to some contrasting characteristics compared with their low-mass counterparts, it is worth speculating on their inferred evolutionary pathways.
 
Contrary to the low-mass counterparts, all qPSB, agnPSB and ePSB tend to have high central concentration, meaning that they could represent different stages of the same evolutionary pathway to the red sequence.  All high-mass post-starburst samples have high fractions of morphologically disturbed galaxies, reaching much higher values in the samples of qPSB and agnPSB, again in contrast with what was observed in the low-mass regime. A striking difference between the low- and high-mass qPSB lies in their pre-burst star-formation histories, which very much resemble those characteristic of the quiescent control sample. In the case of the high-mass agnPSB and ePSB, there is a roughly equal split among their star-formation histories into those that resemble the quiescent population and those that look more like the star-forming galaxies. Further investigation shows that the ePSB with likely star-forming progenitors tend to have higher fractions of recently formed mass and more often disturbed morphologies, compared with those whose star-formation histories imply a quiescent progenitor. The pre-burst star formation histories lead to the suggestion that 40\% of high-mass post-starburst galaxies are rejuvenation events of an already quiescent elliptical galaxy. Another difference between the low and high-mass samples is in the fractions of stellar mass formed in the recent past, which are typically much lower for the high-mass post-starburst galaxies. This is again consistent with minor-mergers or small rejuvenation events rather than major gas-rich mergers. This makes sense, as it is statistically less likely for two high-mass gas-rich galaxies to merge in the local Universe than two low-mass galaxies or a low and high-mass galaxy. 

Based on the above findings we suggest that high-mass galaxies can reach the post-starburst phase through the following two pathways:
\begin{itemize}
\item An already quiescent galaxy experiences a relatively weak starburst through a minor merger with a lower mass gas-rich galaxy, after which the galaxy goes through a brief post-starburst phase only to return back onto the red sequence. Depending on the merger conditions, the ePSB phase may be visible before the galaxy attains the qPSB/agnPSB characteristics. 
\item Due to a violent triggering event, a star-forming spiral galaxy experiences a morphological transition and starburst which fades as the galaxy evolves to the qPSB/agnPSB phase, possibly through an ePSB phase, to reach the red sequence.
\end{itemize}




\section{Summary}

The evolution of galaxies is a complex multistage process, with post-starburst galaxies playing various roles throughout its different stages and channels. In this work we show that the majority of post-starburst galaxies are consistent with representing a transition phase between the blue cloud and the red sequence, following a merger-induced starburst. The quenching of star-formation may be relatively slow, and is likely to be accompanied by flickering AGN activity.
However, we find that this evolutionary pathway is not the only one that can lead to the post-starburst phase.
In the low-mass regime, post-starburst galaxies also represent a cyclic evolution of galaxies within the blue cloud, in which the stochasticity of star formation leads to a weaker burst, with no morphological transformation or complete quenching.  
On the other hand, at higher masses, post-starburst galaxies also occur in the cyclic evolution of galaxies within the red sequence, when an already quiescent galaxy gradually builds up its stellar mass through small starbursts. Such episodes could perhaps be resulting from mergers with small gas-rich companions, moving the galaxy towards the high-mass end of the red sequence. 


Although in this paper we tried to be exhaustive with the currently available data, it is clear that upcoming and future datasets will have a dramatic impact on our understanding of galaxies transitioning between the blue cloud and red sequence. Of particular importance to the next stages of this work are data from IFU surveys, which will help ascertain whether the galaxies in our sample are nuclear or global post-starbursts, (see e.g. \citealt[e.g.][]{Pracy+2014}) and whether global post-starburst galaxies are more likely to be transitioning, whereas nuclear post-starburst galaxies are more likely to be cyclical events \citep{Rowlands+2015}. Resolved spectroscopy may also reveal whether our class of post-starburst galaxies with high ionisation lines are predominantly due to AGN or shocks. Finally, a detailed investigation of merger simulations will help place further constraints on the visibility of post-merger features in post-starburst galaxies, including its dependence on the  mass, gas content and structural properties of the progenitor galaxies as well as dynamical configuration of the merger.

\section{Acknowledgements}
We would like to thank Natalia Vale Asari, Gustavo Bruzual and Stephane Charlot for providing invaluable help with interpreting the STARLIGHT spectral fitting results and providing updated models. We also thank the referee for a thorough review of the manuscript and useful suggestions, which helped us clarify some important points and improve the presentation of our results.

MMP, VW, JM-A, NJ and KR acknowledge support of the European Research Council via the award of a starting grant (SEDMorph; P.I. V. Wild). LTA acknowledges support from the Iraqi Ministry of Higher Education and Scientific Research. NL acknowledges the support of the Jenny $\&$ Antti Wihuri Foundation. NL and PHJ acknowledge the support of the Academy of Finland project 274931. YZ acknowledges support of a China Scholarship Council – University of St Andrews Scholarship. WL acknowledges support from the ECOGAL project, grant agreement 291227, funded by the European Research Council under ERC-2011-ADG.

Funding for the SDSS and SDSS-II has been provided
by the Alfred P. Sloan Foundation, the Participating
Institutions, the National Science Foundation, the U.S. Department of Energy, the National Aeronautics and Space Administration, the Japanese Monbukagakusho, the Max Planck Society, and the Higher Education Funding Council for England. The SDSS Web Site is http://www.sdss.org/.
The SDSS is managed by the Astrophysical Research
Consortium for the Participating Institutions. The Participating Institutions are the American Museum of Natural History, Astrophysical Institute Potsdam, University of Basel, University of Cambridge, Case Western Reserve University, University of Chicago, Drexel University, Fermilab, the Institute for Advanced Study, the Japan Participation Group, Johns Hopkins University, the Joint Institute for
Nuclear Astrophysics, the Kavli Institute for Particle Astrophysics and Cosmology, the Korean Scientist Group, the Chinese Academy of Sciences (LAMOST), Los Alamos National Laboratory, the Max-Planck-Institute for Astronomy (MPIA), the Max-Planck-Institute for Astrophysics (MPA), New Mexico State University, Ohio State University, University of Pittsburgh, University of Portsmouth, Princeton University, the United States Naval Observatory, and the University of Washington.

This publication also makes use of data products from the Wide-field Infrared Survey Explorer, which is a joint project of the University of California, Los Angeles, and the Jet Propulsion Laboratory/California Institute of Technology, funded by the National Aeronautics and Space Administration.




\appendix

\begin{figure*}
  \centering
   \includegraphics[scale=0.82]{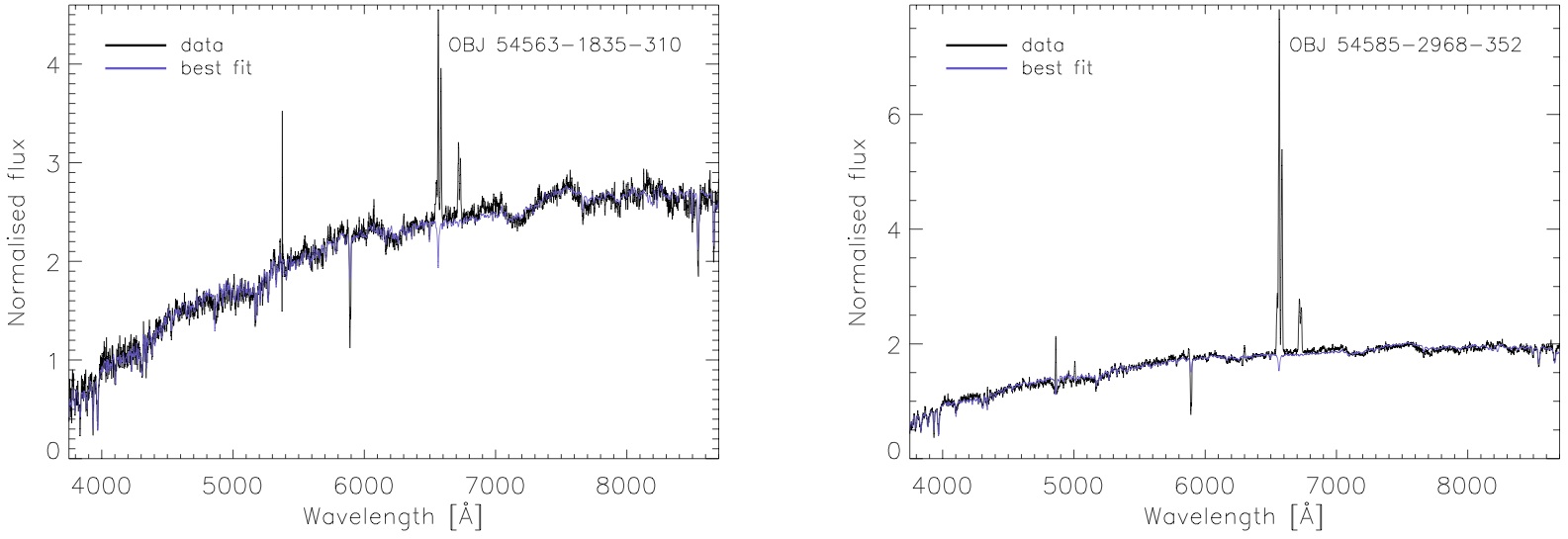}
      \includegraphics[scale=0.82]{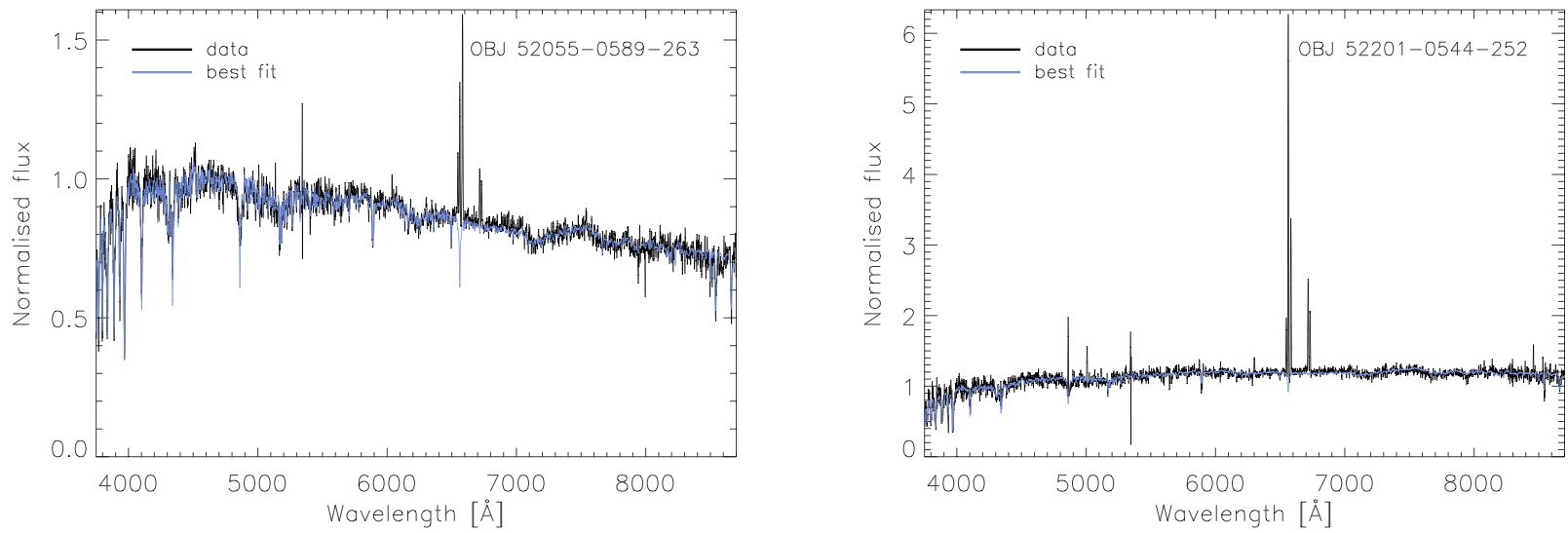}
\caption{SDSS spectra of selected Balmer-strong galaxies (black) with over-plotted best-fit STARLIGHT models (colour). The examples are representative of dPSB (top) and ePSB (bottom) samples and show the difference in the shape of the continuum between the two Balmer-strong galaxy classes clearly.}
\label{fig:sedfits_dpsb}
\end{figure*}

\begin{figure*}
 \includegraphics[scale=1.05]{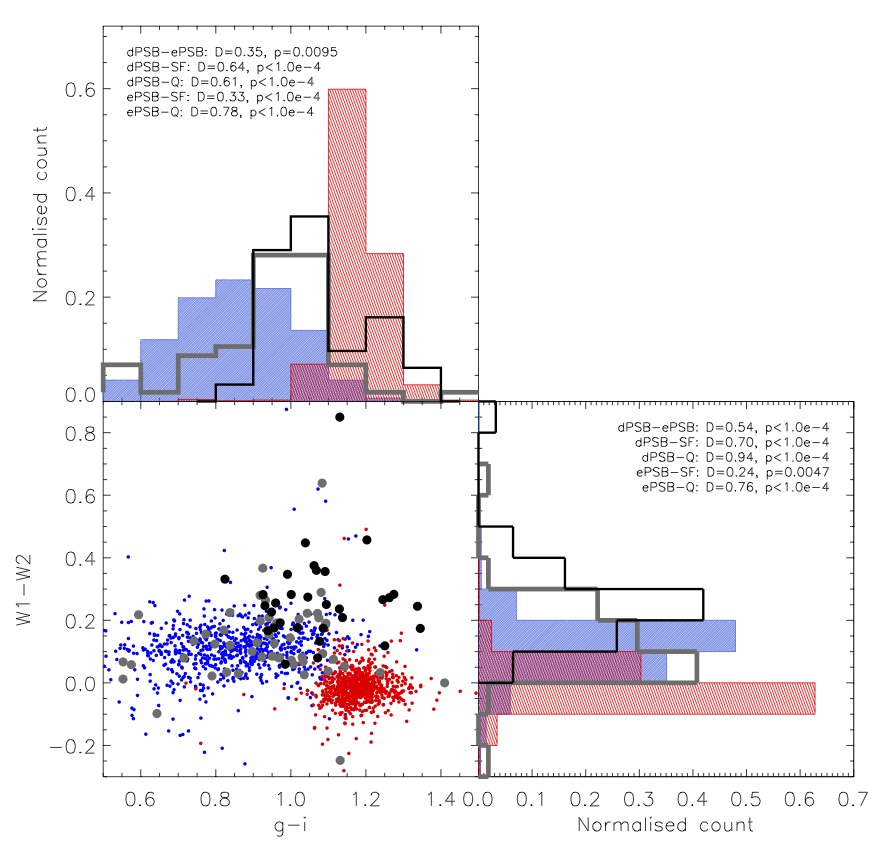}
 \includegraphics[scale=1.05]{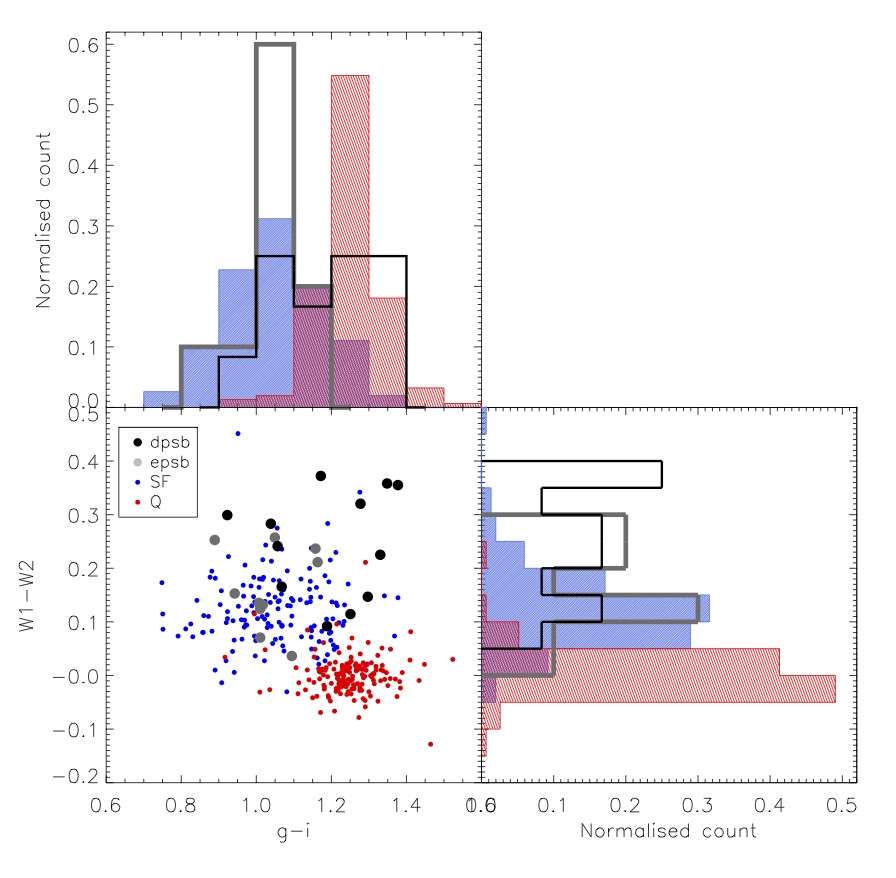}
\caption{Optical (SDSS $g-i$) and infra-red (WISE $W1-W2$) colours measured for the Balmer-strong dPSB and ePSB galaxies and a comparison with control samples.}
\label{fig:gi_w1w2}
\end{figure*}

\begin{figure*}
  \centering
  \includegraphics[scale=1]{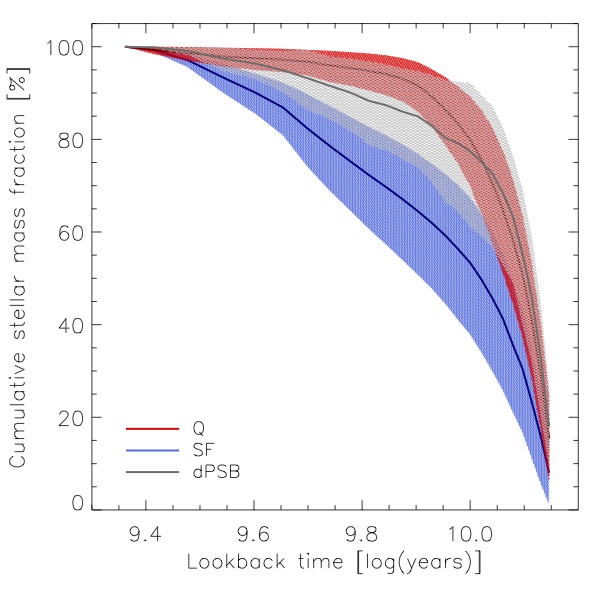}
    \includegraphics[scale=1]{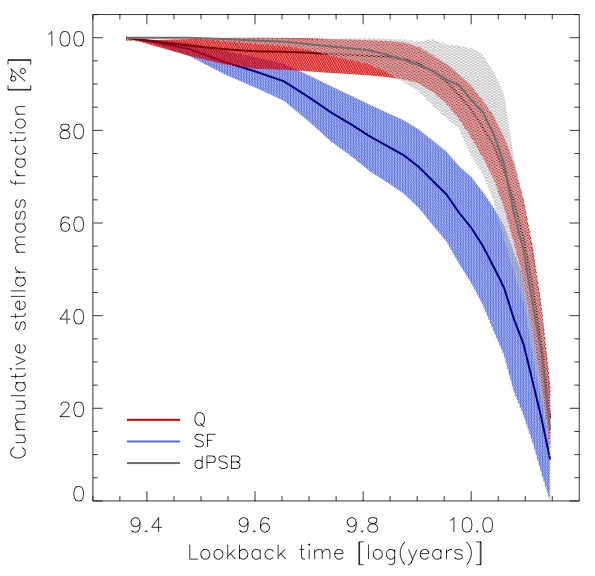}
\caption{The pre-burst star formation histories of the dusty Balmer-strong galaxies (dPSB), compared with control galaxy samples. The figures show the stacked cumulative pre-burst star-formation histories, normalised to unity at 2\,Gyr lookback time, for the low-mass (left) and high-mass (right) samples, with dPSB shown in grey, and the control star-formaing and quiescent galaxies in blue and red, respectively. For each sample, the solid lines represent the mean values and the shaded regions illustrate the spread of values within the sample (measured by the standard deviation from the mean).}
\label{fig:sfh_dpsb}
\end{figure*}

\section{Colours and star-formation histories of the dusty Balmer-strong galaxies}\label{app:dpsb}

\subsection{The $g-i$ and $W1-W2$ colours}\label{app:dpsb_colours}

In Section \ref{sec:discussion_edpsb} we discussed the differences between the ePSB and dPSB classes of the Balmer-strong galaxies. We argued that their discrepant natures are revealed in particular when considering their structural properties, such as the distributions of their S{\'e}rsic indices, as well as their optical colours. The redder colours of the dPSB compared with the ePSB can be inferred from their stacked spectra shown in Figure \ref{fig:SED_stack1}. Additionally, in Figure \ref{fig:sedfits_dpsb} we present examples of individual spectra of some dPSB and ePSB galaxies. These show the differences in the shapes of the continuum of the two classes more clearly, in part due to the linear scaling on the flux axis. 

In Figure \ref{fig:gi_w1w2} we present the photometric colours of the dPSB and ePSB galaxies. In addition to SDSS $g-i$ colour we extend the analysis into the mid-infrared regime, using data from the WISE survey \citep{Wright+2010}. To obtain the W1 and W2 magnitudes (measured in regions centred on 3.4 and 4.6 $\mu$m, respectively) for our Balmer-strong galaxies we used the matched catalogue of 400 million SDSS sources with WISE forced photometry \citep{Lang+2016}. The W1-W2 colour can be used to probe the dust emission in galaxies, and therefore distinguish between galaxies with different dust-content (see eg. \citealt{Nikutta+2014}). As expected, the SDSS $g-i$ colour places the Balmer-strong galaxies in between the star-forming and quiescent control-samples, in the so-called ``green-valley'' with the dPSB galaxies showing a clear tendency towards redder colours, compared with the ePSB galaxies. Moreover, the dPSB galaxies show a strong tendency to have redder $W1-W2$ colour compared with the other samples, which further supports their dusty nature. We note the impact of aperture bias here, which will lead to a difference between the photometric colours and spectroscopic shape and therefore spectral class. For example, the presence of a dust lane through the centre of the galaxy may cause a very red SDSS spectrum, and a ``dPSB'' classification, while the colours obtained from photometry may be more typical of the control samples. 

\subsection{The star-formation histories}\label{app:dpsb_sfh}

While discussing the properties of the dPSB galaxies we pointed out a difficulty with interpreting the star-formation histories, obtained from the spectral fitting for this class of galaxies (Section \ref{sec:res_sfh}). Consequently, we excluded these histories from our analysis in the main body of this paper and present them in Figure \ref{fig:sfh_dpsb}. It appears that the dPSB galaxies follow the history of star-formation characteristic of quiescent galaxies, in complete disagreement with their current star-formation activity, inferred based on their spectra featuring prominent emission-lines and Balmer absorption lines. This is likely a consequence of inaccurate spectral fits rather than a true property of the dPSB galaxies. The high dust content and/or its geometry in these objects might prevent the STARLIGHT code from obtaining reliable fits, leading to unrealistic results. Figure \ref{fig:sedfits_dpsb} shows examples of fits to the spectra of the dPSB and ePSB galaxies. Overall, we found that the residuals associated with the fits to the dPSB spectra are about twice as large as those from the fits to the other Balmer-strong galaxy classes.

\section{The impact of the AGN on the structural parameters}\label{appendix:AGN}

A bright centrally located source within a galaxy, such as an AGN, may significantly affect the measurements of some structural parameters.
The extent of this effect will depend on the type of the AGN, the morphology of the host galaxy, the wavelength at which the measurement is performed and the nature of the measurement itself.
\citet{Pierce+2010} investigated the most extreme case and found that the measurements of the concentration index, $C$ and the $M_{20}$ parameter for quasar-host galaxies are almost certainly unreliable if the quasar contributes at least $20\%$ of the total galaxy light in the B-band. They found that the S{\'e}rsic and Gini indices ($n$, $G$) can also be affected by the emission from the quasar but to a lesser extent, and that the asymmetry parameter, $A$, remains reasonably unaffected, unless the quasar is offset from the galaxy centre. 

In our study we consider only narrow-line AGN, in which the continuum emission is not visible. For such galaxies, the effects on the structural parameters can only be due to strong narrow emission lines, and are therefore much less significant. \citet{Kauffmann+2007} showed that nearby LINERS and Type II Seyfert galaxies are less likely to suffer significant contamination from the nuclear regions, although their study focused on quantifying the UV-optical colours rather than galaxy structure.
Moreover, in our analysis focus primarily on the SDSS $r$-band images, in which we the continuum emission from the AGN should be less dominant, compared with the B-band images considered by \citet{Pierce+2010}. 
We therefore limit ourselves to the following simple check. 
 
\begin{figure*}
  \centering
  \includegraphics[scale=0.67]{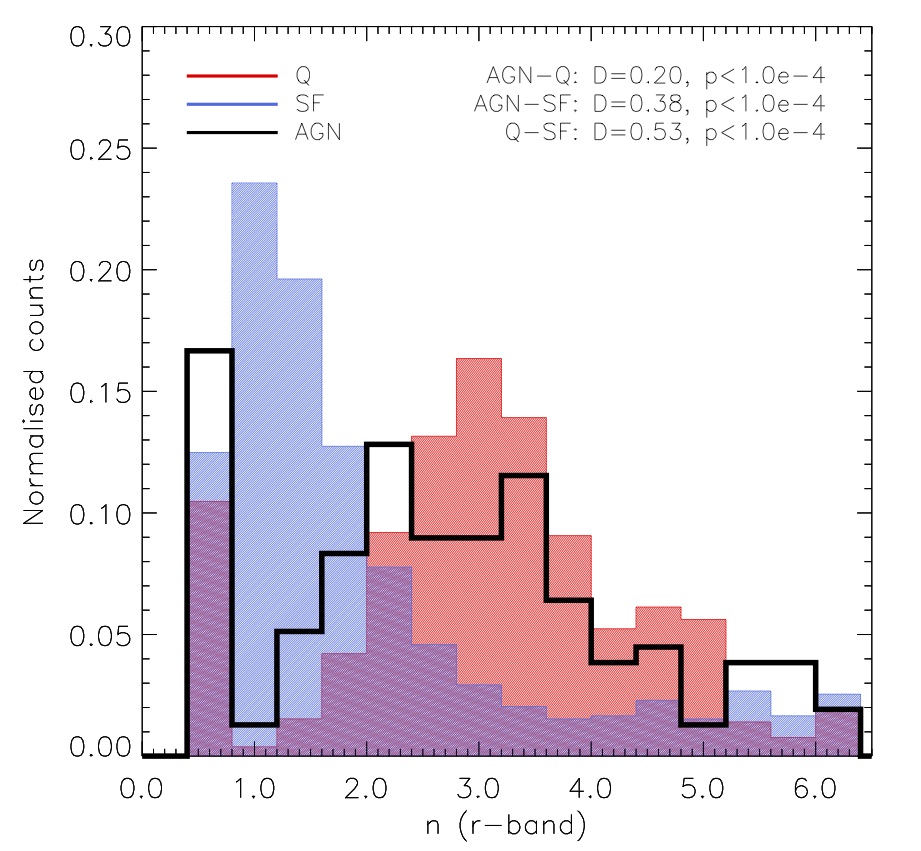}
    \includegraphics[scale=0.67]{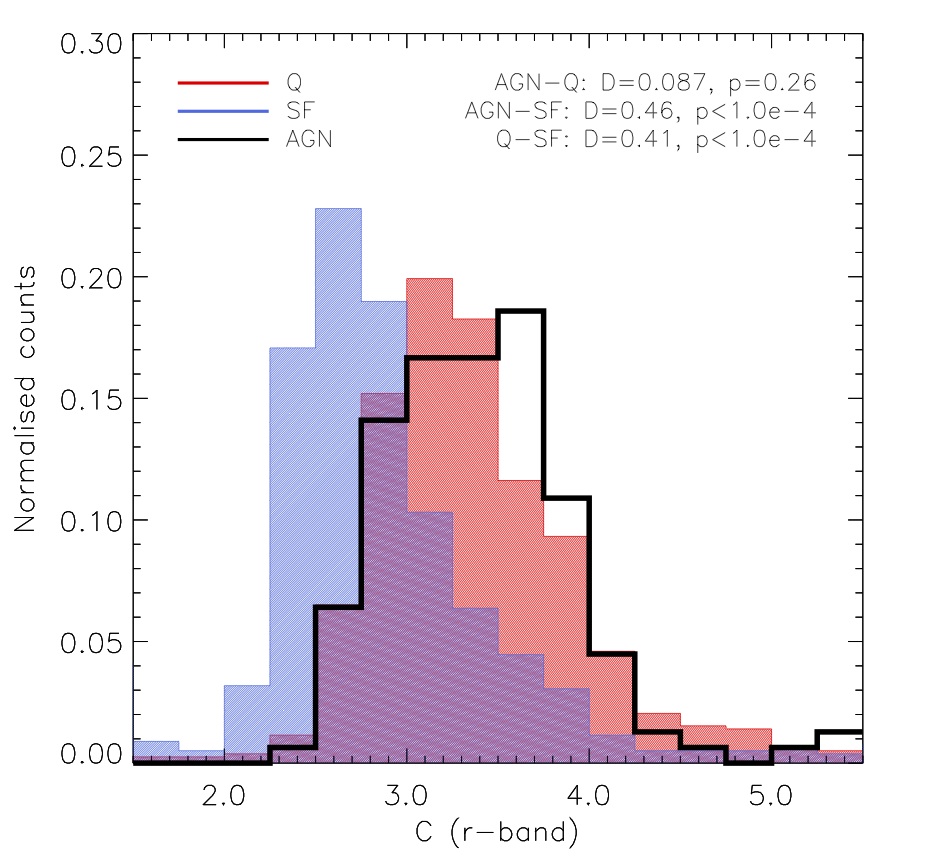}
    \includegraphics[scale=0.67]{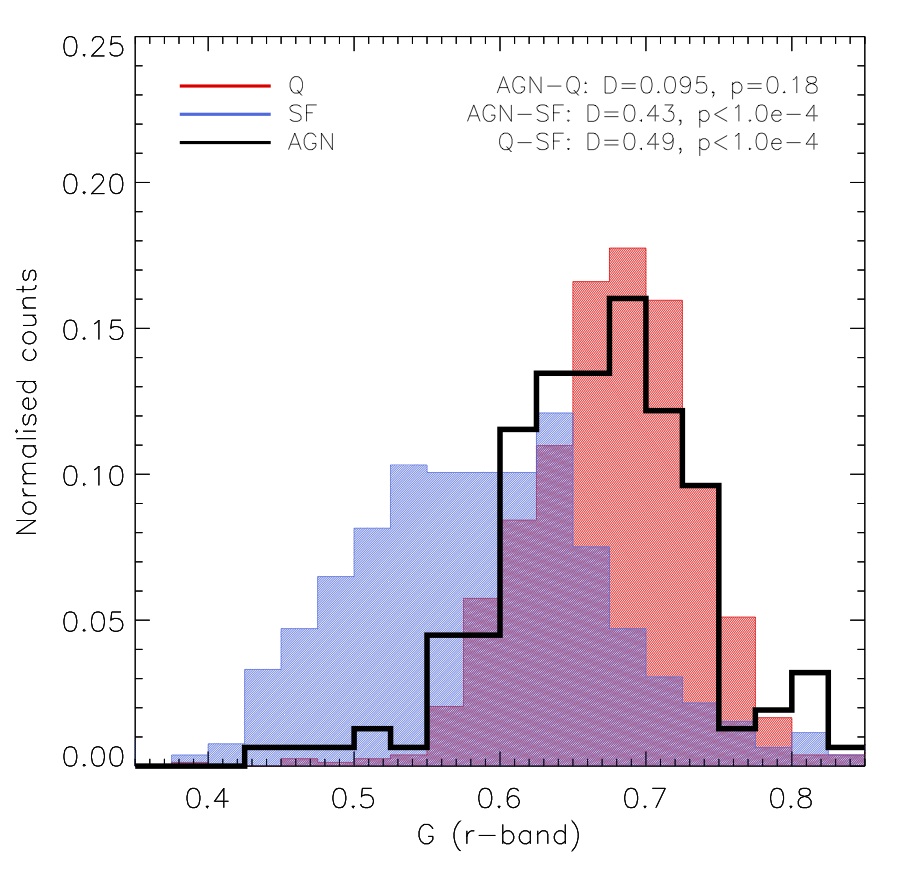}
    \includegraphics[scale=0.67]{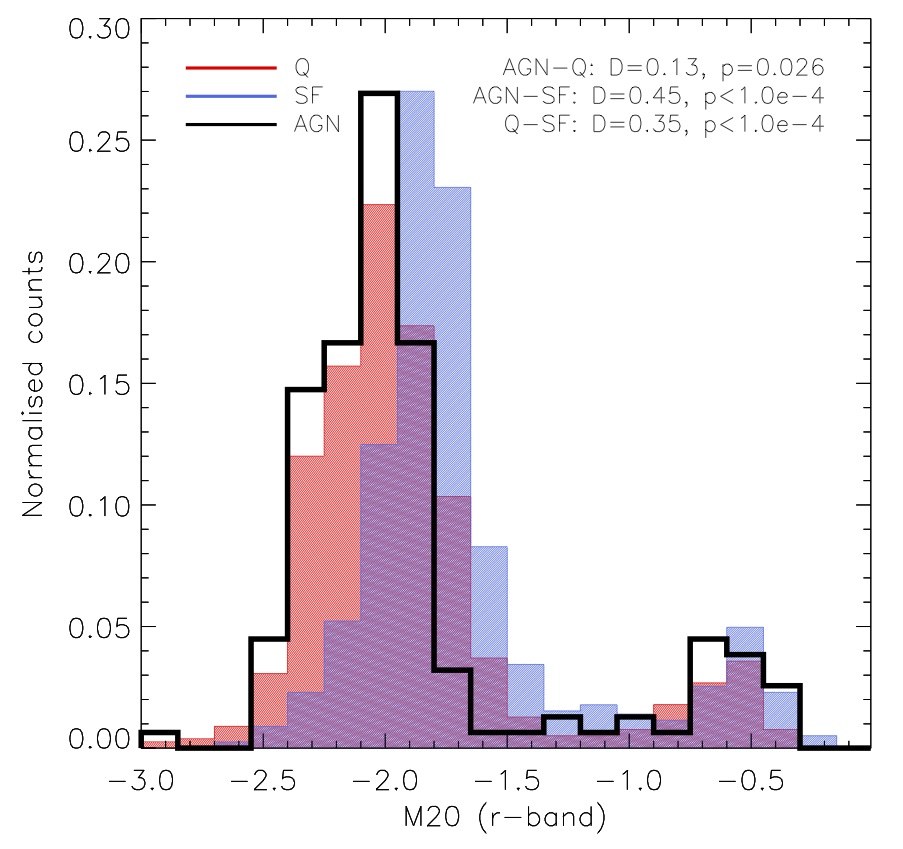}
     \includegraphics[scale=0.67]{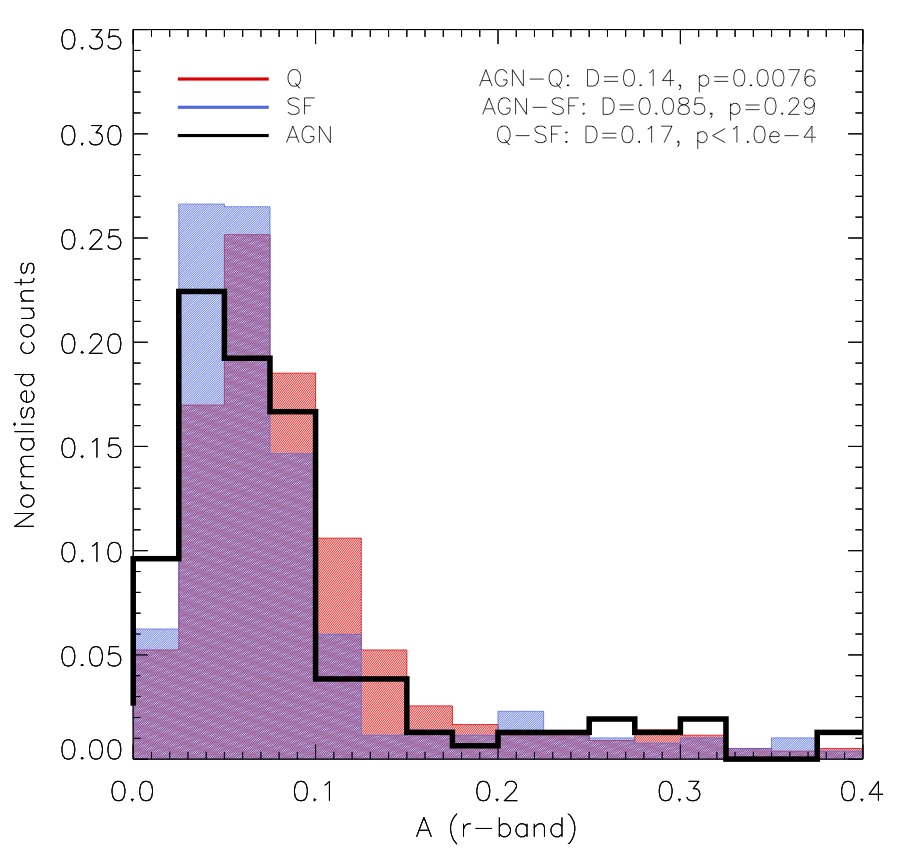}
\caption{A comparison of the $r$-band measurements of the S{\'e}rsic index ($n$), the concentration index ($C$), the Gini index ($G$), the $M_{20}$ statistics and the light-weighted asymmetry parameter ($A$) for quiescent galaxies with and without the presence of a narrow-line AGN (black and red), and star-forming galaxies (blue).}
\label{fig:AGN_testmorph}
\end{figure*}

In Figure \ref{fig:AGN_testmorph} we compare the values of $n$, $C$, $A$, $G$ and $M_{20}$ measured in the $r$-band for a sample of quiescent galaxies hosting an AGN, selected using the same emission-line ratio cuts as for the agnPSB sample (Section \ref{sec:psbselection}), with our control sample of quiescent galaxies. We also show the star-forming galaxy sample to better visualize the dynamic range of the measured parameters. 

There is no significant difference between the quiescent AGN and non-AGN samples in terms of their $r$-band measured values of $C$, $G$, and $M_{20}$ but both are significantly different from the star-forming sample. The  KS test results indicate the distributions of $n$ and $A$ are different, however in both cases there is no obvious tendency for the AGN sample to display either enhanced or diminished values of either of the measures, compared with the quiescent non-AGN sample.
We conclude that the $r$-band measured structural parameters are not significantly affected by the emission from narrow-line AGN and can therefore be used for our purpose of comparison between galaxies which display such nuclear activity and those that do not.

\section{Additional results from the image analysis of post-starburst galaxies}\label{appendix:struct}

\begin{figure*}
  \centering
  \includegraphics[scale=0.77]{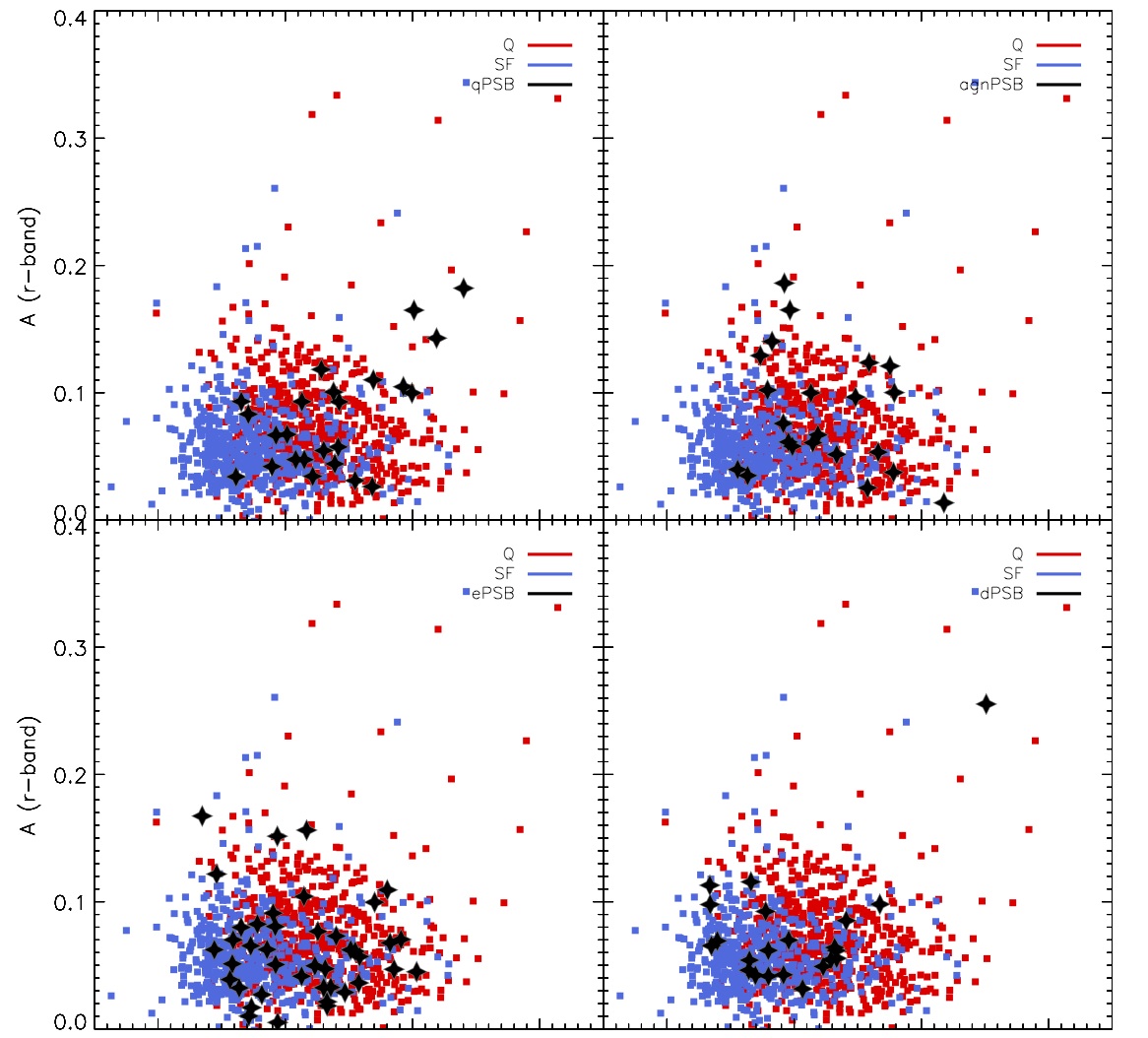}
    \includegraphics[scale=0.77]{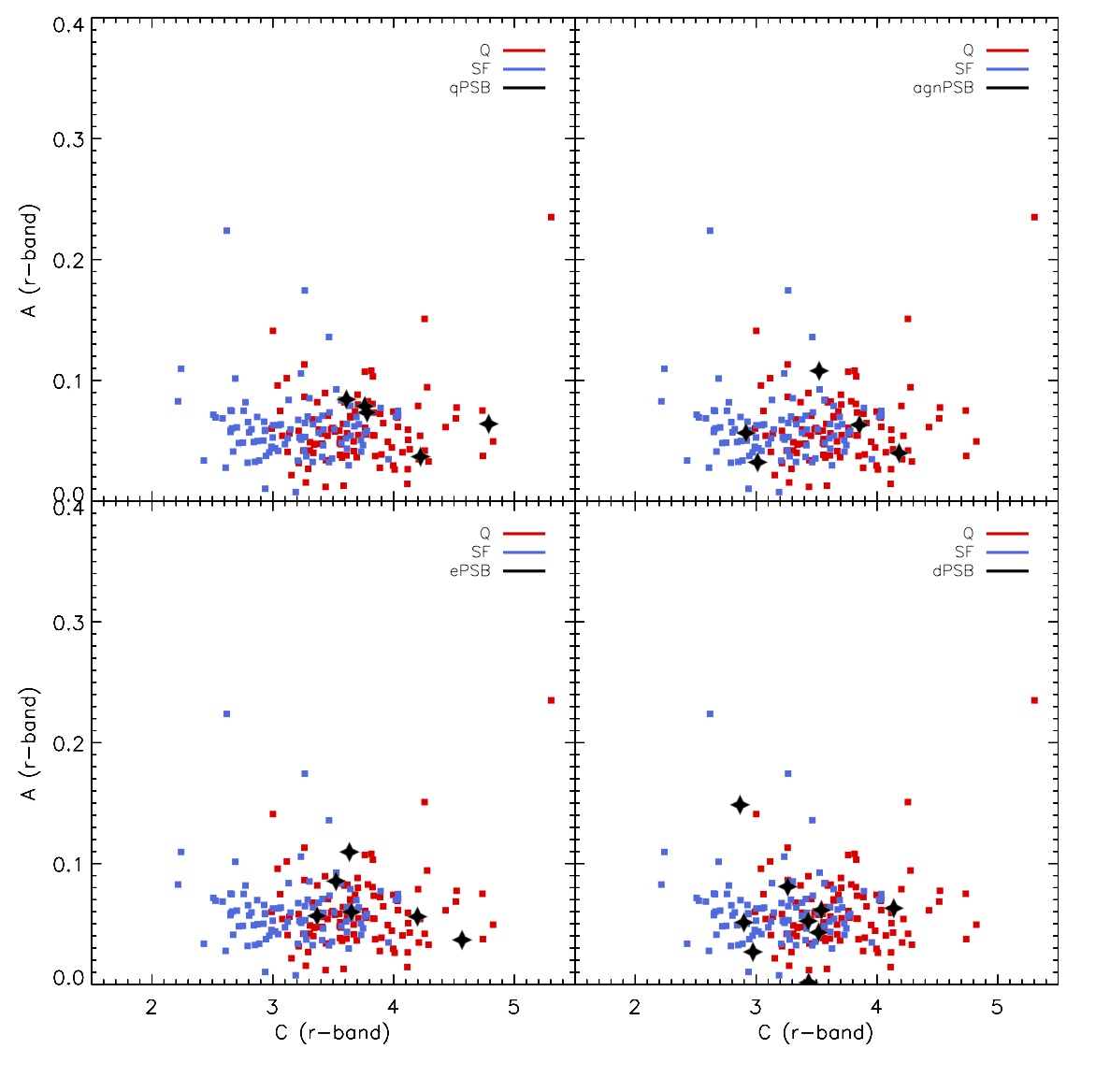}
\caption{The concentration index, $C$ versus the asymmetry parameter, $A$, for the post-starburst galaxies (qPSB, agnPSB, ePSB, dPSB) compared with the control samples of star-forming (SF) and quiescent (Q) galaxies. Both parameters were measured in the $r$-band. The top and bottom panel show results for the low-mass and high-mass samples, respectively.}
\label{fig:morph_CA}
\end{figure*}

\begin{figure*}
  \centering
  \includegraphics[scale=0.77]{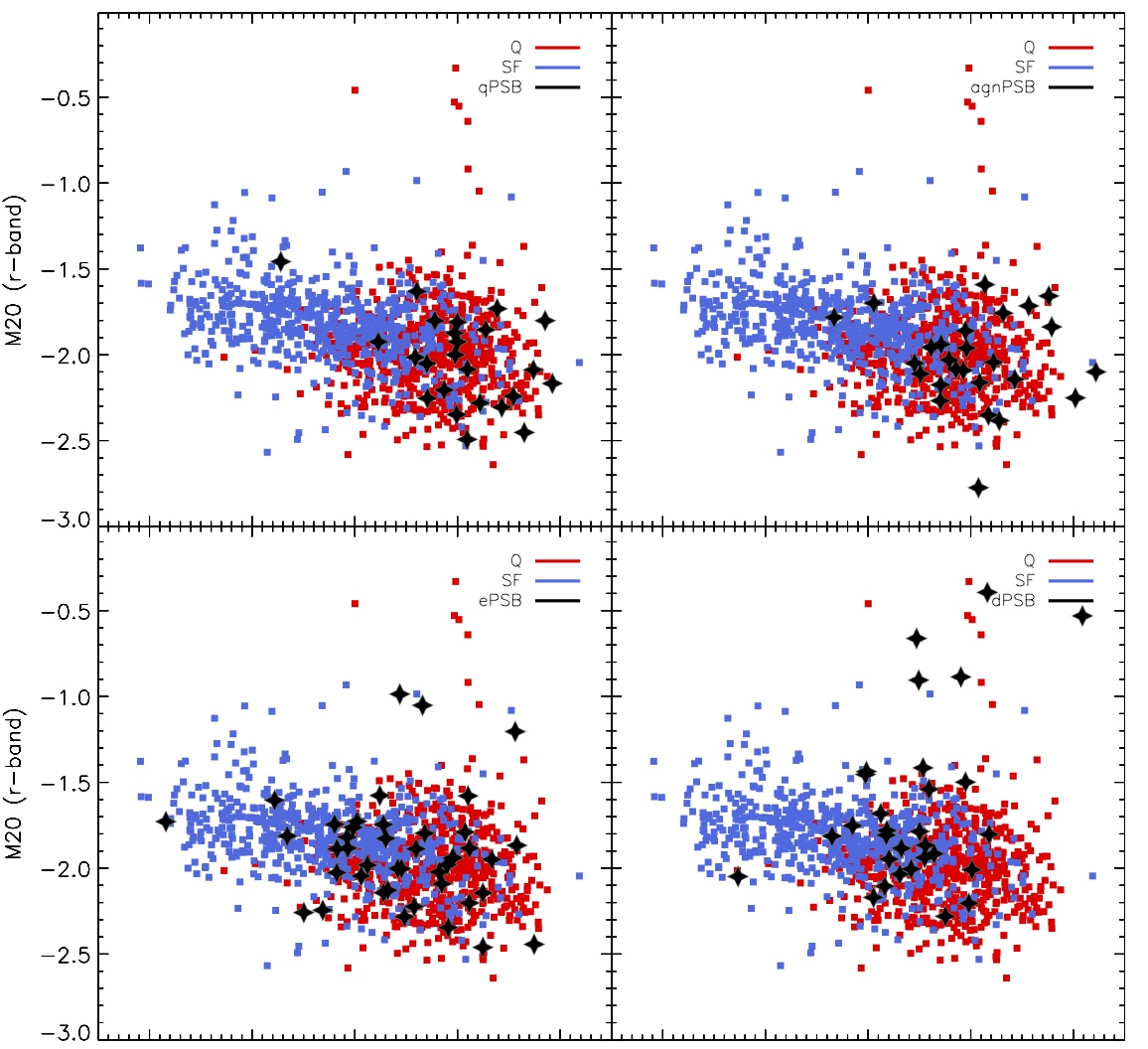}
    \includegraphics[scale=0.77]{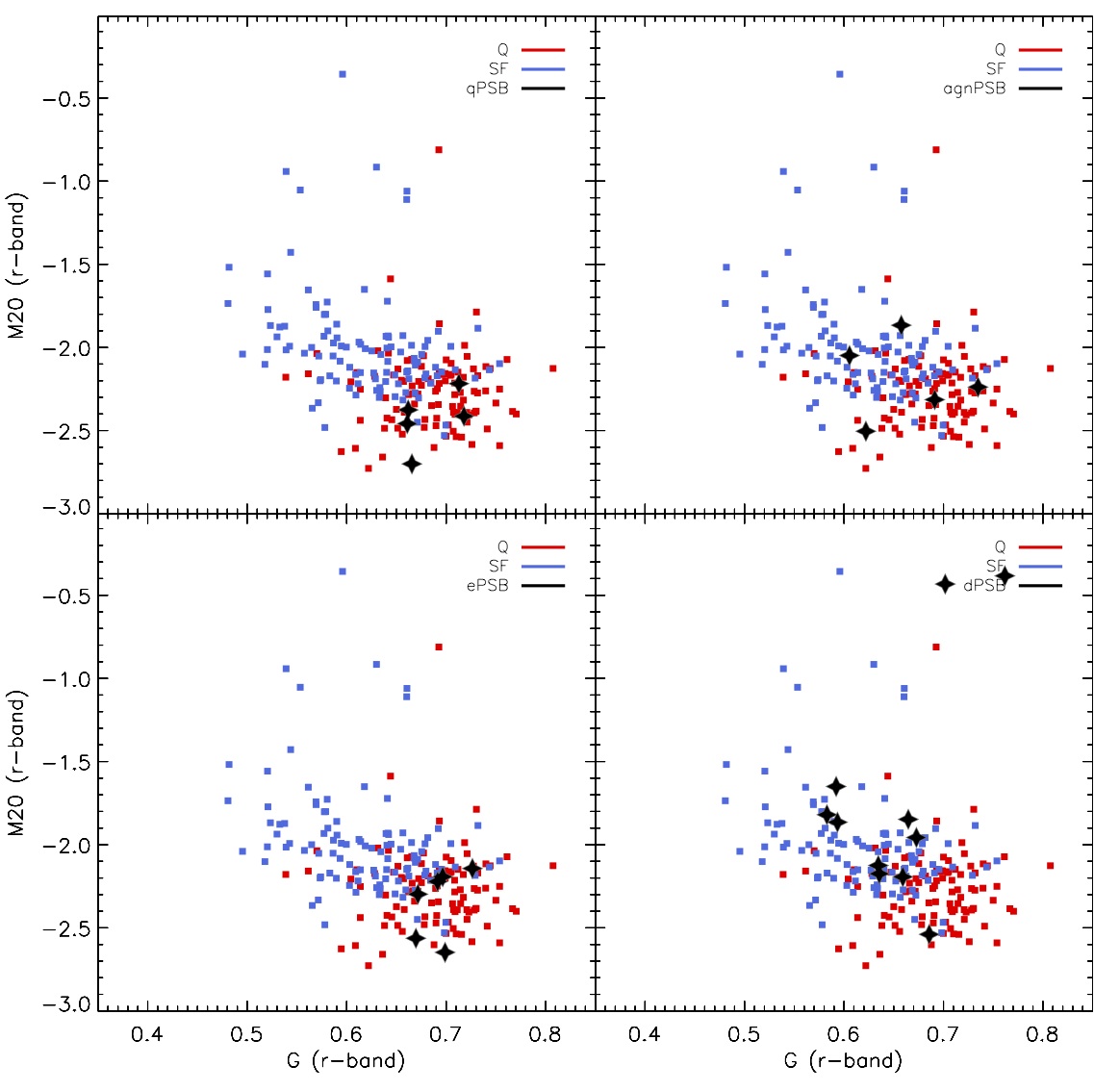}
\caption{The $r$-band $G-M_{20}$ relation for the post-starburst galaxies (qPSB, agnPSB, ePSB, dPSB) compared with the control samples of star-forming (SF) and quiescent (Q) galaxies. The top and bottom panel show results for the low-mass and high-mass samples, respectively.}
\label{fig:morph_GM20}
\end{figure*}

In Figures \ref{fig:morph_CA} and \ref{fig:morph_GM20} we show two further relations between the different morphological measures that we studied, ($C-A$ and $G-M_{20}$, respectively). These relations are often used in quantitative analysis of galaxy morphology and structure, and can help relate between the properties inferred for the post-starburst galaxies and those characteristic of the control star-forming and quiescent samples.

\begin{figure*}
  \centering
  \includegraphics[scale=0.75]{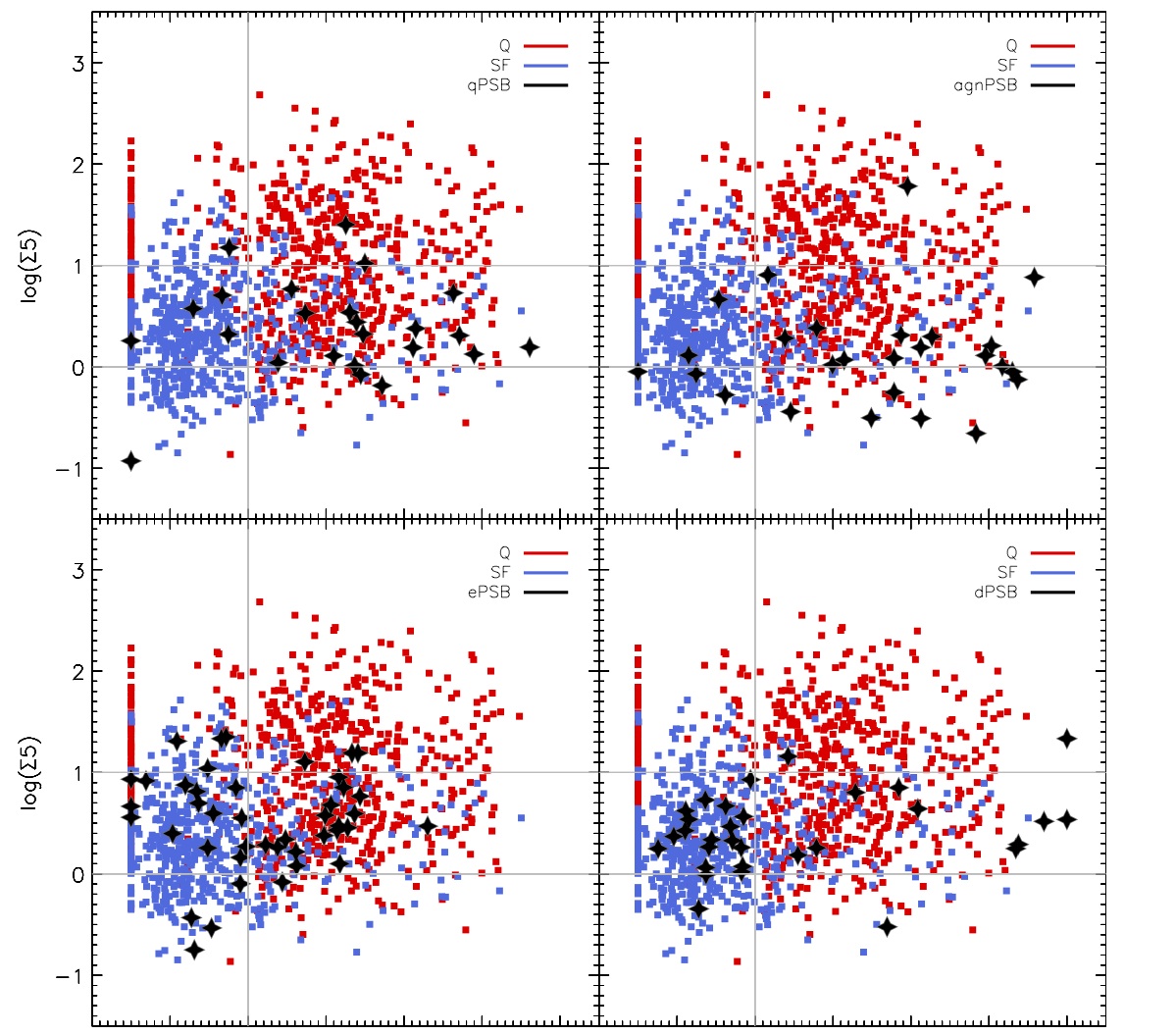}
    \includegraphics[scale=0.75]{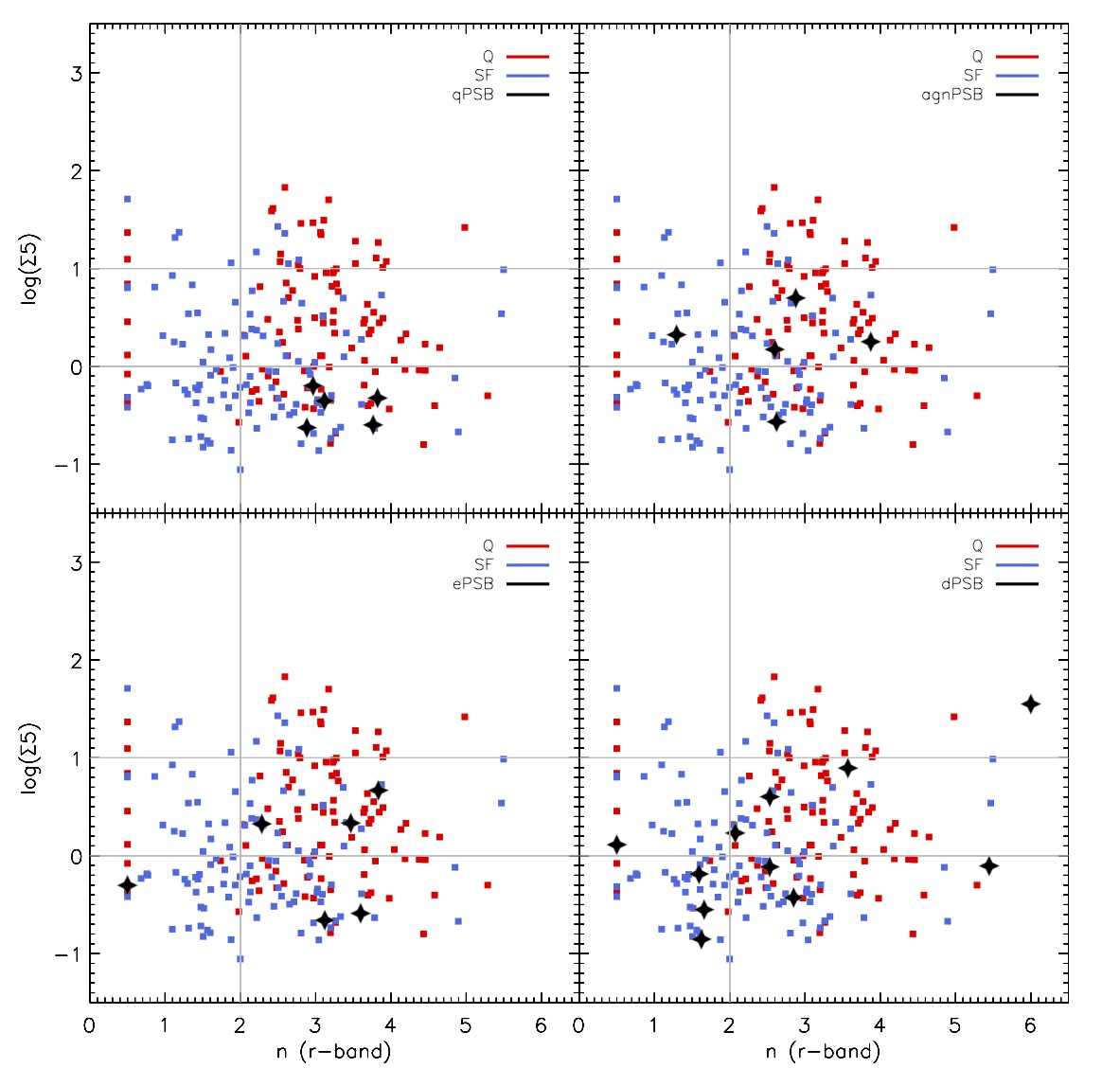}
\caption{The projected number density, log($\Sigma_{5}$) versus the $r$-band S{\'e}rsic index, $n$,  for the post-starburst galaxies (qPSB, agnPSB, ePSB, dPSB) compared with the control samples of star-forming (SF) and quiescent (Q) galaxies. The top and bottom panels show results for the low-mass and high-mass samples, respectively. The vertical line marks the value of $n=2$ used in the analysis to separate between galaxies with steep and shallow light profiles. The horizontal lines show the values of $\Sigma5$ used to differentiate between low-, medium- and high-density environments. }
\label{fig:morph_nE5}
\end{figure*}

In Figure \ref{fig:morph_nE5} we show that there is no trend between morphology, as measured by the S{\'e}rsic index, and environment for the PSB samples.  

\section{Simulated post-mergers}\label{sec:mergers}

The merger simulations used in this work are based on the N-body smoothed particle hydrodynamics (SPH) code \gadget\ \citep{Springel2005}, with improved SPH implementation - SPHGal \citep{Hu+2014,Eisenreich+2017}. 
The improved model includes the use of a pressure-entropy formulation of SPH, a Wendland $C^4$ kernel with $100$ neighbours along with an updated velocity gradient estimator, a modified viscosity switch with a strong limiter, artificial conduction of thermal energy and finally a time step limiter (see \citealt{Lahen+2017} for details).
The cooling of the gas is metallicity-dependent \citep{Wiersma+2009} and takes into account an UV/X-ray background \citep{HaardtMadau2001}. In addition, the sub-resolution astrophysics models include models for star formation, stellar feedback with accompanying metal production, and metal diffusion \citep{Scannapieco+2005,Scannapieco+2006,Aumer+2013}. The stellar metal yields are adopted from \citet{Iwamoto+1999} for SNIa, \citet{WoosleyWeaver1995} for SNII and \citet{Karakas2010} for AGB stars, with both energy and metals being released at rates dependent on the age of the stellar particles and the distance to the neighbouring gas particles (see e.g. \citealt{Nunez+2017}).

\subsection{Galaxy models}\label{sec:mergers_models}

The galaxies were modelled according with the $\Lambda$CDM cosmology ($\Omega_{m}$ = 0.30, $\Omega_{\Lambda}$ = 0.71 and H$_{0}$ =70kms$^{-1}$Mpc$^{-1}$), to resemble the galaxies observed in the local Universe. The physical properties of the galaxies were derived from the parameters describing their host dark matter halos, including the virial mass and velocity ($v_{vir}=160$ km/s, $ M_{vir}=v_{vir}^3/10GH_{0} = 1.34 \times10^{12}M\sun$), with each galaxy consisting of a Hernquist dark matter (DM) halo with mass a $M_\mathrm{DM}=1.286\times 10^{12} \ M_{\sun}$ and a concentration parameter of $c=9$ \citep{Hernquist1990}.
The baryonic mass fraction was set at $m_\mathrm{b}=0.041$ and
each galaxy was set to have a two-component structure, including a bulge and a disk, characterised by an exponential light profile. The scale lengths of the components were determined from the conservation of the angular momentum of the system (with the spin-parameter set to $\lambda=0.033$; for details see \citealt{Springel+2005,Johansson+2009a,Johansson+2009b}). 

In order to produce systems representative of the local star-forming population we chose three types of morphologies, including Sa, Sc and Sd galaxies, characterised by the following parameters:
\begin{itemize}
\item Sa: $B/T=0.5$, $r_{disc}=3.75$ kpc, $f_{gas}=0.17$
\item Sc: $B/T=0.3$, $r_{disc}=3.79$ kpc, $f_{gas}=0.22$
\item Sd: $B/T=0.1$, $r_{disc}=3.85$ kpc, $f_{gas}=0.31$
\end{itemize}
where $B/T$ is the stellar bulge-to-total mass ratio, $r_{disk}$ is the stellar disk scale length and $f_{gas}$ is the disc gas fraction.
In each case the bulge scale length and the stellar disk scale height are to equal to $0.2\times r_{disc}$. 
The gaseous discs were set to be in hydrostatic equilibrium \citep{Springel+2005} with gaseous disc scale lengths that equal the stellar disc scale lengths.

Each galaxy consists of $4\times 10^5$ DM particles with a mass resolution per particle of $m_\mathrm{DM}\approx 3.2\times 10^6 \ M_{\sun}$ and $4\times 10^5$ baryonic particles 
divided according to the mass fraction of each component, as to obtain a mass resolution of $m_\mathrm{b}\approx 1.4 \times 10^5 \ M_{\sun}$ per baryonic particle.

To be fully consistent with the employed sub-resolution models, we have adopted initial metallicity and age distributions
for the gaseous and stellar particles in a fashion similar to the one presented in \citet{Lahen+2017}. Here we adopted a Milky Way-like metallicity gradient of $0.0585$ dex$/$kpc \citep{Zatitsky+1994} and abundances as in \citep{Adelman+1993, Kilian-Montenbruck+1994}, 
yielding roughly solar total metallicities for all galaxies. The initial star formation rates were set iteratively
as SFR$_\mathrm{Sa}=1 \ M_{\sun}/$yr, SFR$_\mathrm{Sc}=2 \ M_{\sun}/$yr and SFR$_\mathrm{Sd}=5 \ M_{\sun}/$yr which define,
together with the simulation start time, the initial ages of the stellar particles to be used in stellar feedback within \gadget.
Additionally, as the galaxies include up to $\sim30\%$ of the disc mass in gas, the galaxies were relaxed by running each galaxy in isolation for $500$ Myr before setting up the actual merger.

\subsection{Major merger simulations}\label{sec:mergers_sims}

The merger simulations were set up set up on collisional trajectories with the following initial orbital configurations, with $i_{1}$ and $i_{2}$ being the angles of inclination of the galactic disks with respect to the orbital planes and $\omega_{1}$ and $\omega_{2}$, the arguments of the orbits' pericentres\footnote{The angular distance between the pericentre and the ascending node, defined as the point of intersection of the plane of the disk with that of the orbit, where the rotation proceeds from `south' to `north' with respect to the orbital plane.} (see \citealt{NaabBurkert2003}):
\begin{itemize}
\item G00 - a symmetric configuration - in-plane prograde-prograde orbits with the angular momenta of the galaxies aligned with the orbital angular momentum
($i_{1}=0^{o}$,$i_{2}=0^{o}$,$\omega_{1}=0^{o}$,$\omega_{2}=0^{o}$);
\item G07 - retrograde-prograde orbits, with both galactic disks inclined with respect to the orbital planes ($i_{1}=-109^{o}$,$i_{2}=71^{o}$,$\omega_{1}=-60^{o}$,$\omega_{2}=-30^{o}$)
\item G13 - retrograde-retrograde orbits with one galaxy inclined and other coplanar with respect to the orbital plane
($i_{1}=-109^{o}$,$i_{2}=180^{o}$,$\omega_{1}=60^{o}$,$\omega_{2}=0^{o}$).
\end{itemize}
Each simulation started at an initial separation given by the mean of the virial radii of the progenitors (160 kpc/h $\sim$225 kpc) and the pericentric separation being the sum of the disc scale lengths ($\sim$2$\times$2.7 kpc/h $\sim$2$\times$3.8 kpc). The galaxies approached each other following nearly parabolic orbits and, in each case, the evolution of the interacting system was modelled along the the whole merger sequence including the pre-merger stages and the post-coalescence evolution of the remnant (total time of 3 Gyr).

\subsection{The mock image and spectral synthesis}\label{appendix:mergers_img}

From the positions, ages and star formation histories of the gas and star particles recorded every 0.02 Gyr we created mock images and integrated spectra of the galaxies that could be compared with the real SDSS imaging and spectroscopic data. Using stellar population synthesis models \citep{BruzualCharlot2003} we assigned a spectral energy distribution (SED) to each particle, based on the information about its age and past star formation history. For the images, these were convolved with the SDSS filter response functions. A simple two-component dust-screen model was used to mimic the attenuation effects from the interstellar medium, assuming an effective optical depth $\tau_V=1.0$ and power law slopes of 0.7 and 1.3 for the diffuse and birth cloud dust respectively (see \citealt{daCuhna+2008} and \citealt{Wild+2007} for details).

For each time step, we projected particle positions onto a two-dimensional grid, matching the SDSS image resolution (with 0.396 arc seconds per pixel) with a distance to the interacting galaxies corresponding to $z=0.04$, as viewed from six different directions characterised by angles of rotation in a three-dimensional coordinate system: $\alpha$ - around the $x$-axis, $\beta$ - around the $y$-axis and $\gamma$ - around the $z$-axis.
They include a face-on orientation ($0$,$0$,$0$), a randomly chosen orientation ($\alpha$,$\beta$,$\gamma$) with each angle between 0 and 2$\pi$, and further five  following orientations: ($\alpha$+$\pi$/2,$\beta$,$\gamma$), ($\alpha$+$\pi$,$\beta$,$\gamma$), ($\alpha$+$3\pi$/2,$\beta$,$\gamma$), ($\alpha$,$\beta$+$\pi$/2,$\gamma$) and ($\alpha$,$\beta$+3$\pi$/2,$\gamma$). 

An image in a given passband and orientation was created by summing up the luminosities of the particles within each pixel, integrated from the SED convolved with a given filter function. The luminosities were converted to the flux at a distance corresponding to $z=0.04$, calculated within the same cosmological framework as that assumed in the simulations.
The flux, $f$, was calibrated to the AB magnitude system, where the zero-point flux density in a given passband is equal to $f_{0}=3631$ Jansky:
\begin{equation}
m_{AB} = -2.5\times \mbox{log}_{10}(f/f_{0})
\end{equation}
The magnitudes were then converted to SDSS counts using the \emph{asinh} magnitude definition \citep{Lupton+1999}:
\begin{equation}
\mbox{counts} = t_{exp}\times f/f_{0}\times 10^{(-0.4(m_{ZP} + kA))},
\end{equation}
\begin{equation}
f/f_{0} = 2b \times \mbox{sinh}\left(-m_{AB}\left(\mbox{ln}(10)/2.5\right)-\mbox{ln}(b)\right),
\end{equation}
where $b$ is the `softening parameter' set to $\sim1\sigma$ of the sky noise, $t_{exp}$ is the exposure time, $m_{ZP}$ is the photometric zero-point in the given passband, $k$ is the extinction coefficient and $A$, the airmass at the given position. More information about the photometric calibration can be found at \url{http://classic.sdss.org/dr7/algorithms/fluxcal.html}.

The obtained idealised images were then convolved with a PSF, built by summing two Gaussian functions with widths and relative maxima matching those characteristic for SDSS images, to mimic the effects of astronomical seeing and camera response on the images. The image synthesis was finalised by addition of Gaussian random noise, with the standard deviation matching the typical error in the photoelectron counts in the SDSS images.

The instantaneous star formation rate output by Gadget was used to estimate the $H\alpha$ line luminosity using the conversion of \citet{K98}, and the 
integrated spectra provided an estimate of the stellar continuum in order to calculate the equivalent width of the $H\alpha$ emission line. To account for the selective extinction by dust of the lines over the stellar continua, the equivalent widths were reduced by $\sim2$.




\bibliographystyle{mn2e} 
\bibliography{ref}   

\label{lastpage}

\bsp	
\end{document}